\documentclass[onecolumn,floats,floatfix,showpacs,prd,superscriptaddress,nofootinbib]{revtex4-1}
\usepackage[utf8]{inputenc}
\usepackage{graphicx,epsfig}
\usepackage{lmodern}
\usepackage{amsmath,amssymb,amsthm}
\usepackage{mathrsfs}
\usepackage{amsfonts}
\usepackage{empheq}
\usepackage{bm}
\usepackage[caption=false]{subfig}
\usepackage[inline]{enumitem}
\usepackage{soul,ulem}
\usepackage{xcolor}
\usepackage{url}
\usepackage{hyperref}
\usepackage{appendix}
\usepackage{lipsum}
\usepackage{xspace}

\hypersetup{
    colorlinks=true,
    linkcolor=blue,
    filecolor=magenta,      
    urlcolor=cyan,
}
\input{CQGformat.input}

\def\RGB{\mathcal{R}_{\rm GB}}

\begin{document}
\title{
Post-Newtonian Gravitational and Scalar Waves in Scalar-Gauss-Bonnet Gravity
}
\author{ Banafsheh Shiralilou$^{1,2}$, Tanja Hinderer$^{1,3}$, Samaya M. Nissanke$^{1,2}$, N\'estor Ortiz$^4$, Helvi Witek$^5$}
\address{$^1$ GRAPPA, Anton Pannekoek Institute for Astronomy and Institute of High-Energy Physics, University of Amsterdam, Science Park 904, 1098 XH Amsterdam, The Netherlands}
\address{$^2$ Nikhef, Science Park 105, 1098 XG Amsterdam, The Netherlands}
\address{$^3$ Institute for Theoretical Physics,
Utrecht University, Princetonplein 5, 3584 CC Utrecht, The Netherlands}
\address{$^4$ Instituto de Ciencias Nucleares (ICN), Universidad Nacional Autónoma de México (UNAM), Circuito Exterior C.U., A.P. 70-543, México D.F. 04510, México.}
\address{$^5$ Illinois  Center  for  Advanced  Studies  of  the  Universe \& Department of Physics, University of Illinois at Urbana-Champaign, Urbana, Illinois 61801, USA}

\begin{abstract}
    Gravitational waves emitted by black hole binary inspiral and mergers enable unprecedented strong-field tests of gravity, requiring accurate theoretical modelling of the expected signals in extensions of General Relativity.
    In this paper we model the gravitational wave emission of inspiralling binaries in
    scalar Gauss-Bonnet gravity theories.
    Going beyond the weak-coupling approximation, we derive the gravitational waveform to first post-Newtonian order beyond the quadrupole approximation and calculate new contributions from nonlinear curvature terms.
    We quantify the effect of these terms and provide ready-to-implement gravitational wave and scalar waveforms as well as the Fourier domain phase for quasi-circular binaries. We also perform a parameter space study, which indicates that the values of black hole scalar charges play a crucial role in the detectability of deviation from General Relativity.
    We also compare the scalar waveforms to 
     numerical relativity simulations to assess the impact of the relativistic corrections to the scalar radiation.
    Our results provide important foundations for future precision tests of gravity.
\end{abstract}
\maketitle
\tableofcontents
\section{Introduction}\label{sec:Intro}
The breakthrough discovery of gravitational waves (GWs) emitted by merging black holes (BHs)~\cite{Abbott:2016blz} has opened a unique new avenue for probing unexplored regimes of the universe.
GWs have already provided unprecedented insights into the origin and population of BHs, including observations closing the mass-gaps~\cite{LIGOScientific:2018mvr, Abbott:2020niy, Abbott:2020khf}. They also hold the key to test gravity in its strong-field, nonlinear regime~\cite{Yunes:2013dva,
Berti:2015itd,Yunes:2016jcc,Yagi:2016jml,
LIGOScientific:2019fpa,Abbott:2020jks,Carson:2020rea, TheLIGOScientific:2016src}.

The GW measurements of BH binaries rely on accurate template waveforms that cover the inspiral, merger and ringdown of compact binaries.
Significant recent progress has been made on constructing accurate templates in General Relativity (GR). However, testing the underlying theory of gravity and searching for signatures of quantum gravity requires theory-specific waveforms beyond GR.
Therefore, GW-based tests of gravity have focused mostly on parameterized or null tests against GR~\cite{Yunes:2016jcc,LIGOScientific:2019fpa,Abbott:2020jks,TheLIGOScientific:2016src,Isi:2019aib}. These tests are performed only on single  coefficients associated to a scaling with frequency~\cite{Yunes:2009ke,Will:2014kxa,LIGOScientific:2019fpa,Abbott:2020jks}
and thus, the interpretation of theoretical constraints remains limited~\citep{Yunes:2016jcc}.
Given the plethora of proposed gravity theories beyond GR, it is not feasible to compute accurate template waveforms for all of them. This makes it important to consider well-motivated classes of theories that capture broadly applicable features for the GW modeling. 

One of the most compelling beyond-GR class of theories is
scalar Gauss-Bonnet (sGB) gravity,
which extends GR by a dynamical scalar field non-minimally coupled to the Gauss-Bonnet~(GB) invariant $\RGB=R_{\mu\nu\sigma\rho}R^{\mu\nu\sigma\rho} - 4 R_{\mu\nu}R^{\mu\nu} + R^2$, with a coupling parameter $\alpha$.
sGB is particularly well motivated for several reasons
(i) it is inspired by  
the low energy limit of quantum gravity paradigms such as string theory after compactification~\cite{Boulware:1985wk,GROSS198741,Kanti:1995vq};
(ii) it can be derived from Lovelock gravity, 
the most general theory of gravity in $D$ spacetime dimensions that yields second order field equations,
via a dimensional reduction~\cite{Charmousis:2011bf,Charmousis:2011ea};
(iii) it corresponds to the lowest order in a series expansion in curvature terms and thus, it is representative of a more general class of higher derivative theories~\citep{PhysRevD.16.953};
and 
(iv) its field equations contain at most second derivatives of the metric, so sGB can potentially be made mathematically well-posed (as shown for weak couplings to the GB invariant in~\cite{Kovacs:2019jqj,Kovacs:2020ywu,Kovacs:2020pns}).

The coupling, $f(\phi)$, between the dynamical scalar field and the GB invariant determines the ``flavour'' of sGB gravity~\cite{Antoniou:2017acq,Antoniou:2017hxj}. For the sake of the summary we label them as ``type~I'' and ``type~II.'' 
In our notation, type~I sGB corresponds to the subclass of theories for which $f'(\phi)$ never vanishes. Representative examples include shift-symmetric ($f=2\phi$) or dilatonic ($f\sim e^{2\phi}$) couplings. In this case the space-time curvature always sources the scalar field and, thus, inevitably yields hairy black holes~\cite{Mignemi:1992nt,Kanti:1995vq,Sotiriou:2014pfa,Benkel:2016rlz,Benkel:2016kcq,Ripley:2019irj,Yunes:2011we,Pani:2009wy,Barausse:2015wia,Kleihaus:2011tg,Kleihaus:2015aje}
that can form dynamically~\cite{Benkel:2016kcq,Benkel:2016rlz,Ripley:2019aqj}.
Type~II sGB corresponds to the subclass of theories for which $f'(\phi)$ can vanish for some values of the scalar field. Examples include a quadratic or Gaussian coupling function. Then, the Schwarzschild and Kerr solutions of GR still exist, and they can spontaneously scalarize~\cite{Silva:2017uqg,Doneva:2017bvd,Berti:2020kgk,Silva:2020omi,Dima:2020yac,Herdeiro:2020wei,Berti:2020kgk,Collodel:2019kkx,Doneva:2021dqn,Doneva:2020nbb,Silva:2020omi,East:2021bqk, Annulli:2021lmn}.
Likewise, neutron stars can spontaneously scalarize in Type~II sGB gravity~\cite{Silva:2017uqg}.
Such scalarized black holes or neutron stars can form dynamically~\cite{Ripley:2019aqj,Kuan:2021lol}.

Type~I sGB gravity has been tested extensively against observations of low-mass x-ray binaries~\cite{Yagi:2012gp} and a Bayesian parameter estimation for GW detections~\cite{Nair:2019iur}. The most recent observational bounds of $\sqrt{\alpha}\lesssim1.7$\,km were obtained by analysing signals of LIGO/VIRGO's first two GW catalogs~\cite{Perkins:2021mhb,Wang:2021yll}. 
Type~II sGB gravity has, so far, remained unconstrained from observations.

To-date, modelling the dynamics of compact binaries in sGB gravity has mainly been restricted to the small coupling regime.
This approximation has enabled the first numerical relativity (NR) simulations of BH binaries with an effective-field theoretical treatment~\cite{Witek:2018dmd,Okounkova:2020rqw} or the decoupling limit~\cite{Silva:2020omi}. 
The numerical study of fully non-linear field equations for general couplings has recently  been advanced in the direction of initial data construction~\cite{Kovacs:2021lgk}, and numerical simulations~\cite{East:2020hgw}
through a modified generalized harmonic formulation of the evolution equations~\cite{Kovacs:2020ywu}. This has also enabled the recent study of spontaneous BH scalarization~\cite{East:2021bqk}. 
The ringdown spectrum has been studied extensively in Ref.~\cite{Blazquez-Salcedo:2016enn,Pierini:2021jxd}. However, the early inspiral waveform, typically first constructed with Post-Newtonian (PN) methods, has only been modelled at leading (Newtonian) order~\cite{Yagi:2011xp}. At this order, the dynamics and the radiation are affected only by the scalar field, and consequently the waveforms are missing the effect of the curvature nonlinearities. Recently, the Lagrangian for the matter dynamics has been computed to 1PN order, where the nonlinearities first appear~\cite{Julie:2019sab}.

In this work we compute, for the first time, analytical waveforms for the inspiral stage of the binary evolution with the effect of nonlinear higher curvature corrections.
We use the PN expansion with the direct integration of the relaxed Einstein equations (DIRE) approach, originally developed by Epstein and Wagoner \citep{EWpaper} 
and extended by Will, Wiseman and Pati  \cite{Will:1996zj, Pati:2000vt,Pati:2002ux}. 
The PN assumption of a gravitationally bound source in the weak-field, low-velocity regime $Gm/rc^2\approx v^2/c^2\ll1$, where $m,r$, and $v$ are the characteristic mass, size, and velocity of the source, enables asymptotic expansions of the solutions to the field equations with $1/c^2$ treated as a formal small expansion parameter. Each factor of $c^{-2}$ then corresponds to one PN order. In the DIRE approach to PN calculations,
the field equations are re-expressed in a \textit{relaxed} form~\cite{1975ctf..book.....L}, namely as an inhomogeneous, sourced wave equation for field perturbations, accompanied by a harmonic gauge condition that is imposed after solving the field equations. 
The equations are solved in two space-time regions, (i) The near-zone, where the separation between source and field point is less than the characteristic wavelength of the GWs; and (ii) The far zone at larger distances, see Fig.~\ref{fig:zones}. In both cases, the solutions are computed as retarded integrals over the past null cone of the field point. This integration method has been shown to provide non-divergent results in agreement with other approaches such as the multipolar post-Minkowskian formalism \citep{Blanchet:2013haa}. We note that this method has also been applied to massless scalar-tensor (ST) gravity theories---which admits scalarized neuron star solutions~\cite{Damour:1993hw}---providing the equation of motion for compact binaries to 3PN order \citep{Bernard:2018hta,Mirshekari:2013vb}, GWs to 2PN~\citep{Lang:2013fna,Sennett:2016klh}, and scalar waves to 1.5PN~\citep{Lang:2014osa} beyond the leading order.

Specifically, in this paper, we calculate scalar and tensor waves of sGB gravity to half and one relative PN-order, respectively. This is the methods paper to our shorter companion paper~\cite{Shiralilou:2020gah} in which we solely concentrated on the application of our results and assessed the detectability of GB deviations in the Fourier-domain GW phase.  In this paper, we present the explicit computations to produce the GW waveforms and phasing as well as the scalar waveform.
Our methods and results are applicable to arbitrary sGB coupling strengths that lie within the theoretical and observational bounds~\cite{Kanti:1995vq,Sotiriou:2014pfa}, as well as general couplings that have remained unconstrained. We also calculate the equations of motion of the sources to 1PN order, based on the results of Ref.~\cite{Julie:2019sab}. 
Having calculated the waveforms, we then focus on quasi-circular inspirals and calculate the energy flux carried off by radiation, as well as the GW phasing to which both the measurements and parametrized test of gravity such as~\cite{Tahura:2018zuq} are very sensitive. Our results include higher order strong-field effects than previously computed, which may mimic biases in fundamental source parameters of BH binaries when analysing with GR-only GW waveforms. We also show that the effect of the GB scalar is distinct from the scalar field of a ST theory due to the presence of explicit 1PN GB coupling dependent terms. This has consequences for interpreting GW signals from BH-neutron star binaries when analyzed by beyond-GR theories~\cite{Carson:2019fxr}. Further, we discuss the   importance of distinguishing different perspectives on the sGB action (i.e. as a phenomenological extension of GR or as resulting from string theory) when interpreting GW constraints as they correspond to different physical frames.

This paper is organized as follows. In Sec.~\ref{sec:II}, we present the field equations for tensor and scalar perturbations and define the matter action for self-gravitating binaries. We cast the field equations into relaxed form in Sec.~\ref{sec:III}, and outline the procedure for finding the solution at large distances from the source.  In Sec.~\ref{sec:IV}, we calculate the near-zone contribution to gravitational and scalar waves for binary systems. We defer the calculation of far-zone contribution to Appendix.~\ref{App:FZ} because, as we show, it does not contribute to the final waveform at 1PN order. In Sec.~\ref{sec:V} we  derive the equations of motion using the two-body Lagrangian from Ref.~\cite{Julie:2020vov}, and transform the waveforms to the center-of-mass (CM) frame. In Sec.~\ref{sec:VI}, we solve for the gravitational and scalar energy loss rate, from which we find the time domain and frequency domain waveforms in Sec.~\ref{sec:VII} and Sec.~\ref{sec8},
respectively. In Sec.~\ref{sec:VII}, we also compare our results for the scalar waveform to the numerical relativity waveforms of Ref.~\cite{Witek:2018dmd} computed in the small coupling approximation. Finally, in Sec.~\ref{Sec:IX}, we present the discussion and conclusions.

In this paper we use the standard notation for symmetrized and antisymmetrized indices, namely $x^{(i} y^{j)}$ denotes symmetrization and $x^{[i} y^{j]}$ anti-symmetrization. We use a multi-index notation for products of vector components:
$x^{i j k} \equiv x^{i} x^{j} x^{k}$, and a capital letter superscript denotes a product of that dimensionality: $x^{L} \equiv x^{k_{1}} x^{k_{2}} \cdots x^{k_l}$.
A subscript preceded by a comma such as $A,_{\mu}$ stands for the partial derivative $\partial_{\mu}A$, while $\nabla_{\mu}A$ indicates the covariant derivative associated to the metric. We label the two bodies in a binary system by $A$, $B$.

\section{Einstein-Scalar-Gauss-Bonnet gravity}\label{sec:II}
\subsection{Gravitational action and the field equations}
The gravitational action of sGB theory, in the presence of matter, is written in the Einstein frame as
\small
\begin{equation}\label{action}
    S=\frac{c^3}{16\pi G}\int_{M} d^4x\sqrt{-g}\left[R-2 (\nabla\phi)^2+\alpha f(\phi)\mathcal{R}^2_{GB}\right]
    +S_{m}[\Psi_m,\mathcal{A}^2(\phi)g_{\mu\nu}]\,,
\end{equation}
\normalsize
where $R$ is the Ricci scalar on the 4-dimensional manifold $M$ with the  metric $g_{\mu \nu}$ in the Einstein frame. $S_{m}$ is the matter action with $\Psi_m$ denoting the matter fields with a generic non-minimal coupling to the metric. The fundamental coupling constant $\alpha$  has dimensions of length squared, and $f(\phi)$ is the dimensionless coupling function. 

In this paper, the choice $f(\phi)=e^{2\phi}/4$ corresponds to \textit{Einstein dilaton Gauss Bonnet} (EdGB) gravity~\cite{Kanti:1995vq}, and $f(\phi)=2\phi$ to \textit{shift symmetric sGB} (ssGB) gravity~\cite{Sotiriou:2014pfa}.
It is useful to rewrite the GB scalar as
\begin{equation}\label{GBscalardef}
\mathcal{R}^2_{GB}=^{*}R_{a b c d}^{*} R^{a b d c}\,,
\end{equation}
with $^{*}R_{c\mu\nu d}^{*} =\frac{1}{4} \epsilon^{a b e f} R_{e f g h} \epsilon^{g h c d}
$ the dual of Riemann tensor and $\epsilon_{\alpha\beta\mu\nu}$  the anti-symmetric Levi Civita tensor.

The field equations resulting from action in Eq.~\eqref{action} are
\small
\begin{subequations}\label{EOMall}
\begin{align}
&\Box\phi=-\frac{\alpha}{4} f'(\phi)\mathcal{R}^2_{GB}+\frac{4\pi G}{c^4\sqrt{-g}} \frac{\partial S_{m}}{\partial \phi}\,,\label{scalarEOM}\\
&G^{\mu\nu}=\frac{8\pi G}{c^4} T^{\mu\nu}+2\nabla^{\mu}\phi \nabla^{\nu}\phi -g^{\mu\nu}(\nabla\phi)^2-4\alpha\left( ^{*}R^{*c\mu\nu d}\nabla_{cd}f(\phi)\right)\,,\label{metEOM}
\end{align}
\end{subequations}
\normalsize
with $\square \equiv g^{\alpha \beta} \nabla_{\alpha} \nabla_{\beta} $ being the d'Alembertian operator, $G^{\mu\nu}$ the Einstein tensor and
$T^{\mu\nu}=(-2/\sqrt{-g})(\partial S_{m}/\partial g_{\mu\nu})$ the distributional matter energy-momentum tensor.

\subsection{Different viewpoints of the action and their consequences}

We note that action~\eqref{action} may be interpreted in two distinct ways:
(i) One may {\textit{postulate}} that the action describes quadratic gravity corrections to GR in the physical (i.e. Jordan) frame, where matter fields are minimally coupled to the metric and results directly correspond to observables. This viewpoint is typically adopted in the literature on testing gravity (see e.g. Refs.~\cite{Yunes:2011we,Yagi:2011xp}) and corresponds to setting the conformal factor $\mathcal{A}(\phi)=1$ in Eq.~\eqref{action}.
(ii) If the action, instead, is regarded as the low-energy effective action of string theories, the conformal factor is nontrivial. From this perspective, Eq.~\eqref{action} denotes the action in the Einstein frame, and one must transform any results to the Jordan (or ``string'') frame in which observables are measured.
For example, the bosonic sector of heterotic string theory is characterized by $f(\phi)=e^{-2\phi}$ for which the conformal factor is $\mathcal{A}(\phi)=e^{\phi}$ (see e.g. Ref.~\cite{Metsaev:1987zx}).
Both of these interpretations of the action become reconciled in the weak-field regime, where $\mathcal{A}(\phi)=e^{\phi} = 1 + \mathcal{O}(\phi)$.
Nevertheless, it is important to remain aware of these assumptions when interpreting observational bounds on higher-curvature effects and their possible implications (or bounds) on underlying quantum gravity theories.
In particular, a non-minimal coupling factor in Einstein frame would lead to additional polarization modes when converted to the Jordan frame.
Ignoring such couplings would lead to the conclusion that a breathing scalar mode is absent in the sGB theory, in contrast to, for example, ST theories.

\subsection{Matter action}
To describe the action of compact self-gravitating binaries, one has to take the internal gravity of each object into consideration. Based on the approach pioneered by Eardley~\cite{Eardley} for 
ST theories~\citep{1975ApJ...196L..59E,Damour:1992we}, and later generalized 
to Einstein-Maxwell-dilaton theories~\cite{Julie:2017rpw}, this can be done by describing the compact objects as point particles but with scalar-dependent total masses. Such a \textit{skeletonized} matter action for binary systems is thus implicitly dependent on the scalar field and is given by the ansatz
\small
\begin{equation}\label{skeletonization}
S_{m}[g_{\mu\nu},\phi,x_{A}^{\mu}]=-c\int M_{A}(\phi)\sqrt{-g_{\mu\nu}dx^{\mu}_{A}dx^{\nu}_{A}}+(A\leftrightarrow B)\,.
\end{equation}
\normalsize
Here $dx_{A}^{\mu}$ is the world line of particle A and the internal self-gravity of the compact objects is incorporated through the scalar-dependent mass $M_{A}(\phi)$.

Note that the point particle description of Eq.~\eqref{skeletonization} does not depend on any gradients of scalar field, which results in neglecting finite size effects (e.g. tidal effects). Inclusion of such terms and calculation of the corresponding scalar-induced tidal effects is subject of future work.

Within the PN approximation, the expression for $M_{A}(\phi)$ can be parametrized by its expansion about the background scalar field $\phi_{0}$:
\small
\begin{equation}
    M_{A}=m_{A}\Big[ 1+\alpha_{A}^{0}\delta\phi+\frac{1}{2}\left((\alpha^{0}_{A})^{2}+\beta_{A}^{0}\right)\delta\phi^2\Big] +\mathcal{O}(\delta\phi^3)\,,
\end{equation}
\normalsize
with $\delta \phi=\phi-\phi_{0}$, and the strong-field parameters 
\begin{equation}
    \left.\alpha_{A}^{0}=\frac{d\, ln\,M_{A}(\phi)}{d\phi}\right|_{\phi=\phi_{0}},
    \qquad\qquad
    \left.\beta_{A}^{0}=\frac{d\, \alpha_{A}(\phi)}{d\phi}\right|_{\phi=\phi_{0}}.
\end{equation}
Also, $m_A$ is the asymptotic value of the particle's mass in the Einstein frame. The parameter $\alpha_A^0$ is called the \textit{scalar charge} and measures the coupling of the physically measurable mass $M_A$ to the background scalar field.

Within the small-coupling approximation, the explicit form of $\alpha_{A}^{0}$ corresponding to static, spherically symmetric BHs has been derived to fourth order in the coupling parameter in Ref.~\citep{Julie:2019sab}.
For example, to first order, the scalar charge is
\begin{equation}\label{BHcharge}
\alpha_{A}^{0}\equiv -\alpha f'(\phi_{0})/2m_{A}^2\,.
\end{equation}
\section{Relaxed-field equations and wave generation}\label{sec:III}
To solve the field equations in the weak-field limit, we use the DIRE approach adapted to modified theories of gravity.
In order to do so, it is common to introduce the tensor-density $\mathfrak{g}^{a b}=\sqrt{-g} g^{a b}$ called \textit{gothic metric}.
By defining 
\small
\begin{equation}\label{H}
H^{\alpha\mu\beta\nu} \equiv \mathfrak{g}^{\mu \nu} \mathfrak{g}^{\alpha \beta}-\mathfrak{g}^{\alpha \nu} \mathfrak{g}^{\beta \mu}\,,
\end{equation}
\normalsize
it can be shown that the following is an identity, valid for
any space-time:
\small
\begin{equation}\label{Hidentity}
\partial_{\mu\nu}H^{\mu \alpha \nu \beta}=(-g)\left(2 G^{\alpha\beta}+\frac{16\pi G}{c^4}\, t^{\alpha\beta}_{LL}\right)\,,
\end{equation}
\normalsize
where $t^{\alpha\beta}_{LL}$ is the Landau-Lifshitz energy-momentum pseudo-tensor as given in Ref.~\cite{1975ctf..book.....L}.

To incorporate sGB gravity into this framework, we rewrite the expression for $G^{\mu\nu}$ given by Eq.~\eqref{metEOM} in terms of the gothic metric. In order to analyze the behaviour of fields outside the sources and to look for the generated tensor and scalar radiation, we expand the gothic metric around Minkowski space-time and the scalar field around the background value $\phi_{0}$. The perturbation variables are thus defined through
\small
\begin{equation}\label{perturbation}
    \mathfrak{g}^{\mu\nu}=\eta^{\mu\nu}+h^{\mu\nu}\,,\quad\quad\mathrm{and}\quad\quad \phi=\phi_{0}+\delta\phi\,.
\end{equation}
\normalsize
By choosing the Harmonic gauge defined as $\partial_{\nu}\mathfrak{g}^{\mu\nu}=0$, 
Eq.~\eqref{Hidentity} can be written as a wave equation for the perturbations, with sources
\small
\begin{equation}\label{wave}
\begin{split}
    \Box h^{\alpha\beta}&= \frac{16\pi G}{c^4} \mu^{\alpha\beta}\,,\\
    \mu^{\alpha\beta}&=(-g)T^{\alpha\beta}_{m}+\frac{c^4}{16\pi G}(\Lambda_{GB}^{\alpha\beta}+\Lambda^{\alpha\beta}_{GR})\,,\\
    \Lambda^{\alpha\beta}_{GR}&=\frac{16\pi G}{c^4}(-g)t_{LL}^{\alpha\beta}+h^{\alpha\nu}_{,\mu}h^{\beta\mu}_{,\nu}-h^{\mu\nu}h^{\alpha\beta}_{\,\,,\mu\nu}\,,\\
    \Lambda_{GB}^{\alpha\beta}&=-8\alpha(-g)\left(^{*}\hat{R}^{*c\alpha\beta d}f(\phi)_{,cd}\right)
    +4\phi_{,c}\phi_{,d}\big(\mathfrak{g}^{\alpha c}\mathfrak{g}^{\beta d}-\frac{1}{2}\mathfrak{g}^{\alpha \beta}\mathfrak{g}^{c d}\big)\,,
\end{split}
\end{equation}
\normalsize
where $\Lambda_{GB}^{\alpha\beta}$ captures the scalar field and the nonlinear GB contributions, with $^{*}\hat{R}^{*c\alpha\beta d}$ the gauge fixed, dual Riemann tensor written in terms of gothic variables.
The explicit expression for $\Lambda_{GB}$ in terms of the metric and scalar perturbations is given in Appendix~(\ref{appAweakfield}).\\
The scalar field equation, Eq.~\eqref{scalarEOM}, is itself a wave equation. In terms of the gothic metric we have
\small
\begin{equation}\label{Sgothic}
\begin{split}
    \Box \delta\phi&=\frac{4\pi G}{c^4}\mu_{s}\,,\\
    \mu_s&=\frac{S_{m,\phi}}{\sqrt{-g}}-\frac{c^4 }{16\pi G}\alpha f'(\phi)\hat{\mathcal{R}}^2_{GB}\,,
\end{split}
\end{equation}
\normalsize
with $\hat{\mathcal{R}}^2_{GB}$ being the gauge-fixed, GB scalar written in terms of gothic variables. For its expression in terms of metric perturbation see Appendix.~(\ref{appAweakfield}).

\begin{figure}
    \includegraphics[width=0.5\columnwidth]{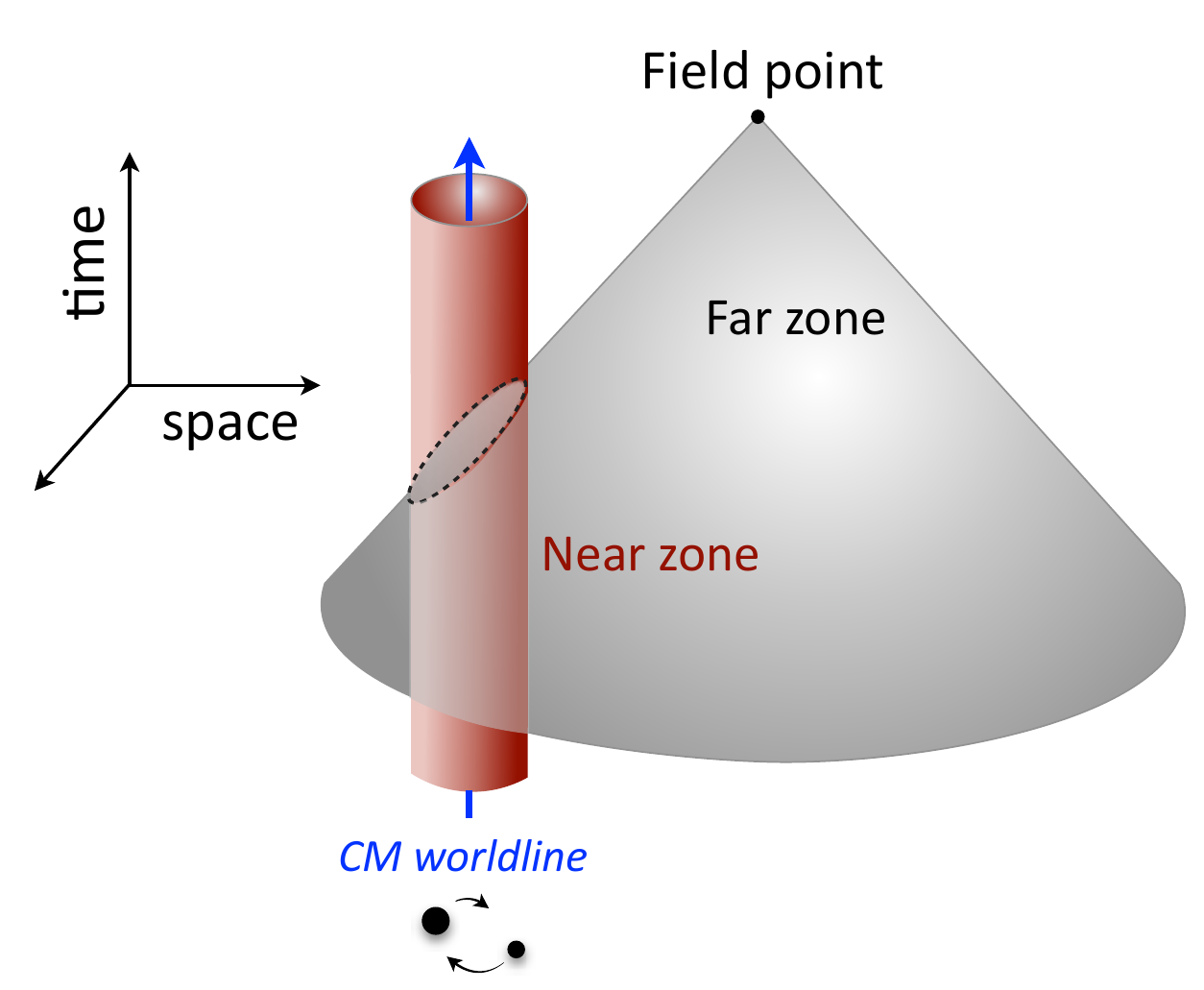}
    \caption{The past null cone of the field point $x$ and its intersection with the near-zone, for a field point located in the far-zone.}
    \label{fig:zones}
\end{figure}

Formal solutions to the wave equations
~\eqref{wave} and~\eqref{Sgothic} 
can be computed in all regions of space-time by using the appropriate retarded Green's function, giving
\small
\begin{subequations}\label{waveall}
\begin{align}
h^{\mu \nu}(x)&=\frac{4G}{c^4} \int \frac{\mu^{\mu \nu}\left(t^{\prime}, \mathbf{x}^{\prime}\right) \delta\left(t^{\prime}-t+\left|\mathbf{x}-\mathbf{x}^{\prime}\right|/c\right)}{\left|\mathbf{x}-\mathbf{x}^{\prime}\right|} d^{4} x^{\prime}\label{Twaveall}\,,\\
\delta \phi(x)&= \frac{G}{c^4}\int \frac{\mu_s\left(t^{\prime}, \mathbf{x}^{\prime}\right) \delta\left(t^{\prime}-t+\left|\mathbf{x}-\mathbf{x}^{\prime}\right|/c\right)}{\left|\mathbf{x}-\mathbf{x}^{\prime}\right|} d^{4} x^{\prime}\,.\label{Swaveall}
\end{align}
\end{subequations}
\normalsize
The integrals here are over the past null cone
of the field point $x=(ct,\mathbf{x})$.\\
In principle, the explicit solution of Eq.~\eqref{waveall} depends on the position of the field point $\mathbf{x}$ relative to the source $\mathbf{x'}$, \textit{i.e.} the size of the quantity $\|\mathbf{x}-\mathbf{x}'\|=R$ relative to other scales in the system. Defining the characteristic size of the source to be $\mathcal{S}$, the near-zone
is defined as the region with $R<\mathcal{R}$, where $\mathcal{R} \sim \mathcal{S}/v$ is the characteristic wavelength of GWs from the system. The region outside the near-zone is called far-zone (see Fig.~\ref{fig:zones}).
Given the position of the field point $\textbf{x}$, one can evaluate each integral in two separate pieces: an integral over the source point $\mathbf{x}'$ in the near-zone
and another over the source point in the far-zone. This leads to four different integrals in total, as described in detail in Refs.~\citep{Will:1996zj,Pati:2000vt}.

We are particularly interested in the two calculations based on a far-zone field point. 
Assuming weak-field and low-velocity approximations, we perturbatively expand the nonlinear terms in $\mu^{\mu\nu}$, and its scalar analogue $\mu_{s}$, using the formal PN expansion parameter $1/c^2$.
We keep terms up to the relative first PN order. 
Our matching conditions between the near and far zone are chosen such that terms depending on the boundary radius that separates the two zones 
cancel,\textit{ i.e.}, the final answer is independent of this parameter.
This property is generally expected and is explicitly shown to be correct within GR~\cite{Will:1996zj}.

In the next section, we first find the near-zone contribution to the waveforms, by defining the tensor and scalar multipole moments and  explicitly calculating them for binary systems. We outline the general formalism for evaluating the far-zone contribution to waveforms in Appendix~\ref{App:FZ}, and show that the results are of higher PN order than considered in this paper. In both cases, we would only be concerned with the results in a subset of the far-zone named the \textit{far-away zone},
where the GW detectors are located. We hold off on presenting the final expressions of waveforms until Sec.~\ref{sec:V}.
\section{Calculating the tensor and scalar waveforms}\label{sec:IV}
\subsection{General structure of near-zone calculation}
Having the field point in the far-away zone, with $R\gg \mathcal{R}$,
it can be shown~\cite{Will:1996zj} that the near-zone contribution to $h^{ij}$ , named $h_{\mathcal{N}}^{i j}$, is given as
\small
\begin{equation}\label{EWstart}
    h^{ij}_{\mathcal{N}}(x)=\frac{2G}{Rc^4}\frac{d^2}{dt^2}\sum_{l=0}^{\infty}\hat{n}_{k_{1}...k_{l}}I_{EW}^{ijk_{1}...k_{m}}(\tau)+\mathcal{O}(R^{-2})\,,\qquad
\end{equation} 
\normalsize
where $\hat{n}_{k_{i}}$ is the unit normal vector pointing from the source to the detector. The tensor multipole moments $I_{\mathrm{EW}}^{ijk_{1}...k_{m}}$
are known as Epstein-Wagoner (EW) moments and are functions of retarded time $\tau=t-R/c$. They are given by
\small
\begin{equation}\label{EW}
\begin{split}
&I_{\mathrm{EW}}^{i j}  \equiv \frac{1}{c^2}\left[\int_{\mathcal{M}} \mu^{00} x^{i j} d^{3} x+I_{\mathrm{EW}(\mathrm{surf})}^{i j}\right]\,,\\
&I_{\mathrm{EW}}^{i j k} \equiv \frac{1}{c^3}\left[\int_{\mathcal{M}}\left(2 \mu^{0(i} x^{j) k}-\mu^{0 k} x^{i j}\right) d^{3} x+I_{\mathrm{EW}(\mathrm{surf})}^{i j k}\right]\,, \\
&I_{\mathrm{EW}}^{i j k_{1} \cdots k_{m}}  \equiv \frac{2}{m !c^2} \frac{d^{m-2}}{d (c t)^{m-2}} \int_{\mathcal{M}} \mu^{i j} x^{k_{1} \cdots k_{m}} d^{3} x \quad(m \geq 2)\,.
\end{split}
\end{equation}
\normalsize
with the so-called surface moments being
\small
\begin{equation}\label{EWsurface}
\begin{split}
&\frac{d^{2}}{d t^{2}} I_{\mathrm{EW}(\mathrm{surf})}^{i j} =\oint_{\partial \mathcal{M}}\left(4 \mu^{l(i} x^{j)}-\left(\mu^{k l} x^{i j}\right)_{, k}\right) \mathcal{R}^{2} \hat{n}^{l} d^{2} \Omega\,,\\
&\frac{d}{d t} I_{\mathrm{EW}(\mathrm{surf})}^{i j k} =\oint_{\partial \mathcal{M}}\left(2 \mu^{l(i} x^{j) k}-\mu^{k l} x^{i j}\right) \mathcal{R}^{2} \hat{n}^{l} d^{2} \Omega\,.\,\,
\end{split}
\end{equation}
\normalsize
These two- and three- index surface moments are the results of using the conservation law $\partial_{\alpha}\mu^{\alpha\beta}=0 $ in the derivation of Eq.~\eqref{EWstart} (see Ref.~\cite{Will:1996zj}).
As before, $\mathcal{R}$ is the radius of the boundary region $\partial \mathcal{M}$ of the near-zone, and we have used the fact that the surface element is $dS^k = \mathcal{R}^2d\Omega^2 \hat{n}^k$ at this boundary region. 

For computing GWs to relative 1PN order higher than the quadrupole, it is sufficient to consider up to the four-index EW moments and to only look for transverse-traceless (TT) parts of the spatial tensor, such that
\small
\begin{equation}\label{1PNFZNZ}
h^{ij}_{\mathcal{N},_{TT}}(x)=\frac{2G}{ Rc^4}\frac{d^2}{dt^2}\left\{I^{ij}+\hat{n}_k I^{ijk}+\hat{n}_{kl} I^{ijkl}\right\}_{_{TT}}.
\end{equation}
\normalsize
In the above equation, the TT projection operator acting on a tensor $A^{kl}$ is such that
\begin{equation}\label{TTdef}
h_{\mathrm{TT}}^{i j}=\left(P^{i k} P^{j l}-\frac{1}{2} P^{i j} P^{k l}\right) A^{k l}\,,\qquad
P^{i j}=\delta^{i j}-\hat{n}^{i} \hat{n}^{j}\,,
\end{equation}
with the properties $P^{i i}=2, P^{i j} P^{i j}=2,$ and $P^{i j} P^{i k}=P^{j k}$.
\subsection{Formal structure of near-zone source}
To compute EW moments requires expanding the components of the source $\mu^{\alpha\beta}$ to 1PN order. The 1PN expansions will depend on the solution of Eq.~\eqref{waveall} found in the near-zone. To 1PN order, these potentials are computed and can be found in Ref.~\cite{Julie:2019sab}.

For the expansion of the source terms, we begin by defining a simplifying notation for the different components of the fields, following Ref.~\citep{Pati:2000vt},
\begin{equation}\label{not}
\begin{split}
& N  \equiv h^{00} \sim \mathcal{O}(c^{-2})\,, \quad\,\,\,
 K^{i}  \equiv h^{0 i} \sim \mathcal{O}(c^{-3})\,, \\
 &B^{i j}  \equiv h^{i j} \sim \mathcal{O}(c^{-4})\,, \quad
 B \equiv h^{i i} \sim \mathcal{O}(c^{-4})\,,\\
 &\Phi \equiv \delta\phi \sim \mathcal{O}(c^{-2})\,.
\end{split}
\end{equation}
Using the same strategies used in GR calculations, we find that the weak-field expansion of the source $\mu^{\alpha\beta}$ to 1PN order (see Appendix.~(\ref{appAweakfield}) for the expansion of the GB terms),
is
\small
\begin{subequations}
\label{sources}
\begin{equation}\label{mu00}
\begin{split}
\mu^{00} &= m_{A}c^2\left[1+\left(\frac{v_{A}^{2}}{2c^2} +
\frac{3}{ 4}N +\alpha^0_A\Phi\right)\right]\delta^{3}(\mathbf{x}-\mathbf{x}_{A}(t)) +(A\leftrightarrow B)\\
&-\frac{c^4}{16\pi G}\left[\frac{7}{8}(\nabla N)^{2}-2(\nabla\Phi)^2\right]
-2\frac{\alpha f'(\phi_{0})c^4}{ 16\pi G}\left(\partial_{ij}\Phi\partial_{ij}N\right)+\mathcal{O}(c^{-2})\,,\qquad\qquad\qquad
\end{split}
\end{equation}
\begin{equation}\label{mu0j}
\mu^{0j} = m_{A}c\, v_{A}^{j} \delta^{3}(\mathbf{x}-\mathbf{x}_{A}(t))+(A\leftrightarrow B)+\mathcal{O}(c^{-1})\,,\qquad\qquad\qquad\qquad\qquad\qquad\qquad
\end{equation}
\begin{equation}
\begin{split}
\mu^{i j} &=m_{A} v_{A}^{i} v_{A}^{j} \delta^{3}(\mathbf{x}-\mathbf{x}_{A}(t))+(A\leftrightarrow B)\\
&+\frac{c^4}{16\pi G}\Bigg[\frac{1}{4}\bigg(\partial^{i}N \partial^{ j}N-\frac{\delta^{i j}(\nabla N)^{2}}{2}\bigg)+4\Big(\partial^{i}\Phi \partial^j\Phi
-\frac{\delta^{ij}}{2}(\nabla \Phi)^2\Big)\Bigg]\\
&-2\frac{\alpha f'(\phi_{0})c^4}{16\pi G }\Big[\Big(\partial_{ki}\Phi \partial_{kj}N-\partial_{kj}\Phi \partial_{ki}N\Big)
+\Big(\partial_{ij}\Phi\nabla^2N+\partial_{ij}N\nabla^2\Phi\Big)\Big]
+\mathcal{O}(c^{-2})\,.\quad\,\,\,\,
\end{split}
\end{equation}
\end{subequations}
\normalsize
As pointed out, for a binary system, the expressions for the fields in the near-zone are given in Ref.~\cite{Julie:2019sab}.
In the above expressions we only need to substitute for $N$ and $\Phi$ to the lowest PN order, which is given by
\small
\begin{equation}\label{leadingNZNZ}
\begin{split}
    &N=4 \frac{G (m_A+m_B)}{rc^2}+\mathcal{O}(c^{-4})\,,\qquad\quad
    \Phi=-\frac{G(m_A \alpha^{0}_A+m_B \alpha^{0}_B)}{rc^2}+\mathcal{O}(c^{-4})\,,
\end{split}
\end{equation}
\normalsize
where $r$ is the relative separation between the masses such that $\mathbf{r}^{i}=\mathbf{x}_{A}^{i}-\mathbf{x}_{B}^{i}$.

\subsection{Evaluation of two-body Epstein-Wagoner moments }\label{sec:EWcalc}
As can be seen from Eqs.~\eqref{EW} and~\eqref{EWsurface}, 
the EW moments are integrals over a sphere of radius $\mathcal{R}$, about the CM of the system, with the variables entering the integrands to be evaluated at retarded time $\tau$.
We repeatedly integrate by parts using the identity (see Ref.~\citep{Will:1996zj}, Eq.~(4.1)\,)
\small
\begin{equation}
\int_{\mathcal{M}} \partial_{k} F^{i j \ldots m} d^{3} x=\left.\oint_{\partial \mathcal{M}} F^{i j \cdots m}\right|_{\mathcal{R}} \hat{n}^{k} \mathcal{R}^{2} d^{2} \Omega\,.
\label{eq:intbyparts}
\end{equation}
\normalsize
Moreover, as mentioned earlier, we are only interested in the physically measurable, TT components of the far-zone tensor perturbation. We therefore make frequent use of the identities below, which follow from the definition of the projection operator $P^{i}_{j}$ in Eq.~\eqref{TTdef}
\begin{equation}
\left(\delta^{i j}\right)_{T T}=0\,, \qquad\quad \left(\hat{n}^{i} F^{j}\right)_{T T}=0\,,
\end{equation}
where the indices $i$ and $j$ apply to the final components of the waveform (and not the integrands), and F denotes a general term.

The general method for evaluating the field integrals of EW moments is explained in detail in Ref.~\citep{Will:1996zj} and can be summarized to the following steps: 
(i) use integration by parts to leave one potential undifferentiated using Eq.~\eqref{eq:intbyparts} and keep track of the resulting surface terms, (ii) change integration variables in such a way to put the center of the differentiated potentials at the origin, (iii) expand the undifferentiated potential in spherical harmonics, (iv) express all unit vector products in terms of symmetric-trace-free (STF) unit vectors and use the relations between spherical harmonics and STF unit vectors (see Appendix.~\ref{CalcTechs}) to integrate over $d\Omega$, (v) use the integration formula (Eq.~\eqref{rint}) for $f(\frac{1}{r})$ to integrate over $dr$.\\
In the following calculations, we only report the calculations that are new to the sGB theory, and we refer the reader to Ref.~\cite{Will:1996zj} for a detailed description of integration methods, as well as the calculation of GR contributions. 
In Appendix~\ref{CalcTechs}, we calculate two of the key integrals and explain their main steps that are then used throughout the rest of the calculations.

We note that the specific form of the near-zone potentials as given in Eq.~\ref{leadingNZNZ} is such that $\Phi_A=-\alpha_A U_A$, with $U_A=Gm_A/rc^2$ the Newtonian-order potential of body A. In the following calculations, we will use this relation to substitute for $\Phi$ so that the source terms simplify and that the functional form of the integrals mimic the usual GR terms.

\subsubsection{Two-index moment $I^{ij}_{EW}$}
We split the calculation into GB coupling dependent and independent terms, such that $I^{ij}_{EW}=I^{ij}+I^{ij}_{GB}$.\\
The calculation of $I^{ij}$ is similar to that of GR at 1PN order. The only difference is the additional scalar charge dependent term in the energy-momentum tensor of matter ($\alpha_{A}^{0}\Phi$) as well as the $(\nabla\Phi)^2$ contribution, which follow the same calculation as the GR ones, resulting in
\small
\begin{equation}\label{2I}
I^{i j}=m_{A} x_{A}^{i j}\left[1+ \frac{1}{c^2}\left(\frac{v_{A}^{2}}{2}- \frac{G m_{B}}{2r}-\frac{G m_B \alpha^0_A\alpha^0_B}{2r}\right)\right]+(A\leftrightarrow B)+ {\cal O}(c^{-3}) .
\end{equation}
\normalsize

For $I^{ij}_{GB}$ to 1PN order, substituting GB terms of Eq.~\eqref{mu00} in Eq.~\eqref{EW}  gives
\small
\begin{equation}
    I^{ij}_{GB}\propto-\frac{c^2}{16\pi G} \int_{\mathcal{M}}
    \partial_{ij}\Phi\partial_{ij}N x^{ij}d^3x\,,
\end{equation}
\normalsize
where we have omitted the coupling factor for simplicity. 
We use integration-by-parts several times on this expression and use Eq.~\eqref{leadingNZNZ} to simplify the potentials. Ignoring constants and scalar-charge factors in front, we find
\small
\begin{equation}\label{2inIGB}
\begin{split}
    I^{ij}_{GB}\propto  \Big\{&\oint_{\partial\mathcal{M}}\partial_{m}U\partial_{lm}U x^{ij} dS^l
    -\oint_{\partial\mathcal{M}}\partial_{ll}U \partial_m U x^{ij}dS^m+\int_{\mathcal{M}}\partial_{ll}U \partial_{m}U \partial_{m}(x^{ij})d^3 x\\
    &-\int_{\mathcal{M}} \partial_{m}U \partial_{lm}U \partial_{l}(x^{ij})d^3 x+\int_{\mathcal{M}} \partial_{ll}U \partial_{mm}U x^{ij}d^3 x
    \Big\}\,,
    \end{split}
\end{equation}
\normalsize
where the Newtonian potential $U=\sum_{A} U_A=\sum_{A} Gm_{A}/rc^2$ is used to simplify the expressions.
It can be shown that the resulting surface integrals vanish as they either do not have $\mathcal{R}$ independent integrands or involve integration of a delta function $\delta^{3}(\mathbf{x_A}-\mathbf{x_B})$ that is zero on the boundary of the near-zone. For the evaluation of the last volume integral, we use the fact that $\partial_{ll}U=\nabla^2 U=-4\pi\sum_A m_A \delta^3 (\mathbf{x}-\mathbf{x_A})$.
This results in a term proportional to $m_A m_B \delta^3 (\mathbf{x_A}-\mathbf{x_B}).$
As the calculations are being done in the inspiral regime, where $\mathbf{x_A} \neq \mathbf{x_B}$ by definition, this term is zero and thus, we are left with
\small
\begin{equation}\label{Igbibp}
    I^{ij}_{GB}\propto  \frac{1}{G c^2}\int_\mathcal{M}\partial_{ll}U \partial_{m}U \partial_{m}(x^{ij})d^3 x
    -\frac{1}{G c^2}\int_\mathcal{M} \partial_{m}U \partial_{lm}U \partial_{l}(x^{ij})d^3 x\,.
\end{equation}
\normalsize
The first terms involves $\nabla^2 U$ and can be readily evaluated. For the second term, we further apply integration by parts and get 
\small
\begin{equation}
    \begin{split}
     -\int_\mathcal{M} \partial_{m}U \partial_{lm}U \partial_{l}(x^{ij})d^3 x=&-\oint\partial_{l}U \partial_m U \partial_{l} (x^{ij}) dS^m +\int\partial_{mm}U \partial_l U \partial_{l}(x^{ij})d^3x\\
     &+\int \partial_{m}U \partial_{l}U \partial_{ml}(x^{ij})d^3x
    \end{split}
\end{equation}
\normalsize
It can be shown again that the surface term vanishes and that the second term is similar to the first term of Eq.~\eqref{Igbibp} . The new contribution is the volume integral in the last line that involves integration of a $U^{,m} U^{,l}$ term for which we can again use integration-by-parts to leave one potential undifferentiated. This gives
\small
\begin{equation}
    \begin{split}
        \int \partial_m U \partial_l U \partial_{ml}(x^{ij})=\oint U \partial_l U \partial_{ml}(x^{ij}) dS^m -\int U \partial_{lm}U \partial_{ml}(x^{ij}) d^3x\,.
    \end{split}
\end{equation}
\normalsize
The surface integral does not have a $\mathcal{R}$-independent term and thus does not contribute. The remaining volume integral can be evaluated using methods within GR and can be found in Appendix.~\ref{CalcTechs}.\\
Putting together the non-vanishing contributions, the final expression for the GB coupling dependent two-index moment is
\small
\begin{equation}\label{2Igb}
  I_{GB}^{i j}=-\frac{\alpha  f'(\phi_{0})}{r^2} \frac{G m_Am_{B}\alpha^{0}_{B}}{rc^2} \left[2r^{(i} x^{j)}_{A}+r^{ij}-2\delta^{ij}r^2\right]
  +(A \leftrightarrow B)+{\cal O}(c^{-4})\,.
\end{equation}
\normalsize

Regarding the two-index surface moment $I^{ij}_{EW\operatorname{(surf)}}$, we can show that the coupling independent terms do not have any TT contribution at 1PN order. The contributions from 
terms depending on the GB coupling also vanish by taking surface integration, as they do not give $\mathcal{R}$-independent integrands.

Dropping the last term  of Eq.~\eqref{2Igb} that does not have a TT contribution and adding Eq.~\eqref{2I}, the final expression for $I_{EW}^{ij}$ becomes
\small
\begin{equation}
\label{2indexEW}
\begin{split}
I_{EW}^{i j}&=  m_{A} x_{A}^{i j}\left[1+\frac{1}{c^2}\left( \frac{v_{A}^{2}}{2}-  \frac{G m_{B}(1+\alpha_A^{0}\alpha_B^{0})}{2r}\right)\right]
-\frac{\alpha f'(\phi_{0})}{r^2}\frac{G m_A m_{B}\alpha^{0}_{B}}{rc^2} \left( 2r^{(i} x_{A}^{j)} +r^{ij}\right)\\
&+(A\leftrightarrow B)+\mathcal{O}(c^{-4})\,.
\end{split}
\end{equation}
\normalsize

An important point to note is the scaling of the GB terms. Since the terms enter at ${\cal O}(c^{-2})$ in the PN expansion, they are a 1PN effect, yet, due to the dimensionless ratio $\alpha/r^2$, the GB contribution is suppressed at large separation compared to the other 1PN terms. This is not surprising as it encapsulates a different physical effect.
This is similar to the case of tidal interactions in GR, where Newtonian tides at ${\cal O}(c^{0})$ have the same scaling with separation $r$ as 5PN point-mass terms~\cite{Flanagan:2007ix}. Calculations of tidal effects in compact binaries involve a double expansion in PN and finite-size corrections~\cite{Flanagan:1997fn}. Likewise, the GB term in Eq.~\eqref{2Igb} is the leading-order gravitational effect of the higher curvature corrections. However, since the GB effect first appears only at 1PN order, we can here keep the full dependence on the GB coupling without requiring an explicit double expansion, nor any assumptions on the coupling strength.

\subsubsection{Three-index moment $I^{ijk}_{EW}$}
In order to obtain the waveform to relative 1PN order, it is sufficient to calculate $I_{E W}^{i j k}$ to 0.5PN order. As there are no GB dependent terms at this order (see Eq.~\eqref{mu0j}), the three-index moment is the same as in GR 
\small
\begin{equation}\label{3indexEW}
I^{i j k}_{EW}= \frac{m_{A}}{c}\left(2 v_{A}^{(i} x_{A}^{j)} x_{A}^{k}-v_{A}^{k} x_{A}^{ij}\right)+(A\leftrightarrow B)+\mathcal{O}(c^{-3})\,.
\end{equation}
\normalsize
Also, the three-index surface moment $I_{EW\mathrm{(surf)}}^{ijk}$ is shown to be zero in GR to 1PN order. 
\subsubsection{Four-index moment $I^{ijkl}_{EW}$}
We again split the calculation into coupling dependent and independent terms, namely $I_{EW}^{ijkl}=I^{ijkl}+I^{ijkl}_{GB}$.\\
The moment $I^{ijkl}$ has a GR part and 
a contribution from the scalar field  that is similar to the GR one \citep{Will:1996zj} up to a factor difference. Including the factors, these terms together give 
\small
\begin{equation}
I^{ijkl}=
\frac{G m_A m_{B}(1+\alpha_A^{0}\alpha_B^{0}) r^{ij}}{12\,r\,c^2}
\left(\frac{r^{kl}}{r^2}-\delta^{k l}-6\frac{ x_{A}^{kl} }{r^2}\right)
+\frac{m_{A}}{c^2} v_{A}^{ij}  x_{A}^{kl}+(A\leftrightarrow B)+{\cal O}(c^{-4})\,,
\end{equation}
\normalsize
where the terms proportional to $\delta^{ij}$ are dropped as they do not produce TT components. For $I^{ijkl}_{GB}$ we have
\small
\begin{equation}
    I^{ijmn}_{GB}\propto-\frac{c^2}{16\pi G} \int_{\mathcal{M}} \Big\{\Big(\partial_{ki}\Phi \partial_{kj}N-\partial_{kj}\Phi \partial_{ki}N\Big)
    +\Big(\partial_{ij}\Phi\nabla^2N+\partial_{ij}N\nabla^2\Phi\Big)
    \Big\}x^{mn}d^3x\,.
\end{equation}
\normalsize
Using Eq.~\eqref{leadingNZNZ} we see that the terms in the first parentheses vanish. For the terms in the second parentheses, we can directly insert the Laplacian of Eq.~\eqref{leadingNZNZ} to remove the integrals. Ignoring the $\delta^{ij}$ dependent terms that do not give TT components we find
\small
\begin{equation}\label{4indexEW}
\begin{split}
 I^{ijkl}_{EW}=&
\frac{1}{c^2}m_{A}v_{A}^{ij} x_{A}^{kl}
+\frac{G m_A m_{B}(1+\alpha_A^{0} \alpha_B^{0}) r^{ij}}{12 \,r\, c^2}\left( \frac{r^{kl}}{r^2}-\delta^{k l}-\frac{6 x_{A}^{kl} }{r^2}\right)\\
&-\frac{3G \alpha f'(\phi_{0})
m_A m_B\alpha^0_{B} r^{ij}}{c^2\, r^{5}}\,x^{kl}_{A}+(A\leftrightarrow B)+{\cal O}(c^{-4}).
\end{split}
\end{equation}
\normalsize
\subsection{Scalar multipole moments and waveform}\label{sec:scalarmultipoles}
The scalar field perturbation,
solving its field equation~\eqref{Swaveall} in the near zone and with the source term given in Eq.~\eqref{Sgothic},
can be written as
\small
\begin{equation}
\label{Swave}
\begin{split}
    &\Phi_{\mathcal{N}}(x)=\sum_{l=0}^{\infty} \Phi_{l}(t,\textbf{x})\,,\qquad \Phi_{l}(x)=\frac{G}{R\,c^4}\frac{\hat{n}_{L}}{l!}\left(\frac{\partial}{c\, \partial t}\right)^l
    I^{L}_{s}+\mathcal{O}(R^{-2})\,,
    \end{split}
\end{equation}
\normalsize
with the scalar multipole moments defined as
\small
\begin{equation}\label{smms}
I_{s}^{L}=\int_{\mathcal{M}} \mu_{s}(\tau,x')x'^{L}d^3x'\,.
\end{equation}
\normalsize
It is important at this stage to discuss the counting of PN orders. By comparing Eq.~\eqref{Swave} to Eq.~\eqref{EWstart}, we see that the lowest order term in the tensor wave, containing the quadrupole moment, is of order $\mathcal{O}(c^{-2})$ higher than that of the scalar wave due to the double time derivative. Thus, by assuming the leading (quadrupole) part of the tensor radiation to be relative 0PN order, the scalar monopole and dipole moments will be relatively -1PN and -0.5PN, respectively. This implies that finding the scalar radiation to 1PN requires the scalar moments to be calculated to relative 2PN order.\\
As we have expanded $\mathcal{R}^2_{GB}$ (see Appendix \ref{appAweakfield}, Eq. \eqref{app:rgbohphi}) to leading order in the weak-field limit, we are limited to having $\mu_{s}$ to $\mathcal{O}(c^{-3})$ at most, and  therefore, our final expression for the scalar waveform will be at most 0.5PN order.

Using the scalings of terms in Eq.~\eqref{not}, it can be easily shown that the PN expansion of $\mathcal{R}^2_{GB}$ does not contain 0.5PN and 1.5PN terms. The 1PN term,  put together with the expansion of the matter source, gives the scalar source
\begin{equation}\label{sourcesS}
\begin{split}
\mu_{s}=&
+ m_A c^2\,\alpha_{A}^{0} \delta^3(\mathbf{x}-\mathbf{x}_A(t))\left[1-\frac{v^2_{A}}{2c^2}-\frac{3N}{4}+\left(\alpha_A^0+\frac{\beta_A^0}{\alpha_A^0}\right)\Phi\right]\\
&-\frac{\alpha f'(\phi_{0})c^4}{32\pi G}\Big[(\partial_{kl}N)^{2}-(\partial_{kk}N)^{2}\Big]+(A\leftrightarrow B)+\mathcal{O}(c^{-2})\,.
\end{split}
\end{equation}

With this result, we can now calculate the scalar multipole moments of Eq.~\ref{smms}. We define $I_{s}^{L}=I_{s,c}^{L}+I_{s,GB}^{L}$ in order to separate the calculation of
GB coupling dependent and independent terms again.
As the 0.5PN expansion of GB source is zero, the scalar quadrupole moment at this order does not have a GB contribution.
Also, integration-by-parts shows that the 1PN GB contribution to the scalar monopole moment vanishes. The 1.5PN contribution to the scalar monopole is also zero as we mentioned that there is no 1.5PN GB source term.
This leaves only the 1PN scalar dipole calculation.
After integrating-by-parts and neglecting the surface terms that do not contribute, we obtain
\small
\begin{equation}\label{1PNGBmultipole}
I_{s,GB}^{i}= -\frac{\alpha f'(\phi_{0})c^4}{32\pi G}\int_\mathcal{M} \Big[(\nabla^2 N)^2 -\partial_{i}N \nabla^2 N
-\partial_{l}N \partial_{il}N\Big]d^3x\,.
\end{equation}
\normalsize
The non-vanishing part of the above equation involves the integral $\int (\nabla N)^2 d^3x$. As already discussed (see discussion below Eq.~\eqref{2Igb}), this term is also zero and thus leaves us with no explicit GB corrections to the multipole moments.

Thus, the only remaining contributions to $I_{s,c}^{L}$ are terms depending on the matter sources. It is straightforward to see that:
\small
\begin{equation}\label{scalarmultipoles}
\begin{split}
I_{s,c}&= m_Ac^2 \alpha_A^{0} \Big\{1-\frac{v^2_A}{2c^2}- \frac{G m_B \alpha_{AB}}{r\, c^2}+\mathcal{O}(c^{-4})\Big\}+(A\leftrightarrow B),\\
I_{s,c}^{i}&= x^{i}_{A} m_Ac^2\alpha_A^{0}  \Big\{1-\frac{v^2_A}{2c^2}-\frac{G m_B \alpha_{AB} }{r\, c^2}+\mathcal{O}(c^{-4})\Big\}\
+(A\leftrightarrow B)\,,\\
I_{s,c}^{ij}&= x^{ij}_{A} m_Ac^2 \alpha_A^{0}\{1+\mathcal{O}(c^{-2})\}+(A\leftrightarrow B)\,,\\
I_{s,c}^{ijk}&= x^{ijk}_{A} m_Ac^2\alpha_A^0\{1+\mathcal{O}(c^{-2})\}+(A\leftrightarrow B)\,,
\end{split}
\end{equation}
\normalsize
where we have defined $\alpha_{AB}=(1+\alpha_A\alpha_B+\beta_A\alpha_B\alpha_A^{-1})$\,.



\section{Two-body waveforms in the center of mass frame}\label{sec:V}

\subsubsection{Center of mass and relative acceleration to 1 Post-Newtonian order}
In the previous subsections, we explicitly derived the EW moments and the scalar multipoles for a binary system. In this section, we calculate their corresponding waveforms in the CM frame.

The conservative dynamics of a two-body system to 1PN order is governed by the Lagrangian derived in Ref.~\citep{Julie:2019sab}, where the authors constructed the Fokker Lagrangian using the near-zone solution to the field equations [Eq.~\eqref{EOMall}], finding
\small
\begin{equation}
\label{LagrAB}\begin{split}
&L_{A B} =-m_{A}c^2+\frac{1}{2} m_{A} \mathbf{v}_{A}^{2}+\frac{ G\bar\alpha\, m_{A} m_{B}}{2r}+\frac{1}{8c^2} m_{A} \mathbf{v}_{A}^{4} \\
&+\frac{G\bar\alpha\, m_{A} m_{B}}{rc^2}\left[-\frac{G\bar\alpha m_{A}}{2 r}\left(1+2 \bar{\beta}_{B}\right)+\frac{3}{2}\left(\mathbf{v}_{A}^{2}\right)-\right.\\
&\left.\frac{7}{4}\left(\mathbf{v}_{A} \cdot \mathbf{v}_{B}\right)-\frac{1}{4}\left(\hat{\mathbf{n}}_{AB} \cdot \mathbf{v}_{A}\right)\left(\hat{\mathbf{n}}_{AB} \cdot \mathbf{v}_{B}\right)+\frac{\bar{\gamma}}{2}\left(\mathbf{v}_{A}-\mathbf{v}_{B}\right)^{2}\right] \\
&+\frac{\alpha f^{\prime}\left(\varphi_{0}\right)}{r^{2}} \frac{G^2 m_{A} m_{B}}{r^{2} c^2}\left[m_{A}\left(\alpha_{B}^{0}+2 \alpha_{A}^{0}\right)\right]+(A\leftrightarrow B)\,.
\end{split}\end{equation}
\normalsize
As before, $r=\|\mathbf{r}\|=\|\mathbf{x}_{A}-\mathbf{x}_{B}\|$ so that $\hat{\mathbf{n}}_{AB} =\mathbf{r}/ r$ and  in anologous to ST theories~\cite{Damour:1992we}, we define the binary parameters
\small
\begin{equation}\label{EOMparams}\begin{split}
\bar{\alpha} \equiv \left(1+\alpha_{A}^{0} \alpha_{B}^{0}\right)\,,\quad
\bar{\gamma}\equiv-2 \frac{\alpha_{A}^{0} \alpha_{B}^{0}}{\bar{\alpha}}\,,\quad
\bar{\beta}_{A} \equiv \frac{1}{2} \frac{\beta_{A}^{0} (\alpha_{B}^{0})^{ 2}}{\bar{\alpha}^{2}}.
\end{split}\end{equation}
\normalsize
The parameters $\bar{\gamma}$ and $\bar{\beta}$ are generalizations of 1PN Eddington parameters, and $\bar{\alpha}$ appears as a re-scaling factor for the gravitational constant $G$. 
It is useful to also introduce the combinations
\small
\begin{equation}
\mathcal{S}_{\pm}=\frac{\alpha_A^0\pm\alpha_B^0}{2\sqrt{\bar{\alpha}}},\qquad\qquad \beta_{\pm}=\frac{\bar{\beta}_{A}\pm\bar{\beta}_{B}}{2}\,.
\end{equation} 
\normalsize

We see that the Lagrangian $L_{A B}$ decomposes into a piece that is structurally equivalent to the two-body Lagrangian in massless ST theories~\citep{Damour:1992w} plus a distinct GB-coupling dependent contribution. The effects of the scalar field are thus entirely contained in the charge-dependent binary parameters of Eq.~\eqref{EOMparams}.
In massless ST theories, the gravitational constant $G$ always appears in combination with ST analogous parameter of $\bar{\alpha}$, which implies that the effective gravitational coupling in these theories is $G\bar{\alpha}$. However in sGB gravity, the
GB higher curvature terms break this rescaling, as made explicit in the last line of Eq.~\eqref{LagrAB}. Thus, starting at the 1PN relative order, the effect of the scalar on the two-body dynamics in the two theories can in fact be distinguished.

Our final aim in this section is to express the waveforms in terms of relative variables, by transforming to the CM frame with $X^{i}_{CM}=0$ defined as
\small
\begin{equation}\begin{split}
    &X^{i}_{CM}=\frac{1}{m}\int \mu^{00}r^i d^3\mathbf{r}\,.\\
\end{split}
\end{equation}
\normalsize
By computing the above integral with the same methods as in Sec.~\ref{sec:EWcalc}, we find that, to 1PN order, the coordinates of each body in the CM frame are
\small
\begin{equation}\label{CMx}\begin{split}
&\mathbf{x}_{A}=\left[\frac{m_{B}}{m}+\frac{\mu\Delta m} { 2 m^{2}c^2}\left(v^{2}-\frac{G m\bar{\alpha}}{r}\right) \right]\mathbf{r}+\delta+{\cal O}(c^{-3})\,, \\
&\mathbf{x}_{B}=\left[-\frac{m_{A}}{m}+\frac{\mu\Delta m}{2 m^{2}c^2} \left(v^{2}-\frac{G m\bar{\alpha}}{r}\right) \right]\mathbf{r}+\delta+{\cal O}(c^{-3}),
\end{split}\end{equation}
\normalsize
where $m=m_A+m_B$ is total mass of the system, $\eta=\mu/m=m_A m_B/m^2$ is the symmetric mass ratio, $\mathbf{v}=\mathbf{v}_{A}-\mathbf{v}_{B}$ is the relative velocity, $\Delta m =m_A-m_B$, and $\dot{r}=(\mathbf{v}\cdot\mathbf{r})/r$.\\ 
The GB coupling dependent term is given by
\small
\begin{equation}
    \delta=-2\eta\left( \frac{G m\bar{\alpha}}{r\, c^2} \frac{\alpha f'(\phi_{0})}{\sqrt{\bar{\alpha}}r^2}\mathcal{S}_{+}\right)\mathbf{r}\,.
\end{equation}
\normalsize
Taking a time derivative of Eq.~\eqref{CMx}, we obtain the velocities 
\small
\begin{equation}\label{CMv}
\begin{split}
\mathbf{v}_{A} &=\frac{m_{B}}{m}\mathbf{v} +
\frac{\mu\Delta m}{2 m^2c^2}\left[\left(v^{2}-\frac{G m\bar{\alpha} }{r}\right) \mathbf{v}-\frac{G m\bar{\alpha}}{r^{2}} \dot{r} \mathbf{r}\right]+
\dot{\delta}+{\cal O}(c^{-3})\,, \\
\mathbf{v}_{B}&=-\frac{m_{A}}{m}\mathbf{v}+
 \frac{\mu\Delta m}{2m^2\, c^2}\left[\left(v^{2}-\frac{G m\bar{\alpha} }{r}\right) \mathbf{v}-\frac{G m\bar{\alpha}}{r^{2}} \dot{r} \mathbf{r}\right]+
\dot{\delta}+{\cal O}(c^{-3})
\,,
\end{split}
\end{equation}
\normalsize
with the GB coupling dependent contribution being
\small
\begin{equation}
    \dot{\delta}=2\eta \frac{G m\bar{\alpha}}{rc^2}\frac{\alpha f'(\phi_{0})}{\sqrt{\bar{\alpha}}r^2}\mathcal{S}_{+}(3\dot{r}\mathbf{r}-\mathbf{v})\,.
\end{equation}
\normalsize

We substitute these relations in the Euler-Lagrange equations corresponding to Eq.~\eqref{LagrAB}, and find the 1PN relative matter equation of motion
\small
\begin{equation}\label{relEOM}
\begin{split}
    \frac{d^2 \mathbf{r}}{dt^2}=&\boldsymbol{a}=-\frac{G \bar{\alpha}m}{r^2}\hat{\mathbf{n}}_{AB}
    +\frac{G \bar{\alpha}m}{r^2\, c^2}\Big\{2\dot{r}\mathbf{v}\Big[ 2-\eta+\bar{\gamma} \Big]+ \Big [-(1+3\eta +\bar{\gamma})\mathbf{v}^2+\frac{3}{2}\eta \dot{r}^2\\
    &+\frac{2G\bar{\alpha}m}{r}\Big(2+\eta+\bar{\gamma}+\beta_{+}-\frac{\Delta m}{m}\beta_{-}
    +\frac{2 \alpha f'(\phi_{0})}{\bar{\alpha}^{3/2}r^2}(3\mathcal{S}_{+}+\frac{\Delta m}{m}\mathcal{S}_{-})\Big)\Big]\hat{\mathbf{n}}_{AB}
    \Big\}+\mathcal{O}(c^{-3})\,,
\end{split}
\end{equation}
\normalsize

From the Lagrangian, we can also read the conserved binding energy of the system to 1PN order,
\small
\begin{equation}\label{binding energy}
\begin{split}
    E&=m_{A}\left(\frac{v_A^2}{2}+\frac{3v_A^4}{8c^2}\right)+m_{B}\left(\frac{v_B^2}{2}+\frac{3v_B^4}{8c^2}\right)-\frac{G\mu m\bar{\alpha}}{r}\\
    &+\frac{G\mu m\bar{\alpha}}{rc^2}\left[\frac{3+2\bar{\gamma}}{2}(v_A^2+v_B^2)-\frac{1}{2}(v_A\cdot n)(v_B\cdot n)-\frac{7+4\bar{\gamma}}{2}(v_A\cdot v_B)\right]\\
    &+\frac{\mu}{c^2}\left(\frac{G m\bar{\alpha}}{r}\right)^2\left[\frac{1}{2}+\beta_{+}-\frac{\Delta m}{m}\beta_{-}\right] -\frac{\alpha f'(\phi_{0})\mu }{\bar{\alpha}^{3/2}r^2c^2}\left(\frac{G m\bar{\alpha}}{r}\right)^2\left(3\mathcal{S}_{+}+\frac{\Delta m}{m}\mathcal{S}_{-}\right)\,.
\end{split}
\end{equation}
\normalsize
In the CM frame, we have
\small
\begin{equation}\label{bindingCM}
\begin{split}
E=& \mu\Bigg\{\frac{1}{2} v^{2}-\frac{G \bar{\alpha} m}{r} +\frac{3}{8c^2}(1-3\eta)v^4+\frac{G\bar{\alpha} m}{2rc^2}\left[(3+2 \bar{\gamma}+\eta) v^{2}+\eta \dot{r}^{2}\right]\\
&+ \frac{1}{c^2}\left(\frac{G\bar{\alpha} m}{r}\right)^{2}\left[\frac{1}{2}+\beta_{+}-\frac{\Delta m}{m} \beta_{-}-\frac{\alpha f'(\phi_{0})}{\bar{\alpha}^{3/2}r^2}\left(3\mathcal{S}_{+}+\frac{\Delta m}{m}\mathcal{S}_{-}\right)\right]\Bigg\}\,.
\end{split}
\end{equation}
\normalsize
\subsubsection{Final expressions for the waveform}
By substituting the expressions for EW moments (see Eqs.~\eqref{2indexEW}, \eqref{3indexEW}, and \eqref{4indexEW}) in Eq.~\eqref{1PNFZNZ},
and simplifying further using Eqs.~\eqref{CMx} and \eqref{CMv}, we find to $\mathcal {O}(c^{-3})$ that
\small
\begin{equation}
\begin{split}
    &h^{ij}_{TT}=\frac{2G\mu }{R
    c^4}\frac{d^2}{dt^2}\Bigg\{\left[1+ \frac{1}{c^2}\left(1-\frac{3\mu}{m}\right)\tilde{E}+\frac{G m\bar{\alpha}}{3r\,c^2}\left(1-\frac{6\mu}{m}\right) \right]r^{ij}\\&
    -\frac{\Delta m}{m\,c^2}\left( 2v^{(i}r^{j)}(\hat{\mathbf{n}}\cdot\mathbf{r})-(\hat{\mathbf{n}}\cdot\mathbf{v})r^{ij}\right)
    +\frac{1}{c^2}\left(1-\frac{3\mu}{m}\right)(\mathbf{r}\cdot\hat{\mathbf{n}})^2\left(v^{ij}-\frac{G m\bar{\alpha} r^{ij}}{3r^3}\right)\\
    &-\frac{\alpha f'(\phi_{0})}{\sqrt{\bar{\alpha}}r^2}\frac{G \bar{\alpha}mr^{ij}}{rc^2}\left[ 4\left(\mathcal{S}_{+}+\frac{\Delta m}{m}\mathcal{S}_{-}\right)+3\big[\mathcal{S}_{+}(1-2\eta)+\mathcal{S}_{-}(1+2\eta)\big]\frac{(\mathbf{r}\cdot\hat{\mathbf{n}})^2}{r^2}\right] \Bigg\}_{TT},
\end{split}
\end{equation}
\normalsize
with $\tilde{E} = v^2/2 - G m\bar{\alpha}/r + \mathcal{O}(c^{-2})$ being the orbital binding energy per reduced mass at Newtonian order.\\
Taking time derivatives and perturbatively using the relative acceleration when needed (see Eq.\eqref{relEOM}), we end up with the final expression of the near-zone contribution to the gravitational waveform. As shown in Appendix~\ref{App:FZ}, the far-zone contribution is beyond 1PN order, and thus the overall expression for the gravitational waveform becomes
\small
\begin{equation}\label{waveformfinale}
\begin{split}
    &h^{ij}_{_{TT}}=
    \frac{2G\mu}{Rc^4}\{Q^{ij}\}_{_{TT}}=
    \frac{2G\mu}{Rc^4}\left\{\tilde{Q}^{ij}+\frac{1}{c}P^{1/2} \tilde{Q}^{i j}+\frac{1}{c^2}\left(P \tilde{Q}^{i j}+P \tilde{Q}^{ij}_{GB}\right)+{
    \cal O}(c^{-3})\right\}_{_{TT}}\,,\\
    \\
   &\tilde{Q}^{ij}=2\left[v^{ij}-\frac{G m\bar{\alpha}r^{ij} }{r^{3}} \right]\,,\\
   \\
   &P^{1 / 2} \tilde{Q}^{i j}=\frac{\Delta m }{ m}\left[3\frac{G m\bar{\alpha}} { r^{3}}(\hat{\mathbf{n}} \cdot \mathbf{r})\left(2 v^{(i} r^{j)}-\frac{\dot{r}r^{ij}}{r}\right)-(\hat{\mathbf{n}} \cdot \mathbf{v})\left(2 v^{ij}-\frac{G m\bar{\alpha} r^{ij}}{ r^{3}}\right)\right]\,,\\
   \\
  &P \tilde{Q}^{i j}=\frac{1-3 \eta}{3}\Bigg\{(\mathbf{r}\cdot\hat{\mathbf{n}})^{2}\, \frac{G\bar{\alpha }m}{r^3}\left[\left(6\bar{E}-15 \dot{r}^{2}+13 \frac{G\bar{\alpha} m}{r}\right)\frac{r^{i j}}{r^2}+30\dot{r}\frac{r^{(i} v^{j)}}{r}-14 v^{i j}\right]\\
  &\quad+(\hat{\mathbf{n}}\cdot \mathbf{v})^{2}\left[6 v^{i j}-2 \frac{G\bar{\alpha} m}{r^3}r^{i j}\right]+(\mathbf{r}\cdot \hat{\mathbf{n}})(\hat{\mathbf{n}} \cdot \mathbf{v}) \frac{G\bar{\alpha} m}{r^2}\left[12\frac{\dot{r} r^{i j}}{r^2}-32 \frac{r^{(i} v^{j)}}{r}\right]\Bigg\}
  \\
  &\quad+\frac{1}{3}\left\{\left[3(1-3 \eta) v^{2}-2(2-3 \eta) \frac{G \bar{\alpha} m}{r}\right] v^{i j}+4 \frac{G \bar{\alpha} m}{r}(5+3 \eta+3 \bar{\gamma})\frac{\dot{r}}{r} r^{(i} v^{j)}\right.\\
  &\quad \left.+\frac{G \bar{\alpha} m}{r^3}r^{ij}\left[3(1-3 \eta) \dot{r}^{2}-(10+3 \eta+6 \bar{\gamma}) v^{2}+\left(29+12 \bar{\gamma}+12 \beta_{+}-12 \frac{\Delta m}{m} \beta_{-}\right) \frac{G \bar{\alpha} m}{r}\right]\right\}\,,
  \\
  \\
  &P\tilde{Q}_{GB}^{ij}=4\frac{G \bar{\alpha}m}{r}\frac{\alpha f'(\phi_{0})}{\sqrt{\bar{\alpha}}r^2}\left[\left(\mathcal{S}_{+}+\frac{\Delta m}{m}\mathcal{S}_{-}\right)\left(\frac{6\dot{r}r^{(i}v^{j)}}{r}+\frac{r^{ij}}{r^2}\left(6\tilde{E}+\frac{5G \bar{\alpha}m}{r}-15\dot{r}^2\right)\right.\right.\\
  &\quad\left.\left.-2v^{ij}\right)+
  2\frac{G\bar{\alpha}m}{r}\frac{r^{ij}}{r^2}\left(3\mathcal{S}_{+}+\frac{\Delta m}{m}\mathcal{S}_{-}\right)\right]+3\frac{G \bar{\alpha}m}{r}\frac{\alpha f'(\phi_{0})}{\sqrt{\bar{\alpha}}r^4}\bigg(\mathcal{S}_{+}(1-2\eta)+\mathcal{S}_{-}(1+2\eta)\bigg)\\
  &\quad\Bigg\{(\mathbf{r}\cdot\hat{\mathbf{n}})^{2}\left[\frac{r^{ij}}{r^2}\left(10\tilde{E}-35\dot{r}^2+\frac{9 G \bar{\alpha}m}{r}\right)+\frac{10\dot{r}r^{(i}v^{j)}}{r}-2v^{ij}\right]-2(\hat{\mathbf{n}}\cdot \mathbf{v})^{2}r^{ij}\\
  &\qquad+(\mathbf{r}\cdot \hat{\mathbf{n}})(\hat{\mathbf{n}} \cdot \mathbf{v}) \left[20\frac{\dot{r} r^{i j}}{r}-4r^{(i}v^{j)}\right]
  \Bigg\}\,,
\end{split}
\end{equation}
\normalsize
where $P$ is a book keeping parameter with its superscripts denoting the PN order of each expression.

For the scalar waveform \eqref{Swave}, we first find the various contributions $\Phi_{l}$, including up to 0.5PN terms. By rewriting scalar multipoles (Eq.~\eqref{scalarmultipoles}) in the CM frame using Eqs.~\eqref{CMx}~-~\eqref{CMv}, and taking time derivatives, we find
\small
\begin{equation}\label{Swavemodes}
\begin{split}
&\Phi_{0}=\frac{2Gm\sqrt{\bar{\alpha}}}{Rc^2}\left(\mathcal{S}_{+}+\frac{\Delta m}{m} \mathcal{S}_{-}\right)\\
&\quad+\frac{2G\mu\sqrt{\bar{\alpha}}}{R c^4}\left\{-\frac{v^2}{2}\left(\mathcal{S}_{+}-\frac{\Delta m}{m} \mathcal{S}_{-}\right)+\left(\frac{8}{\bar{\gamma}}\left(\mathcal{S}_{+} \beta_{+}+\mathcal{S}_{-} \beta_{-}\right)-2\mathcal{S}_+\right)\frac{G\bar{\alpha}m}{r}\right\}\,,\\
 &\Phi_{1}=\frac{2G\mu\sqrt{\bar{\alpha}}}{Rc^3}\Bigg\{(\hat{\mathbf{n}} \cdot \mathbf{v})\Bigg[2\mathcal{S}_{-}+\frac{v^2}{c^2}\left(\frac{\Delta m}{m} \mathcal{S}_{+}-\eta \mathcal{S}_{-}\right)\\
 &\quad+\frac{G  \bar{\alpha}m}{r c^2}\Bigg(\frac{\Delta m}{2m} \mathcal{S}_{+}+\left(2 \eta-\frac{3}{2}\right) \mathcal{S}_{-}
 -\frac{4}{\bar{\gamma}} \frac{\Delta m}{m}\left(\mathcal{S}_{+} \beta_{+}
 +\mathcal{S}_{-} \beta_{-}\right)
 +\frac{4}{\bar{\gamma}}\left(\mathcal{S}_{-} \beta_{+}+\mathcal{S}_{+} \beta_{-}\right)\\
 &\quad-
 2\eta\mathcal{S}_{+}\frac{\Delta m}{m}\left(\mathcal{S}_{+}+\frac{\Delta m}{m}\mathcal{S}_{-}\right)\frac{\alpha f'(\phi_0)}{\sqrt{\bar{\alpha}}r^2}\Bigg)\Bigg]
 +\frac{G \bar{\alpha}m }{c^3 r^2} \dot{r}(\hat{\mathbf{n}} \cdot \mathbf{r})\Bigg[-\frac{5}{2} \frac{\Delta m}{m} \mathcal{S}_{+}+\frac{3}{2} \mathcal{S}_{-}\\
 &\quad+\frac{4}{\bar{\gamma}} \frac{\Delta m}{m}\left(\mathcal{S}_{+} \beta_{+}+\mathcal{S}_{-} \beta_{-}\right)-\frac{4}{\bar{\gamma}}\left(\mathcal{S}_{-} \beta_{+}+\mathcal{S}_{+} \beta_{-}\right)+
 6\eta\mathcal{S}_{+}\frac{\Delta m}{m}\left(\mathcal{S}_{+}+\frac{\Delta m}{m}\mathcal{S}_{-}\right)\frac{\alpha f'(\phi_0)}{\sqrt{\bar{\alpha}}r^2}
 \Bigg]\Bigg\}\,,\\
&\Phi_{2}=\frac{2G\mu\sqrt{\bar{\alpha}}}{Rc^4}\left(\mathcal{S}_{+}-\frac{\Delta m}{m} \mathcal{S}_{-}\right)\left\{(\hat{\mathbf{n}} \cdot \mathbf{v})^{2}-\frac{G \bar{\alpha}m }{r}\left(\frac{\hat{\mathbf{n}} \cdot \mathbf{r}}{r}\right)^{2}\right\}\,,\\
&\Phi_{3}=\frac{2G\mu\sqrt{\bar{\alpha}}}{R c^5}\left((1-2 \eta) \mathcal{S}_{-}-\frac{\Delta m}{m} \mathcal{S}_{+}\right)\left\{\frac{3}{2} \frac{G \bar{\alpha} m}{r^4} \dot{r}(\hat{\mathbf{n}} \cdot \mathbf{r})^{3}-\frac{7}{2} \frac{G\bar{\alpha} m}{r^3}(\hat{\mathbf{n}} \cdot \mathbf{v})(\hat{\mathbf{n}} \cdot \mathbf{r})^{2}+(\hat{\mathbf{n}} \cdot \mathbf{v})^{3}\right\}\,.
    \end{split}
\end{equation}
\normalsize
As expected, the scalar dipole moment does not vanish by choosing CM coordinates. This is a direct consequence of sGB theories violating the strong equivalence principle. Also, up to the order we are considering, the terms that explicitly depend on GB coupling only lead to scalar dipole radiation. Higher scalar multipoles, such as the quadrupole radiation, appear at higher PN orders. These modes are the dominant contribution for equal-mass binaries, for which the dipole radiation is suppressed. 
We will come back to this point in Sec.~\ref{sec:VIID}, where we compare the PN scalar waveform with scalar waves from numerical relativity simulations.

Next, we consider the scalar waves, arranging terms together based on their PN order. The final expression for the scalar waveform is given by
\small
\begin{equation}\label{scalarwavefinale}
\begin{split}
    &\Phi=\frac{2 G\mu \sqrt{\bar{\alpha}}}{R c^3}\Big\{P^{-1/2}\tilde{\Phi}+\frac{1}{c}\tilde{\Phi}+\frac{1}{c^2}P^{1/2}\tilde{\Phi}+{
    \cal O}(c^{-3})\Big\}\,,\\
    &P^{-1/2}\tilde{\Phi}=2 \mathcal{S}_{-}(\hat{\mathbf{n}} \cdot \mathbf{v})\,,\\
    &\tilde{\Phi}=\left(\mathcal{S}_{+}-\frac{\Delta m}{m} \mathcal{S}_{-}\right)\left[-\frac{G \bar{\alpha}m }{r}\left(\frac{\hat{\mathbf{n}} \cdot \mathbf{r}}{r}\right)^{2}+(\hat{\mathbf{n}} \cdot \mathbf{v})^{2}-\frac{1}{2} v^{2}\right]\\
    &\quad+\frac{G\bar{\alpha} m}{r}\left[-2 \mathcal{S}_{+}+\frac{8}{\bar{\gamma}}\left(\mathcal{S}_{+} \beta_{+}+\mathcal{S}_{-} \beta_{-}\right)\right]\,,\\
   &P^{1/2}\tilde{\Phi}=\left(-\frac{\Delta m}{m} \mathcal{S}_{+}+(1-2 \eta) \mathcal{S}_{-}\right)\left[\frac{3}{2} \frac{G \bar{\alpha} m}{r^4} \dot{r}(\hat{\mathbf{n}} \cdot \mathbf{r})^{3}-\frac{7}{2} \frac{G\bar{\alpha} m}{r^3}(\hat{\mathbf{n}} \cdot \mathbf{v})(\hat{\mathbf{n}} \cdot \mathbf{r})^{2}+(\hat{\mathbf{n}} \cdot \mathbf{v})^{3}\right]\\
   &\quad+(\hat{\mathbf{n}} \cdot \mathbf{v})\left\{\left(\frac{\Delta m}{m} \mathcal{S}_{+}-\eta \mathcal{S}_{-}\right) v^{2}+\frac{ G\bar{\alpha} m}{r}\left[\frac{1}{2} \frac{\Delta m}{m} \mathcal{S}_{+}+\left(2 \eta-\frac{3}{2}\right) \mathcal{S}_{-}\right.\right.\\
   &\quad\left.\left.-\frac{4}{\bar{\gamma}} \frac{\Delta m}{m}\left(\mathcal{S}_{+} \beta_{+}+\mathcal{S}_{-} \beta_{-}\right)+\frac{4}{\bar{\gamma}}\left(\mathcal{S}_{-} \beta_{+}+\mathcal{S}_{+} \beta_{-}\right)\right]\right\}\\
    &\quad+\frac{G \bar{\alpha}m }{r^2} \dot{r}(\hat{\mathbf{n}} \cdot \mathbf{r})\left[\frac{3}{2} \mathcal{S}_{-}-\frac{5}{2} \frac{\Delta m}{m} \mathcal{S}_{+}+\frac{4}{\bar{\gamma}} \frac{\Delta m}{m}\left(\mathcal{S}_{+} \beta_{+}+\mathcal{S}_{-} \beta_{-}\right)-\frac{4}{\bar{\gamma}}\left(\mathcal{S}_{-} \beta_{+}+\mathcal{S}_{+} \beta_{-}\right)\right]\\
    &\quad+2\eta\frac{\Delta m}{m}\frac{G\bar{\alpha}m}{r}\frac{\alpha f'(\phi_0)}{\sqrt{\bar{\alpha}}r^2}\mathcal{S}_{+}\left(\mathcal{S}_{+}+\frac{\Delta m}{m}\mathcal{S}_{-}\right)\left[3\frac{\dot{r}}{r}(\hat{\mathbf{n}}\cdot \mathbf{r})-(\hat{\mathbf{n}}\cdot\mathbf{v})\right]\,,
\end{split}
\end{equation}
\normalsize

For both the scalar and tensor waveforms, we obtain similar results in structure to those of ST gravity \cite{Lang:2013fna, Lang:2014osa} (\textit{i.e.,} through redefined parameters in~\eqref{EOMparams}), apart from new terms that explicitly depend on the GB coupling parameter. 
This feature allows to potentially distinguish the two theories in BH-neutron star binaries, where in ST theories only neutron stars can develop scalar hair. For GWs up to 0.5PN order, the only difference with respect to GR is the presence of the factor $\bar{\alpha}$ multiplying the gravitational constant. At 1PN order, we see dependency on the new set of parameters and explicit GB coupling-dependent terms.

In Sec.~\ref{sec:scalarmultipoles}, we saw that the scalar multipoles consist only of compactly supported terms at 0.5PN order, and that the GB contribution to multipoles vanishes (see Eq.~\eqref{1PNGBmultipole} and its subsequent paragraph). Thus, the deviation of the GB scalar waveform from that of ST gravity arises only from the novel GB coupling dependent terms of matter equation of motion (see~Eq.~\eqref{relEOM}) at 1PN order. Also note that, in the GB coupling-dependent terms of the gravitational waveform, we see dependency on the parameters $\mathcal{S}_{+}$ and $\mathcal{S}_{-}$, which are absent in the gravitational waveform of ST theory at 1PN order and start to appear at 1.5PN order.

Overall, most of the terms have the same form as in GR with modified coefficients. We expect more complicated structures to arise at 1.5PN order and beyond, with contributions from far-zone integrals (\textit{i.e.}, hereditary terms), surface EW moments, as well as dipole radiation-reaction terms in the matter equation of motion.

\section{Energy loss rate}\label{sec:VI}
In this section, we compute the energy dissipation due to tensor and scalar radiation from BH binary systems. As we will see in Sec.~\ref{sec:VII}, the energy flux rate is used to determine the phase evolution of GWs.

\subsection{Tensor mode}
The energy loss from tensor waves is given by
\small
\begin{equation}\dot{E}_T=\frac{c^3\,R^{2}}{32 \pi G} \oint \dot{h}_{\mathrm{TT}}^{i j} \dot{h}_{\mathrm{TT}}^{i j} d^{2} \Omega\,.
\end{equation}
\normalsize
We simplify this expression starting from the effects of TT projection operators. Based on the definition given in Eq.~\eqref{TTdef}, it is straightforward to show that
\small
\begin{equation}
\begin{split}
    \left(P^{i k} P^{j l}-\frac{1}{2} P^{i j} P^{k l}\right)&\left(P^{i m} P^{j n}-\frac{1}{2} P^{i j} P^{m n}\right)= P^{k m} P^{l n}-\frac{1}{2} P^{k l} P^{m n}\,.
\end{split}
\end{equation}
\normalsize
Using this identity, the calculation simplifies to
\small
\begin{equation}\label{edotsimp}
\begin{split}
    \dot{E}_{T}=\frac{\mu^2c^3}{32\pi G}\oint&\big(4\dot{Q}^{ij}\dot{Q}^{ij}-8n^{ln}\dot{Q}^{kl}\dot{Q}^{kn}+2n^{klmn}\dot{Q}^{kl}\dot{Q}^{mn}\big)d^2\Omega\,,
\end{split}
\end{equation}
\normalsize
where we use the same notation as introduced in the introduction.
Taking a time derivative and evaluating angular integrals using Eq.\eqref{spatialintegral}, we find that
\small
\begin{equation}\label{Et}
\begin{split}
\dot{E}_T&=\frac{8}{15}\frac{ \eta^2}{G \bar{\alpha}^2 c^5\, }\left(\frac{G\bar{\alpha}m}{r}\right)^4 \Bigg\{(12 v^{2}-11\dot{r}^{2})\\
&+\frac{1}{28 c^2}\bigg[-16\left(170-10 \eta+63 \bar{\gamma}+84 \beta_{+}-84 \frac{\Delta m}{m} \beta_{-}\right) v^{2} \frac{G \bar{\alpha}m }{r}\\
&+(785-852 \eta+336 \bar{\gamma}) v^{4}-2(1487-1392 \eta+616 \bar{\gamma}) v^{2} \dot{r}^{2}+3(687-620 \eta+280 \bar{\gamma}) \dot{r}^{4}\\
&+8\left(367-15 \eta+140 \bar{\gamma}+168 \beta_{+}-168 \frac{\Delta m}{m} \beta_{-}\right) \dot{r}^{2} \frac{G \bar{\alpha}m }{r}+16(1-4 \eta)\left(\frac{G \bar{\alpha}m }{r}\right)^{2}\bigg]\\
&+\frac{3 f'(\phi_{0})\alpha}{4\sqrt{\bar{\alpha}}r^2c^2}\bigg[4\left(\mathcal{S}_{+}+\frac{\Delta m}{m}\mathcal{S}_{-}\right)\Big(-4v^2 \left(18\tilde{E}+13\frac{G \bar{\alpha}m}{r}-45\dot{r}^2\right)\\
&-\dot{r}^2\left(108\tilde{E}+85\frac{G \bar{\alpha}m}{r}-150\dot{r}^2\right)
+54\dot{r}^4\Big)
\frac{ \bar{G \alpha}m}{r}\left(3\mathcal{S}_{+}+\frac{\Delta m}{m}\mathcal{S}_{-}\right)\left(32v^2+56\dot{r}^2\right)\bigg]\\
&+\frac{45f'(\phi_{0})\alpha}{ 8\sqrt{\bar{\alpha}}r^2\, c^2}\left[\mathcal{S}_{+}(1-2\eta)+\mathcal{S}_{-}(1+2\eta)\right]\bigg[-\frac{4}{7}v^2\left(22\tilde{E}+77\frac{G\bar{\alpha}m}{r}-\frac{199}{3}\dot{r}^2+\frac{18}{5}v^2\right) \\
&+\frac{24}{5}\dot{r}^4-\frac{16}{5}v^2\dot{r}^2-\frac{   \dot{r}^2}{7}\left(992\tilde{E}+737\frac{G\bar{\alpha}m}{r}-\frac{2404}{3}\dot{r}^2+\frac{8}{5}v^2\right)\bigg]+\mathcal{O}(c^{-3}) \Bigg\}\,.
\end{split}
\end{equation}
\normalsize
The explicit calculations of GB contributions to the above result can be found in Appendix \ref{GBL}.
Following the PN convention of the waveforms, we call the lowest order piece of the tensor flux to be 0PN, which results from multiplying the $0 \mathrm{PN}$ piece of $h_{TT}^{ij}$ by itself. At $0.5 \mathrm{PN}$ order, there is no tensor flux as the product of $0.5\mathrm{PN}$ and $0\mathrm{PN}$ terms has an odd number of unit vectors and thus vanishes upon angular integration. The remaining terms are at $1 \mathrm{PN}$ order, comprising $0 \mathrm{PN}-1 \mathrm{PN}$, and $0.5 \mathrm{PN}-0.5 \mathrm{PN}$ products.

\subsection{Scalar mode}
For the scalar field, the energy loss is evaluated from
\small
\begin{equation}
\dot{E}_{s}=\frac{c^3 R^{2}}{32 \pi G} \oint \dot{\Phi}^{2} d^{2} \Omega\,.\end{equation}
\normalsize
Since we have defined the lowest-order tensor flux term as a $0\mathrm{PN}$ term, the lowest-order piece of the scalar flux is a
-1PN term, resulting from multiplying the -0.5 PN piece of $\Phi$ by itself. We find the scalar energy flux to 0PN order to be
\small
\begin{equation}\label{Es}
\begin{split}
&\dot{E_{S}}=\frac{\eta^2}{G\bar{\alpha}c^3}\left(\frac{ G\bar{\alpha} m}{r}\right)^{4}
\Bigg\{\frac{4}{3}\mathcal{S}_{-}^{2}+
\frac{8}{15\, c^2} \left(\frac{G \bar{\alpha}m }{r}\left[\left(-23+\eta-10 \bar{\gamma}-10 \beta_{+}+10 \frac{\Delta m}{m} \beta_{-}\right)  \mathcal{S}_{-}^{2}\right.\right.\\
&\left.\left.-2 \frac{\Delta m}{m} \mathcal{S}_{+} \mathcal{S}_{-}\right]\right.+v^{2}\left[2  \mathcal{S}_{+}^{2}+2 \frac{\Delta m}{m}  \mathcal{S}_{+} \mathcal{S}_{-}+(6-\eta+5 \bar{\gamma})  \mathcal{S}_{-}^{2}-\frac{10}{\bar{\gamma}} \frac{\Delta m}{m}  \mathcal{S}_{-}\left(\mathcal{S}_{+} \beta_{+}+\mathcal{S}_{-} \beta_{-}\right)\right.\\
&\left.+\frac{10}{\bar{\gamma}}  \mathcal{S}_{-}\left(\mathcal{S}_{-} \beta_{+}+\mathcal{S}_{+} \beta_{-}\right)\right] 
+\dot{r}^{2}\left[ \frac{23}{2}  \mathcal{S}_{+}^{2}-8 \frac{\Delta m}{m}  \mathcal{S}_{+} \mathcal{S}_{-}+\left(9\eta-\frac{37}{2}-10 \bar{\gamma}\right)  \mathcal{S}_{-}^{2}\right. -\frac{80}{\bar{\gamma}}  \mathcal{S}_{+}\left(\mathcal{S}_{+} \beta_{+}\right.\\
&\left.+\mathcal{S}_{-} \beta_{-}\right)
+\frac{30}{\bar{\gamma}} \frac{\Delta m}{m}  \mathcal{S}_{-}\left(\mathcal{S}_{+} \beta_{+}+\mathcal{S}_{-} \beta_{-}\right) 
\left.\left.-\frac{10}{\bar{\gamma}}  \mathcal{S}_{-}\left(\mathcal{S}_{-} \beta_{+}+\mathcal{S}_{+} \beta_{-}\right)+\frac{120}{\bar{\gamma}^{2}} \left(\mathcal{S}_{+} \beta_{+}+\mathcal{S}_{-} \beta_{-}\right)^{2}\right]\right)\\
& -\frac{\Delta m}{m}\frac{\eta}{6\, c^2}\left(\frac{\alpha f'(\phi_{0})\mathcal{S}_{-}\mathcal{S}_{+}}{\sqrt{\bar{\alpha}}r^2}\right)\left(\mathcal{S}_{+}+\frac{\Delta m}{m}\mathcal{S}_{-}\right)\left[-9\dot{r}^2+3v^2-\frac{2G \bar{\alpha}m}{r}\right]+\mathcal{O}(c^{-3})\Bigg\}\,.
\end{split}
\end{equation}
\normalsize
In the above expression, there is no contribution to the flux at $-0.5 \mathrm{PN}$ order because the product of the $-0.5 \mathrm{PN}$ and $0 \mathrm{PN}$ pieces of $\Phi$ has an odd number of $\hat{n}^{i}$. At 0 PN order, we have $(-0.5 \mathrm{PN})-(+0.5 \mathrm{PN})$, and $0 \mathrm{PN}-0 \mathrm{PN}$ contributions.
\section{Orbit equations and waveforms in the time domain}\label{sec:VII}

Having the energy flux and the conserved energy at hand, we determine the evolution of the orbital frequency and phase, which is needed to generate inspiral-waveform templates.
Such templates are important for parameter inference studies to search for deviations from GR and quantify possible biases in source parameters that may mimic beyond-GR effects.

In this section, we first present the time-domain evolution of tensor and scalar waveforms for arbitrary GB coupling parameters but focusing on quasi-circular binary systems.
In Sec.~\ref{sec:VIID}, we assume the small coupling limit and compare our results for scalar waveforms against the NR results of Ref.~\citep{Witek:2018dmd}.
\subsection{Dynamics of quasi-circular inspirals}

Here, we focus on the dynamics of orbits that are quasi-circular when they enter the sensitivity band of GW detectors. For such orbits, the only departure from circular motion is induced by radiation reaction, which, as we saw earlier in Sec.~\ref{sec:V}, 
does not explicitly appear in the equations of motion to 1PN order. 

Using the relative acceleration (see Eq.~\eqref{relEOM}) to solve for the circular condition $\Ddot{r}=\dot{r}=0$, we derive the angular velocity $\omega$ in terms of the relative separation to be
\small
\begin{equation}\label{omega2}
    \omega^2=\frac{G \bar{\alpha}m}{r^3}\Bigg\{1-\frac{G \bar{\alpha}m}{rc^2}\Bigg[3-\eta+\bar{\gamma}+2\beta_{+}-2\frac{\Delta m}{m}\beta_{-}+\frac{4\alpha f'(\phi_{0})}{\bar{\alpha}^{3/2}r^2}\left(3\mathcal{S}_++\frac{\Delta m}{m}\mathcal{S}_-\right)\Bigg]+\mathcal{O}(c^{-4})\Bigg\}\,.
\end{equation}
\normalsize
It is useful at this stage to distinguish between the two commonly used PN parameters, namely
\small
\begin{equation}
    \begin{split}
      \gamma_{PN} =\frac{G \bar{\alpha}m}{c^2 r}\,,\qquad\quad
      x=\left(\frac{G \bar{\alpha}m\omega}{c^3}\right)^{2/3}\,,
    \end{split}
\end{equation}
\normalsize
which differ from their GR definition by an additional factor of $\bar{\alpha}$.
At leading order, one has $r^3\omega^2=Gm\bar{\alpha}+\mathcal{O}(c^{-2})$. From Eq.~\eqref{omega2}, we find the relation between PN parameters to next-to-leading order,
\small
\begin{equation}\label{gammatox}
    x=\gamma_{PN}\Bigg\{1-\frac{\gamma_{PN}}{3}\Bigg[3-\eta+\bar{\gamma}+2\beta_{+}-2\frac{\Delta m}{m}\beta_{-}+\gamma_{PN}^2\frac{4c^4}{G^2}\frac{\alpha f'(\phi_{0})}{\bar{\alpha}^{7/2}m^2}\left(3\mathcal{S}_++\frac{\Delta m}{m}\mathcal{S}_-\right)\Bigg]+\mathcal{O}(c^{-4})
    \Bigg\}\,,\qquad
\end{equation}
\normalsize
where we have substituted $r=Gm\bar{\alpha}/(c^2 \gamma_{PN})+\mathcal{O}(c^{-4})$ in the GB coupling dependent term. As mentioned earlier, despite the fact that the GB term here renders a term having a similar scaling with the orbital parameters as 3PN terms, it is a 1PN correction and indicates a different physical effect.\\
Using Eq.~\eqref{gammatox} we can express the total binding energy (see Eq.~\eqref{bindingCM}) of circular orbits in terms of the orbital frequency,
\small
\begin{eqnarray}\label{bindingenergy_r}
    E&&=-\frac{\mu c^2 x}{2}\Bigg\{1+x\left[-\frac{3}{4}-\frac{\eta}{12}-\frac{2\bar{\gamma}}{3}+\frac{2\beta_{+}}{3}-\frac{\Delta m}{m}\frac{2\beta_{-}}{3}\right.\nonumber\\
    &&\left.\qquad+\frac{22c^4}{3G^2}\frac{\alpha f'(\phi_{0})}{\bar{\alpha}^{7/2}m^2}x^2\left(3\mathcal{S}_{+}+\frac{\Delta m}{m}\mathcal{S}_-\right)\right]+\mathcal{O}(c^{-4})\Bigg\}.\quad
\end{eqnarray}
\normalsize
\subsection{Gravitational wave phase evolution}\label{timephase}
In order to compute the full time-dependent waveforms, we need the evolution of the orbital phase angle $\varphi$, obtained from the angular velocity.
In the adiabatic approximation, {\it i.e.} $\dot{\omega}/ \omega^2 \ll 1$, the luminosity of GWs is equal to the change in orbital energy averaged over a period, leading to the energy balance equation
\small
\begin{equation}\label{balance}
    \frac{dE(x)}{dt}=-\mathcal{F}(x)\,,
\end{equation}
\normalsize
with $\mathcal{F}(x)$ being the total energy flux rate. Using $\dot{\varphi}=\omega$, the balance equation can be reformulated to
\small
\begin{equation}\label{orbphase_t}
\frac{d \varphi}{d t}-\frac{c^3 x^{3}}{G\bar{\alpha}m} =0\,, \qquad
\frac{d x}{d t}+\frac{\mathcal{F}(x)}{ E^{\prime}(x)} =0\,,
\end{equation}
\normalsize
where $E'(x)$ is the binding energy derivative with respect to the PN parameter $x$.
In a PN approximation, this pair of differential equations can be solved in different ways, depending on how one chooses to expand the ratio $ \mathcal{F}/ E'$. Here, we choose the so-called \textit{Taylor T4} approximant, which is obtained from expanding the aforementioned ratio to the consistent PN order as a whole. For a review on different approximants, see Ref.~\cite{Buonanno:2009zt}.

From Eq.~\eqref{bindingenergy_r}, we derive $E'(x)$ to relative 1PN to be
\small
\begin{equation}\label{Ev}
    \begin{split}
      &E'(x)=-\frac{\mu c^2}{2} \left[1+E'_{2}x+\mathcal{O}(c^{-4})\right]\,,\\
     &E'_2=-\frac{3}{2}-\frac{\eta}{6}-\frac{4\bar{\gamma}}{3}+\frac{4}{3}\left(\beta_{+}-\frac{\Delta m}{m}\beta_{-}\right)+\frac{88 c^4}{3 G^2}\frac{\alpha f'(\phi_{0})}{
    m^2\bar{\alpha}^{7/2}}\, x^2\left(3\mathcal{S}_++\frac{\Delta m}{m}\mathcal{S}_-\right)\,.
    \end{split}
\end{equation}
\normalsize
The total energy flux, constituting Eqns.~\eqref{Et} and~\eqref{Es}, has the overall structure
\small
\begin{equation}\label{flx}
    \mathcal{F}=\mathcal{F}_{-1,S}+\mathcal{F}_{0,S}+\mathcal{F}_{0,T}+\mathcal{F}_{1,T}\,,
\end{equation}
\normalsize
where the lower-index indicates the PN order of each term. Note that we calculated 0.5PN corrections to the leading order scalar waveform, which
corresponds to a scalar energy flux that is complete at 0PN order, as the leading term is of -1PN order. Since we calculated the tensor flux to 1PN order, the 0.5 and 1PN scalar contributions to the total flux remain undetermined. 
The inclusion of these terms would increase the overall energy flux and thus the phase differences between sGB and GR.

To proceed with the expansion of $\mathcal{F}(x)/E'(x)$, we shall distinguish between the regime where the scalar dipole flux dominates over the leading-order tensor flux, and vice versa. These regimes are commonly referred to as the scalar dipolar driven (DD) regime with $\mathcal{F}_{-1,S}$ as the dominant term, and the tensor quadrupolar driven (QD) regime where, instead, $\mathcal{F}_{0,T}$ dominates. The DD regime is relevant for low frequencies
\small
\begin{equation}\label{DDtoQG}
    x_{\rm DD}\ll \frac{5}{24}{\cal S}_-^2\bar\alpha,\quad {\rm or} \quad f_{\rm GW}^{\rm DD}\ll \left(\frac{5}{24}\right)^{3/2} \frac{c^3\, {\cal S}_-^2 \, \sqrt{\bar\alpha}}{\pi G m}.
\end{equation}
\normalsize
At much higher frequencies than this condition, the system is in the QD regime. Below, we find the 1PN expansion of $\mathcal{F}(x)/E'(x)$ in the two regimes.
\subsubsection{Dipolar driven regime}
For systems with large scalar dipole or very large separation, factoring out the dipolar scalar flux in Eq.~\eqref{flx} gives,
\small
\begin{equation}
\begin{split}
    \mathcal{F}^{DD}(x)=\frac{4\eta^2\mathcal{S}^{2}_{-} c^5x^4}{3\bar{\alpha}G }&\left\{ 1+f^{DD}_{2}x+f^{DD}_{3}x^{3/2}+f^{DD}_{4}x^2+\mathcal{O}(c^{-5})\right\}\,.
\end{split}
\end{equation}
\normalsize
For 1PN corrections to the phase at this regime, it is sufficient to keep the factor $f_2^{DD}=(\mathcal{F}_{0,S}+\mathcal{F}_{0,T})/\mathcal{F}_{-1,S}$, being
\small
\begin{equation}
\begin{split}
    &f_2^{DD}=\frac{24}{5\bar{\alpha}\mathcal{S}_{-}^2}+\frac{4}{5}\left(\frac{\mathcal{S}_{+}}{\mathcal{S}_{-}}\right)^2-\frac{20}{3}\left(\beta_{+}-\frac{\Delta m}{m}\beta_{-}\right)-\frac{54}{5}+\frac{4}{3}\eta
    -\frac{10}{3}\bar{\gamma}+\frac{4}{\bar{\gamma}}\frac{\mathcal{S}_{+}}{\mathcal{S}_{-}}\left(\beta_{-}-\frac{\Delta m}{m}\beta_{+}\right)\\
    &+\frac{4}{\bar{\gamma}}\left(\beta_{+}-\frac{\Delta m}{m}\beta_{-}\right)
    -x^2\frac{c^4}{G^2}\frac{2\alpha f'(\phi_0)}{m^2\bar{\alpha}^{5/2}}\left[\frac{13}{3\bar{\alpha}}\left(3\mathcal{S}_{+}+\frac{\Delta m}{m}\mathcal{S}_-\right)
    +\eta\frac{\Delta m}{16m}\frac{\mathcal{S}_{+}}{\mathcal{S}_{-}}\left(\mathcal{S}_{+}+\frac{\Delta m }{m}\mathcal{S}_{-}\right)\right]\,.
\end{split}
\end{equation}
\normalsize
The last term in the above equation comes from the leading order term of the tensor flux. The rest of the terms are the 0PN scalar flux terms. Expanding the ratio $\mathcal{F}^{DD}(x)/E'(x)$ to 1PN order we find
\small
\begin{equation}\label{DDF/E}
    \frac{\mathcal{F}^{DD}(x)}{E'(x)}=-\frac{8\eta c^3\mathcal{S}_{-}^{2}x^4}{3 G\bar{\alpha}m}\left[1+(f_{2}^{DD}-E'_{2})x+\mathcal{O}(c^{-4})\right]\,.
\end{equation}
\normalsize
Note that this expression is complete at the relative 1PN order.
\subsubsection{Quadrupolar driven regime}
For quadrupolar driven systems, the flux is expanded about the Newtonian-order term $\mathcal{F}_{0,T}$ such that
\small
\begin{equation}
  \mathcal{F}^{QD}(x)=\frac{32\eta^2 c^5 x^5 }{5G\bar{\alpha}^2}\left[\xi+f_{2,T}^{nd}x+\mathcal{O}(c^{-3})\right]+x^4 \frac{4\eta^2 \mathcal{S}_{-}^2}{3G\bar{\alpha}}\,.
\end{equation}
\normalsize
The factor $\xi=5\bar{\alpha}f_{2}^{DD}\mathcal{S}_{-}^2/24$, captures the 0PN flux terms.
For the coefficient $f_{2}^{nd}=\mathcal{F}_{1,T}/\mathcal{F}_{0,T}\equiv f_{2,T}^{nd}+f_{2,GB}^{nd}$ we have
\small
\begin{eqnarray}\label{f2nd}
f_{2,T}^{nd}=&&
\frac{-1}{336}\left[1247+980\eta+448\bar{\gamma}+896\left(\beta_{+}-\frac{\Delta m}{m}\beta_{-}\right)\right]\nonumber\\
f_{2,GB}^{nd}=&&-\frac{c^4}{G^2}\frac{f'(\phi_{0})\alpha\mathcal{S}_{+}}{m^2\bar{\alpha}^{5/2}} x^{2}\Bigg[12+\frac{495}{28}(1-2\eta)+\frac{32}{3}\frac{\mathcal{S}_{-}}{\mathcal{S}_{+}}\frac{\Delta m}{m}+\frac{495}{28}\frac{\mathcal{S}_{-}}{\mathcal{S}_{+}}(1+2\eta)\Bigg]\,,
\end{eqnarray}
\normalsize
As mentioned above, these expansions miss a  contribution from $\mathcal{F}_{1,S}$ which we are unable to compute within our approximations, having $\dot{E}_S$ to 0PN order. The inclusion of the missing term would, in principle, increase the overall energy flux, and thus the difference between sGB and GR waveforms. In the following, we set these contributions to zero, similar to what was done in a similar situation for tidal effects in GR~\cite{Damour:2012yf}.\\
Overall, the expansion of $\mathcal{F}^{QD}(x)/E'(x)$ to 1PN order becomes
\small
\begin{eqnarray}\label{QDF/E}
\frac{\mathcal{F}^{QD}(x)}{E'(x)}=&&-\frac{64\eta c^3 x^5}{5Gm\bar{\alpha}^2}\left[\left(\xi-E'_{2}\frac{5\bar{\alpha}\mathcal{S}_{-}^2}{24}\right)+(f_{2,T}^{nd}-\xi E'_{2})x+\frac{5\bar{\alpha}\mathcal{S}_{-}^2}{24}x^{-1}+\mathcal{O}(c^{-4})\right]\,.
\end{eqnarray}
\normalsize

Having Eqns.~\eqref{DDF/E} and~\eqref{QDF/E}, together with an adequate choice of parameters, we numerically solve Eq.~\eqref{orbphase_t} to determine the time evolution of $\varphi$ and $\omega$ in different regimes, and therefore, time-domain waveforms.
\subsection{Gravitational waveforms in the time domain}
\begin{figure}
\includegraphics[width=.25\columnwidth]{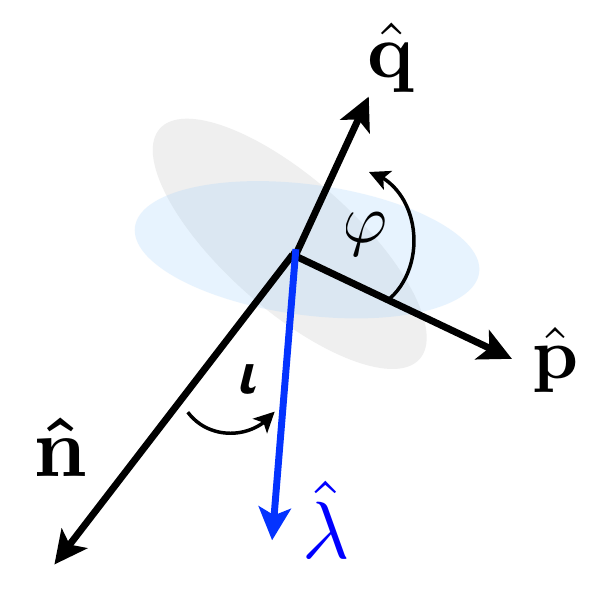}
\caption{Orientation of the orbital plane and the sky plane based on the orthonormal triad $\{\hat{\mathbf{n}},\hat{\mathbf{p}},\hat{\mathbf{q}}\}$. $\mathbf{\hat{p}}$ lie along the lines of nodes and defines an origin for the orbital phase angle $\varphi$. $\hat{\lambda}$ is the unit vector of the orbital angular momentum.}
\label{fig:triad}
\end{figure}

To derive time-domain tensor and scalar waveforms, we parametrize the orbital motion by choosing the standard convention for the direction and orientation of the orbits, namely
the orthonormal triad $\{\hat{\mathbf{n}},\hat{\mathbf{p}},\hat{\mathbf{q}}\}$, with $\hat{\mathbf{n}}$ being the radial direction to the observer, $\hat{\mathbf{p}}$ lying along the intersection of the orbital plane with the plane of the sky, and $\hat{\mathbf{q}}=\hat{\mathbf{n}}\times \hat{\mathbf{p}}$.
A schematic view of this choice can be found in Fig.~\ref{fig:triad}. 

The normal to the orbit is inclined at an angle $i$ relative to $\hat{\mathbf{n}}$. The orbital phase angle of body $A$ is measured from the line of nodes in a positive sense. We have that
\small
\begin{equation}\label{convwave}
\begin{split}
&\hat{\mathbf{n}}_{AB}=\hat{\mathbf{p}} \cos \varphi+(\hat{\mathbf{q}} \cos i+\hat{\mathbf{n}} \sin i) \sin \varphi\,,\\ &\hat{\boldsymbol{\lambda}}=-\hat{\mathbf{p}} \sin \varphi+\left(\hat{\mathbf{q}}\cos  i+\hat{\mathbf{n}} \sin i\right) \cos \varphi\,,
\end{split}  
\end{equation}
\normalsize
where $\boldsymbol{v}= r \omega \hat{\boldsymbol{\lambda}}$ for circular orbits.

Gravitational wave detectors measure a 
linear combination of polarization waveforms $h_{+}(t)$ and $h_{\times}(t)$ that are defined by the projections
\small
\begin{equation}\label{polariz}
\begin{split}
&h_{+}=\frac{1}{2}\left(\hat{p}_{i} \hat{p}_{j}-\hat{q}_{i} \hat{q}_{j}\right) h^{i j}\,,\qquad h_{\times}=\frac{1}{2}\left(\hat{p}_{i} \hat{q}_{j}+\hat{q}_{i} \hat{p}_{j}\right) h^{i j}\,.
\end{split}
\end{equation}
\normalsize
Applying Eqns.~\eqref{convwave} and~\eqref{polariz} on Eq.~\eqref{waveformfinale} and simplifying combinations such as $\hat{n}^{i} \hat{n}^{j}, \hat{\lambda}^{i} \hat{\lambda}^{j},$ and $\hat{n}^{(i} \hat{\lambda}^{j)}$ , we find the 1PN polarization waveforms as functions of angular phase and orientation. The corresponding ready-to-implement waveforms can be found in Appendix \ref{polarizations}.

\begin{figure*}[hbt!]
\centering
\includegraphics[width=1.\textwidth]{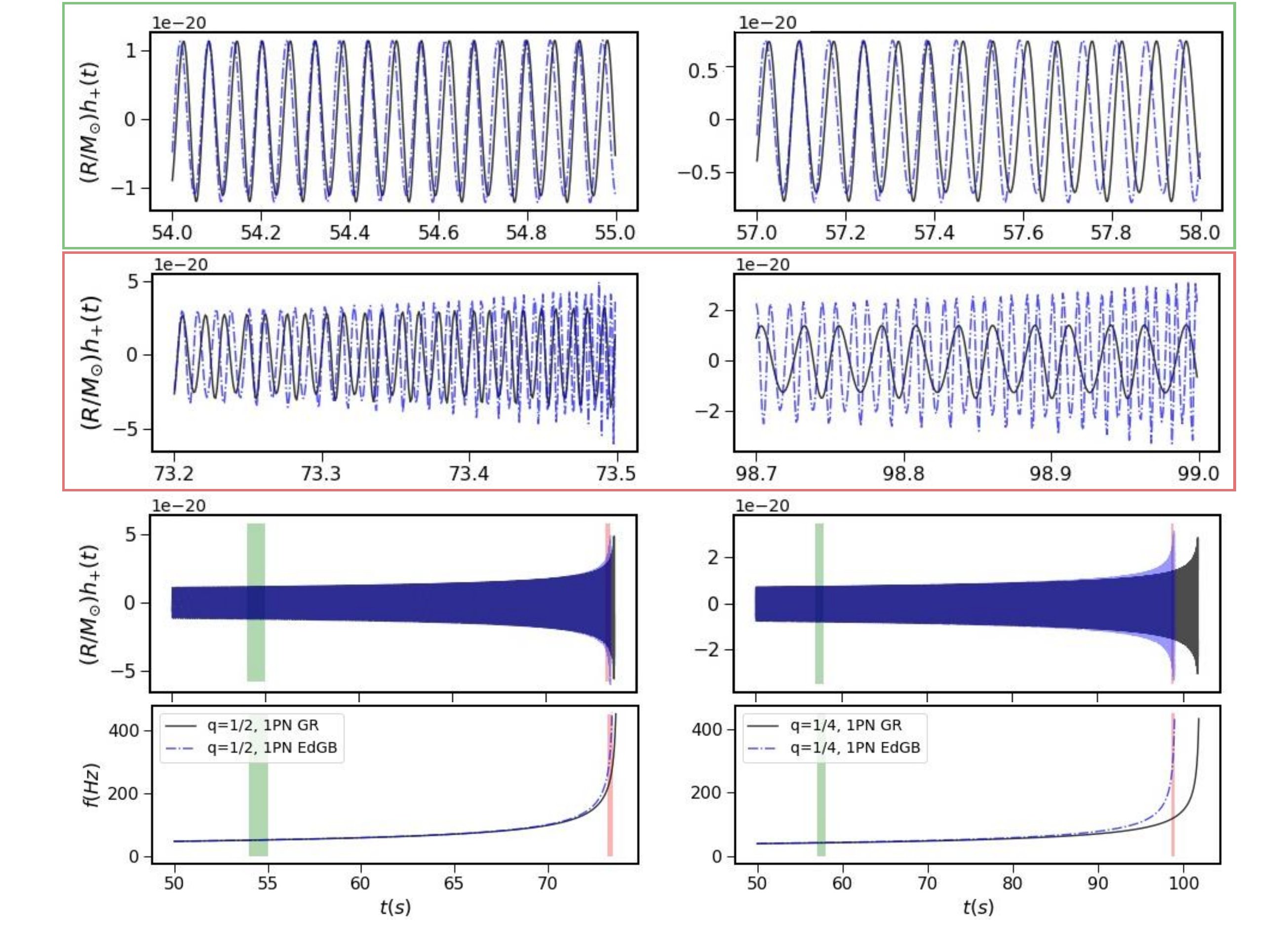}
\caption{Time evolution of gravitational waveform and orbital frequency for a $m=15M_{\odot}$ binary with $q=1/2$ (left) and $q=1/4$ (right), and $\alpha=0.01m^2$. Blue dashed curves correspond to EdGB waveforms and black curves correspond to 1PN GR waveforms. Orbits are viewed edge-on ($i=\pi/2$), and $t=0$ corresponds to $f\approx 50 Hz$.  The green shaded regions (top) show $1s$ interval of the waveforms at intermediate times, whilst red shaded regions (middle) show the $0.3s$ intervals where EdGB binaries are close to the merger.}
\label{waveformm15}
\end{figure*}

We first display our results
in Fig.~\ref{waveformm15}
by plotting analytical GWs of BH binary systems in EdGB theory and comparing them with 1PN GR waveforms. Using Eq.~\eqref{BHcharge}, we write BH scalar charges, and the related parameters, in terms of $\alpha$. The GB coupling parameter $\alpha$ is thus the only parameter that we have to specify.
Exemplarily, we choose $\alpha=0.01m^2$. 
Also note that, as shown In Sec.~\ref{Swaveall}, the difference between the ssGB and EdGB phase evolution is relatively small compared to their overall deviations from the GR phasing. As a result, we only present the waveforms for the EdGB case.

By its very nature, the deviation from GR is most prominent in the high curvature regime, i.e., for small-mass BHs. Therefore, we consider quasi-circular BH binaries with a total mass of $m=15M_{\odot}$ and mass ratios $q=1/2$ and $q=1/4$.
For $q=1$ binaries, the deviations from GR waveforms are expected to be smaller due to the suppressed dipole radiation, as can also be seen in Fig.2 of Ref.~\cite{Shiralilou:2020gah}.

Figure~\ref{waveformm15} shows waveforms in the aforementioned two cases. The observer is viewing the orbit edge on, so that $i=\pi/2$, and thus only the $h_{+}(t)$ polarization is present. 
Due to the dissipation of energy in scalar field radiation, the inspiral of BH binaries in sGB is accelerated as compared to their GR counterpart and results in a GW phase shift. The scalar charge, and hence the scalar radiation, increases as the mass ratio decreases for fixed total mass of the binary. 
As can be seen in Fig.~\ref{waveformm15}, the mass ratio has a high impact on the GW signal, which exhibits a larger phase shift for smaller mass ratios.\\
Throughout the early inspiral evolution of the binaries,
in both cases, the amplitudes of EdGB waves do not deviate significantly from the GR waveforms. This is because the GB coupling dependent terms are suppressed at large distances compared to other 1PN terms (see the discussion below Eq.~\eqref{2indexEW}). Despite the amplitudes, the dephasing between EdGB and GR waveforms starts early in the evolution and could thus be a detectable feature.

\subsection{Scalar waveforms in time domain}\label{sec:VIID}
For scalar waveforms, similar steps to the ones described in the previous subsection should be taken in order to rewrite Eq.~\eqref{scalarwavefinale} in terms of the orbital phase and inclination angle. The final ready-to-implement 0.5PN order scalar waveform is reported in Appendix~\ref{polarizations}.

In Fig.~\ref{swave} we display the scalar waveform emitted by quasi-circular binaries, corresponding to the gravitational waveforms shown in Fig.~\ref{waveformm15}.
In these examples, for the mass ratio $q=1/2$, the scalar wave amplitude is suppressed with respect to the gravitational wave amplitude by an order of magnitude. 
As we decrease the mass ratio to $q=1/4$, the increase of the scalar field amplitude during the late inspiral becomes stronger and, in fact, the amplitude becomes comparable to that of the gravitational radiation. In general, one expects the scalar radiation to increase with decreasing the mass ratio while keeping the total mass fixed. However, we should note that the PN approach is not reliable for extreme-mass-ratio systems.


We next compare PN scalar waveforms against those resulting from NR simulations of BH binary systems during the inspiral.
For this purpose,
we use spherical harmonics $Y_{lm}$ to decompose the scalar radiation into its radiative modes
\small
\begin{equation}\label{decomp}
\Phi_{lm}(t,R)=\int d\Omega\, \Phi(t,R,\hat{\Theta},\hat{\Phi})Y^*_{lm}(\hat{\Theta},\hat{\Phi})\,,
\end{equation}
\normalsize
where, in order to express $\Phi$ in terms of spherical coordinate variables, we choose to work with $\hat{\boldsymbol{p}}=-\bf{e}_{\hat{\Phi}}$, $\hat{\boldsymbol{q}}=\bf{e}_{\hat{\Theta}}$, such that $\hat{\Theta}=i$ and $\hat{\Phi}=\pi/2-\omega t$.
It is easy to verify that the leading-order contribution to the radiation comes from $\Phi_{l=m}$ modes, $i.e.$,
\small
\begin{equation}\label{integmodes}
   \Phi_{ll}(t,R)=\int d\Omega\,\Phi_{l}Y^{*}_{ll} (\hat{\Theta}, \hat{\Phi})\,.
\end{equation}
\normalsize
We use Eq.~\eqref{BHcharge} to re-write the scalar-charge dependent parameters in terms of the GB coupling. For ssGB gravity we have
\small
\begin{eqnarray}
      \alpha_A^0&&=-\frac{\alpha}{m_A^2}+\mathcal{O}(\alpha^2)\,,\quad
      \bar{\alpha}=1+\frac{\alpha^2}{2m_A^2 m_B^2}+\mathcal{O}(\alpha^3)\,,\nonumber\\
      \mathcal{S}_{+}&&=-\frac{\alpha}{2\sqrt{\bar{\alpha}}}\left(\frac{1}{m_A^2}+\frac{1}{m_B^2}\right)+\mathcal{O}(\alpha^2)\,,\nonumber\\
      \mathcal{S}_{-}&&=-\frac{\alpha}{2\sqrt{\bar{\alpha}}}\left(\frac{1}{m_A^2}-\frac{1}{m_B^2}\right)+\mathcal{O}(\alpha^2)\,,
\end{eqnarray}
\normalsize
with the rest of the parameters being identically zero.\\
Specifying Eq.~\eqref{Swavemodes} to quasi-circular orbits with the above-mentioned parameters, we find the various $\Phi_{l}$ modes, required for Eq.~\eqref{integmodes}, to be
\small
\begin{equation}
    \begin{split}
     \Phi_1&=\frac{G\alpha}{c^2 R}\left(\frac{\Delta m}{m_Am_B}x_0^{1/2}-\frac{m\Delta m }{2m_A m_B}x_0^{3/2}\right)\,,\\
     \Phi_2&=-\frac{G\alpha}{c^2 R}\frac{(m_B^2-m_Am_B+m_A^2)}{m m_A m_B}x_0\,,\\
     \Phi_3&=-\frac{G\alpha}{c^2R}\frac{\Delta m (m_A^2+m_B^2)}{8m^2m_Am_B}x_0^{3/2}\,,
    \end{split}
\end{equation}
\normalsize
where the parameter $x_{0}=(Gm\omega/c^3)^{2/3}+\mathcal{O}(\alpha)$ is the leading order term of the PN parameter $x$.

\begin{figure*}[hbt!]
\centering
\includegraphics[width=1.\textwidth]{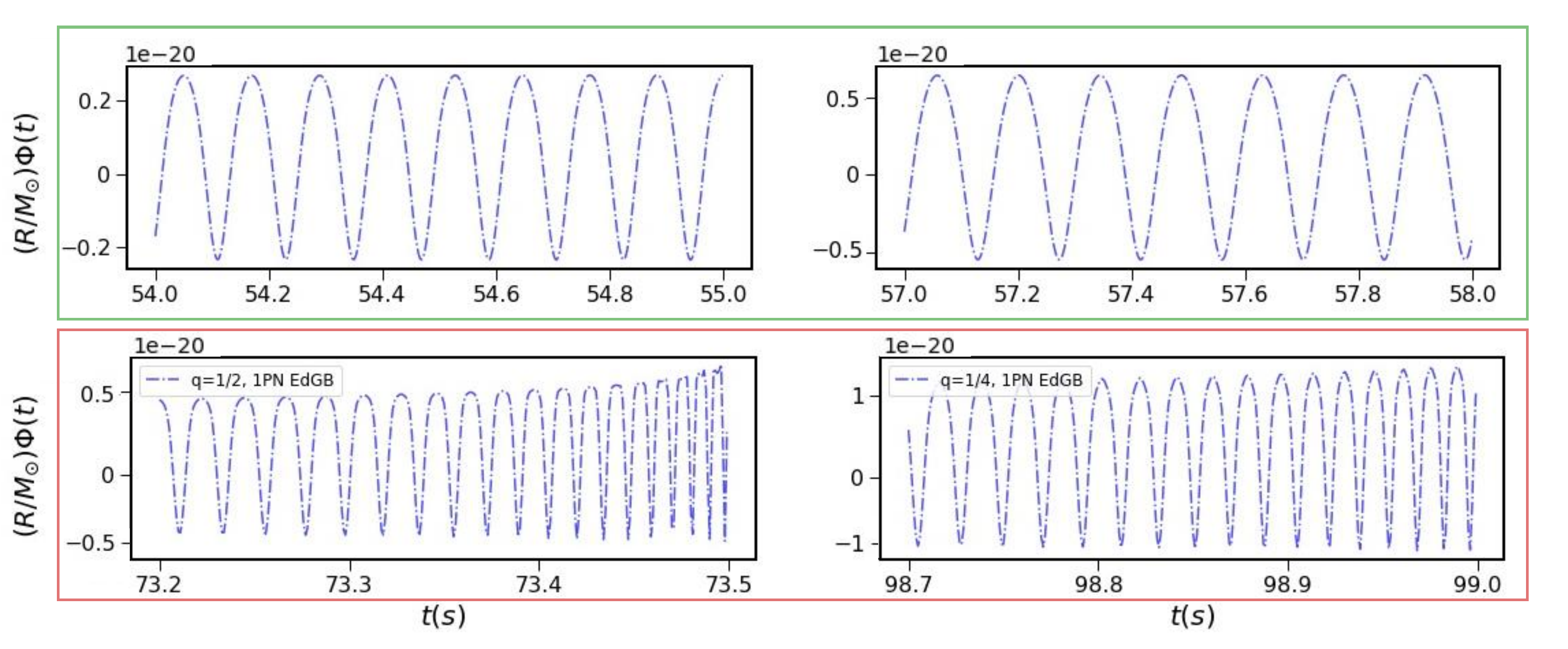}
\caption{The time evolution of scalar field waveform for $m=15M_{\odot}$ binary with (left) $q=1/2$ and (right) $q=1/4$, and $\alpha=0.01m^2$, corresponding to Fig.~\ref{waveformm15} binaries and time intervals. Orbits are viewed edge-on ($i=\pi/2$) and thus the waves deviate from a full sinusoidal.}
\label{swave}
\end{figure*}

We compare our PN scalar waveforms against previous PN calculations of Ref.~\cite{Yagi:2011xp} and also the results from a numerical relativity simulation reported in Ref.~\cite{Witek:2018dmd}, with data kindly provided by the authors. This NR simulation implements an effective-field theoretical approach, which is valid to first order in the GB coupling parameter $\alpha$ in ssGB theory, and it covers
about ten orbits before the binary BHs merge.
As expected, our results agree with those
of Ref.~\citep{Witek:2018dmd}.
for the leading PN terms of each mode. Also, our calculations expand these results to next-to-leading order,
deriving a 0.5PN correction to the dipole amplitude $\Phi_{11}$ at order $\mathcal{O}(\alpha)$. In general, we also derive 0.5PN corrections to $\Phi_{11}$ that are of higher order in the GB coupling constant. We neglect these corrections here as the comparison versus NR results only requires the $\mathcal{O}(\alpha)$ corrections. 

In order to have a meaningful comparison between the new PN dipolar radiation terms and the results of Ref.~\citep{Witek:2018dmd}, we obtain the orbital frequency evolution from the derivative of the GW phase of numerical data, instead of using the analytical results of Sec.~\ref{timephase}. This means that the comparison between PN and NR results concerns only the amplitudes of the waveforms, while their phases agree by definition.

\begin{figure}[htp!]
\centering
    \includegraphics[width=1.03\linewidth]{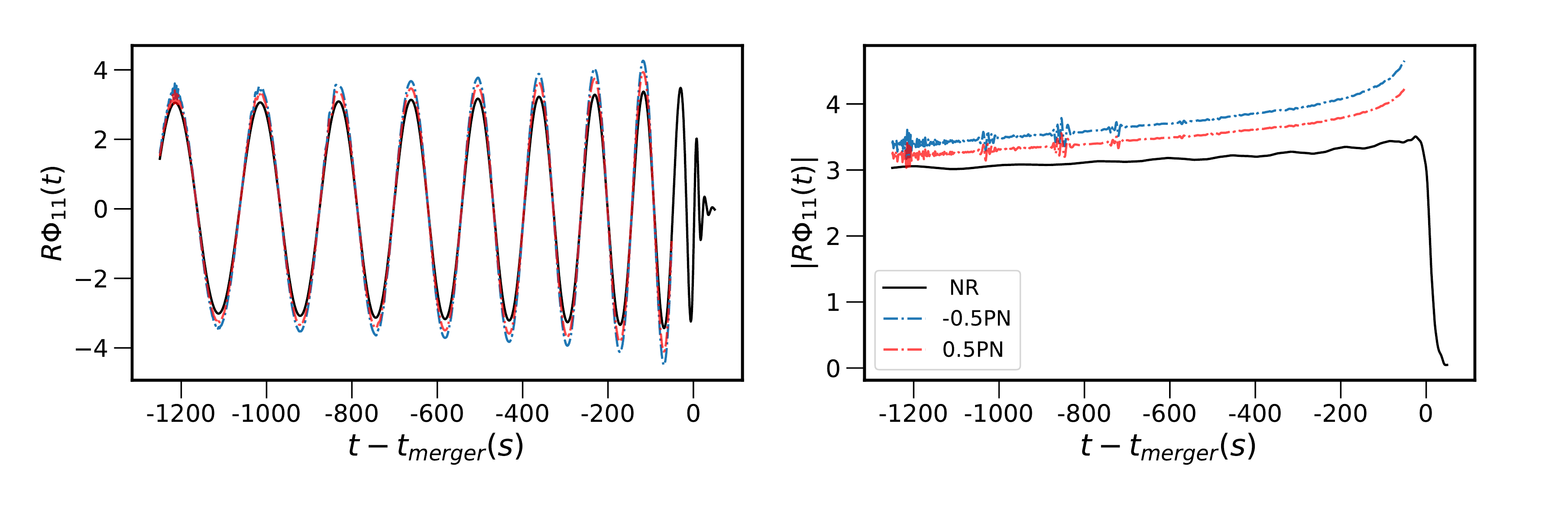}
    \caption{Time evolution of the scalar mode $\Phi_{11}$ (left) and its absolute value (right), re-scaled by the extraction radius $R=100 m$. Black curves indicate results from NR simulations. The blue dashed curve indicates the analytical PN inspiral results of Ref.~\cite{Yagi:2011xp} to -0.5 PN order. The red dashed curve shows the 0.5PN order results of this paper.
    }
    \label{fig:dipoleSq12}
\end{figure}

In Fig.~\ref{fig:dipoleSq12} we show the NR scalar waveform for the $l=m=1$ mode during the inspiral phase, and its comparison to the analytical results of this paper (red dashed curve) and that of Ref.~\citep{Yagi:2011xp} (blue dashed curve). The binary has a mass ratio of $q=1/4$ and extraction radius $R=100\, m$. The waveforms are shifted in time such that $t=0$ marks the merger time. We initially align the waves by minimizing the squared difference between the amplitude of the waveforms over an extended window in time,
$t-t_{\text{merger}}:\{-1410,-900\}$. In this way, we avoid spliting the waves by introducing arbitrary fudge factors.

Overall, we see an approximate factor of $1.5$ improvement in the matching of analytical results to numerical data due to the new 0.5PN contributions, as compared to the results of Ref.~\cite{Yagi:2011xp}.
Comparing the $q=1/2$ waveforms shows a similar overall factor of improvement, and thus we do not show the plot here.
The noise in the analytical amplitude is due to the precision of numerical data from which the phase evolution is extracted and does not affect the overall results.
Based on these results, we claim that our analytical waveforms can thus be used as benchmarks for NR simulations.

\section{Phase evolution in Fourier domain }\label{sec8}
For the purpose of data analysis, and since GW measurements are mainly sensitive to the phase evolution of waveforms, it is useful to provide
waveforms in the Fourier domain. For instance, in theory-agnostic tests of gravity, the parametrized models used by the LVC in \textit{e.g.}, Refs.~\cite{LIGOScientific:2019fpa,TheLIGOScientific:2016src},
are also based on the analysis of GWs in the Fourier domain.

Here, we derive the 1PN Fourier phase analytically by following roughly the same steps as in Sec.~\ref{timephase}. Using the stationary phase approximation (SPA) and focusing on waveforms with GW Fourier phase $\psi$ at frequency $f=\omega/\pi$, the waveform in the frequency domain can be written as
\small
\begin{equation}
\begin{split}
\tilde{h}^{\mathrm{spa}}(f)&=a\left(t_{f}\right)\sqrt{\frac{2\pi}{m \dot{\omega}\left(t_{f}\right)}} e^{-i\left[\psi_{f}\left(t_{f}\right)+\pi / 4\right]}\,,\qquad \psi_{f}(t)\equiv 2 \phi(t)-2 \pi f t\,,
\end{split}
\end{equation}
\normalsize
where $t_{f}$ is the time when the GW frequency becomes equal to the Fourier variable $f$, by solving $d \psi_{f}(t) / d t=0$. $a(t_{f})$ represents the amplitude of the mode with frequency $f$.
Using the so-called \textit{Taylor F2}
approximation, we can solve for the time and orbital phase through
\small
\begin{equation}\label{TaylorF2}
\begin{split}
t(\bar{v}) &=t(\bar{v}_{\mathrm{ref}})+ \int_{\bar{v}}^{\bar{v}_{\mathrm{ref}}} \frac{E^{\prime}(\bar{v})}{\mathcal{F}(\bar{v})} d \bar{v}\,, \\
\phi(\bar{v}) &=\phi(\bar{v}_{\mathrm{ref}})+\frac{c^3}{G\bar{\alpha}m} \int_{\bar{v}}^{\bar{v}_{\mathrm{ref}}}\bar{v}^{3} \frac{E^{\prime}(\bar{v})}{\mathcal{F}(\bar{v})} d \bar{v}\,,
\end{split}
\end{equation}
\normalsize
where $\bar{v}=x^{1/2}=(G \bar{\alpha}m\omega/c^3)^{1/3}$ and the subscript $ref$ refers to the choice of reference point in the evolution. The gravitational phase is thus given by
\small
\begin{equation}
    \psi_{f}(t_{f})=2\left.\left(\varphi(v)-\frac{c^3}{G  \bar{\alpha}m} \bar{v}^{3} t(\bar{v})\right)\right|_{\bar{v}=\bar{v}_{f}}\,,
\end{equation}
\normalsize
where $\bar{v}_{f} \equiv (\pi G m\bar{\alpha} f/c^3)^{1 / 3}$.
As before, we split the calculation into DD regime and QD regime, and evaluate Eq.~\eqref{TaylorF2}  term-by-term by re-expanding the expressions in the PN variable $\bar{v}$ and truncating them at 1PN order.
 
\subsection{Dipolar-driven regime}
In the DD regime, expanding the ratio $E'(\bar{v})/\mathcal{F}^{DD}(\bar{v})$ to 1PN order gives
\small
\begin{equation}\label{ratioDDfreq}
    \frac{E'(\bar{v})}{\mathcal{F}^{DD}(\bar{v})}=-\frac{3G \bar{\alpha}m}{4 c^3\eta \mathcal{S}_{-}^2\bar{v}^7}\left[1+\left(E'_{0}-f_2^{DD}\right)\bar{v}^2+\mathcal{O}\left(c^{-4}\right)\right]\,.
\end{equation}
\normalsize
By substituting Eq.~\eqref{ratioDDfreq} into Eq.~\eqref{TaylorF2}, and integrating term-by-term we find the phasing to be
\small
\begin{equation}\label{FphDD}
\begin{split}
    \psi(t_f)&=-\frac{1}{4\eta \mathcal{S}_{-}^2 \bar{v}_{f}^3}
    \Bigg[1+\frac{9}{2}\rho^{DD}\bar{v}_f^2+6\rho_{GB}^{DD}\log(\bar{v}_{f})\bar{v}_{f}^6+\left(\frac{\bar{v}_f}{\bar{v}_{ref}}\right)^6\Big(1+\frac{3}{2}\bar{v}_{ref}^2 -6\rho_{GB}^{DD}\bar{v}_{ref}^6 \log(\bar{v}_{ref})\Big)\\
    &-2\left(\frac{\bar{v}_f}{\bar{v}_{ref}}\right)^3\Big(1+3\rho^{DD}\bar{v}_{ref}^2-\rho^{DD}_{GB}\bar{v}_{ref}^6\Big)
    \Bigg]+\phi(\bar{v}_{ref})-2\pi f t(\bar{v}_{ref})\,,
\end{split}
\end{equation}
\normalsize
where one can easily verify that
\small
\begin{eqnarray}
\rho^{DD}&&=\frac{93}{10}-\frac{3}{2}\eta+2\bar{\gamma}+\frac{24}{3}\left(\beta_{+}-\frac{\Delta m}{m}\beta_{-}\right)-\frac{4}{5}\left(\frac{\mathcal{S}_+}{\mathcal{S}_-}\right)^2\nonumber\\
&&-\frac{24}{5\bar{\alpha}\mathcal{S}_-^2}-\frac{4}{\bar{\gamma}}\frac{\mathcal{S}_{+}}{\mathcal{S}_-}\left(\beta_{-}-\frac{\Delta m}{m}\beta_{+}\right)-\frac{4}{\bar{\gamma}}\left(\beta_{+}-\frac{\Delta m}{m}\beta_{-}\right)\,,\nonumber\\
\rho^{DD}_{GB}&&=\frac{c^4}{G^2}\frac{\alpha f'(\phi_{0})}{m^2 \bar{\alpha}^{5/2}}\left[\frac{104}{3\bar{\alpha}}\left(3\mathcal{S}_{+}+\frac{\Delta m}{m}\mathcal{S}_{-}\right)+\eta\frac{\Delta m}{16m}\frac{\mathcal{S}_{+}}{\mathcal{S}_-}\left(\mathcal{S}_{+}+\frac{\Delta m}{m}\mathcal{S}_{-}\right)\right]\,.
\end{eqnarray}
\normalsize
\subsection{Quadrupolar-driven regime}
For the QD regime, by definition (see Eq.~\eqref{DDtoQG}), we expect the dipole radiation to be negligible. This means that we can use the parameter $\mathcal{S}_{-}$ as an identifier of small terms, and thus, split the flux into two pieces as follows,
\small
\begin{equation}
\begin{split}
&\mathcal{F}^{\mathrm{QD}}=\mathcal{F}_{\mathrm{non}-\mathrm{dip}}+\mathcal{F}_{\mathrm{dip}}\,,\\
\\
&\mathcal{F}_{\text {non-dip }}  \equiv \lim _{\mathcal{S}_{-} \rightarrow 0} \mathcal{F}\,, \qquad\mathcal{F}_{\text {dip }}  \equiv \mathcal{F}-\mathcal{F}_{\text {non-dip}}\,.\end{split}\end{equation}
\normalsize
where $\mathcal{F}_{\rm dip}$ indicates the tensor and scalar flux terms that depend on $\mathcal{S}_{-}$ and $\mathcal{F}_{\rm{non-dip}}$ denotes the terms that do not depend on $\mathcal{S}_{-}$.

With this division, to first order in the small quantity $\mathcal{F}_{\text {dip }} / \mathcal{F}_{\text {non-dip}}$, the ratio $E'(\bar{v})/\mathcal{F}(\bar{v})$ can be approximated as
\small
\begin{equation}\label{approxQD}\frac{E'(\bar{v})}{\mathcal{F}(\bar{v})} \simeq\frac{E'(\bar{v})}{\mathcal{F}_{\text {non-dip }}(\bar{v})  }\left(1-\frac{\mathcal{F}_{\text {dip }}(\bar{v})}{\mathcal{F}_{\text {non-dip }}(\bar{v})}\right)\,.
\end{equation}
\normalsize
We obtain the dipolar and non-dipolar parts to be
\small
\begin{eqnarray}
    &&\mathcal{F}_{\text {non-dip }}(\bar{v})=\frac{32\eta^2 \bar{\xi} c^5}{5G\bar{\alpha}^2}\bar{v}^{10}\left[1+f_{2}^{nd}
    \bar{v}^2+\mathcal{O}\left(c^{-3}\right)\right],\nonumber\\
  &&\mathcal{F}_{\text {dip }}(\bar{v})=\frac{4\mathcal{S}_{-}^2\eta^2 c^5 }{3G\bar{\alpha}}\bar{v}^8\left[1+f_{2}^{d} \bar{v}^2+\mathcal{O}\left(c^{-3}\right)\right]\,,
\end{eqnarray}
\normalsize
where $\bar{\xi}=(1+\mathcal{S}_{+}^2\bar{\alpha}/6)$ comes from the
non-dipolar flux terms at Newtonian order. Note that this factor differs from the previously defined factor $\xi$ by the fact that it only includes the flux terms that do not depend on $\mathcal{S}_{-}$ explicitly.
Similarly, the factor $f^{nd}_{2}$ equals those terms in Eq.~\eqref{f2nd} that do not depend on $\mathcal{S}_{-}$, replacing $\xi$ by $\bar{\xi}$. Also, $f^{d}_{2}$ can be found from the factor $f^{DD}_{2}$, and it equals the terms of $f^{DD}_{2}$ that do not depend on $1/\mathcal{S}_-^2$.
Overall, for Eq.~\eqref{approxQD} we find
\small
\begin{equation}
\begin{split}
&\frac{E'(\bar{v})}{\mathcal{F}(\bar{v})} \simeq\frac{5G m\bar{\alpha}^2 }{32 c^3\eta\bar{\xi}\bar{v}^9}\left[1+\left(E'_0-f^{nd}_2\right)\bar{v}^2+\mathcal{O}\left(c^{-4}\right)\right]\\
&\qquad-\frac{25 Gm\bar{\alpha}^3\mathcal{S}_{-}^2}{768 c^3\bar{\xi}^2 \eta\bar{v}^{11}}\left[1+\left(E'_0-2f^{nd}_{2}+f^d\right)\bar{v}^2+\mathcal{O}\left(c^{-4}\right)\right]\,.
\end{split}
\end{equation}
\normalsize
Integrating Eq.~\eqref{TaylorF2} by using the above expression we obtain
\small
\begin{eqnarray}\label{FphQD}
\Psi(t_f)&&=\psi_{\text {non-dip}}+\psi_{\text {dip}}+\phi(\bar{v}_{ref})-2\pi f t(\bar{v}_{ref})\,\qquad
\end{eqnarray}
\begin{eqnarray}
&&\psi_{\text {non-dip}}(t_{f})=-\frac{6\bar{\alpha}}{256\eta \bar{\xi} \bar{v}_f^5}\Bigg[1+\frac{20}{9}\rho^{nd}\bar{v}_{f}^2-20\rho^{nd}_{GB}\bar{v}_{f}^6
\nonumber\\
&&+\frac{5}{3}\left(\frac{\bar{v}_f}{\bar{v}_{ref}}\right)^8\left(1+\frac{4}{3}\rho^{nd}\bar{v}_{ref}^2+4\rho^{nd}_{GB}\bar{v}_{ref}^6\right)-\frac{8}{3}\left(\frac{\bar{v}_f}{\bar{v}_{ref}}\right)^5\left(1+\frac{5}{3}\rho^{nd}\bar{v}_{ref}^2-5\rho^{nd}_{GB}\bar{v}_{ref}^6\right)\Bigg]\,,\qquad
\end{eqnarray}
\normalsize
\small
\begin{eqnarray}
&&\psi_{\text {dip}}(t_{f})=\frac{10\mathcal{S}_-^2\bar{\alpha}^2}{3584\eta\bar{\xi}^2\bar{v}_f^7}\left[1+\frac{7}{4}\rho^{d}\bar{v}_f^2 +\frac{70}{4}\rho^{d}_{GB}\bar{v}_f^6 \right.\nonumber\\
&&+\left.\frac{7}{3}\left(\frac{\bar{v}_f}{\bar{v}_{ref}}\right)^{10}\left(1+\frac{5}{4}\rho^{d}\bar{v}_f^2+\frac{5}{2}\rho_{GB}^d \bar{v}_f^6\right)\frac{10}{3}\left(\frac{\bar{v}_f}{\bar{v}_{ref}}\right)^{7}\left(1+\frac{7}{3}\rho^d \bar{v}_f^2+7\rho_{GB}^d \bar{v}_f^6\right)\right]\,,\qquad\quad
\end{eqnarray}
\normalsize
with the coefficients being
\small
\begin{equation}
\begin{split}
&\rho^{nd}=\frac{1247}{336\bar{\xi}}-\frac{3}{2}+\left(\frac{980}{336\bar{\xi}}-\frac{1}{6}\right)\eta+\left(\frac{448}{336\bar{\xi}}-\frac{4}{3}\right)\bar{\gamma}+\left(\frac{896}{336\bar{\xi}}+\frac{4}{3}\right)\left(\beta_{+}-\frac{\Delta m}{m}\beta_{-}\right)\,,\\
&\rho^{nd}_{GB}=\frac{c^4}{G^2}\frac{f'(\phi_{0})\alpha}{ m^2\bar{\alpha}^{5/2}} \left[\frac{12\mathcal{S}_{+}}{\bar{\xi}}+\frac{495\mathcal{S}_{+}}{28\bar{\xi}}(1-2\eta)
+\frac{88}{3\bar{\alpha}}\left(3\mathcal{S}_++\frac{\Delta m}{m}\mathcal{S}_-\right) \right]\,,\\
&\rho^{d}=\frac{1247}{168\bar{\xi}}-\frac{123}{10}+\left(\frac{980}{168\bar{\xi}}+\frac{7}{6}\right)\eta+\left(\frac{448}{168\bar{\xi}}-\frac{14}{3}\right)\bar{\gamma}+\left(\frac{896}{168\bar{\xi}}-\frac{32}{3}\right)\left(\beta_{+}-\frac{\Delta m}{m}\beta_{-}\right)\\
&\qquad-
\frac{4}{\bar{\gamma}}\left(\frac{\mathcal{S}_{+}}{\mathcal{S}_{-}}\right)\left(\beta_{-}-\frac{\Delta m}{m}\beta_{+}\right)-\frac{4}{\bar{\gamma}}\left(\beta_{+}-\frac{\Delta m}{m}\beta_{-}\right)\,,\\
&\rho^{d}_{GB}=\frac{c^4}{G^2}\frac{f'(\phi_{0})\alpha}{ m^2\bar{\alpha}^{5/2}} \left[\frac{24\mathcal{S}_{+}}{\bar{\xi}}+\frac{495\mathcal{S}_{+}}{14\bar{\xi}}(1-2\eta)\right.\\
&\qquad\left.+\frac{104}{3\bar{\alpha}}\left(3\mathcal{S}_++\frac{\Delta m}{m}\mathcal{S}_-\right) +
\frac{\Delta m}{16m}\frac{\mathcal{S}_{+}}{\mathcal{S}_{-}}\eta\left(\mathcal{S}_{+}+\frac{\Delta m }{m}\mathcal{S}_{-}\right)\right]\,.
\end{split}
\end{equation}
\normalsize
Note, that the current mapping between ppE templates and waveforms in sGB theory is known for the waveform at Newtonian order (corresponding to $b_{ppE}=-7$)~\cite{Yunes:2016jcc,Yagi:2011xp}. The results presented here can be directly mapped to the ppE template with $b_{ppE}=-5$, and thus potentially further improve the bounds on the coupling parameter 
e.g., by extending the work of~\cite{Nair:2019iur,Perkins:2021mhb}.

\begin{figure*}[hbt!]
\centering
\includegraphics[width=1.02\textwidth]{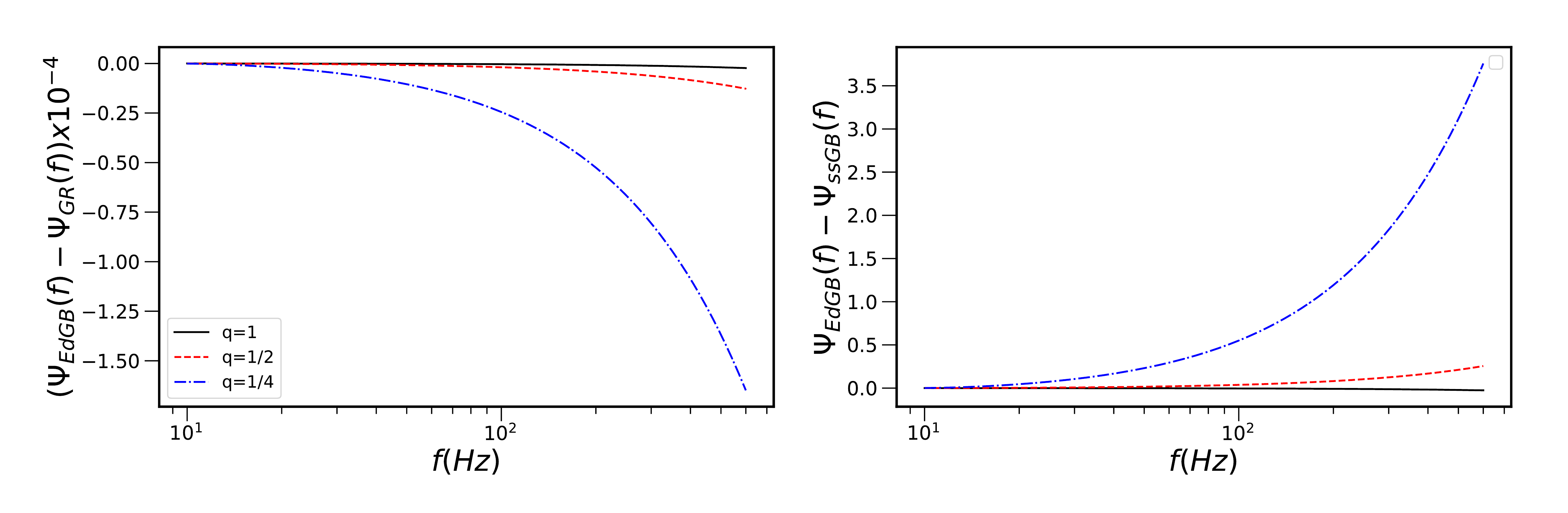}
\caption{Left: The GW phase difference between EdGB $\psi_{\text{EdGB}}$ and 1PN GR $\psi_{\text{GR}}$ shown for the systems with $\alpha=0.01 m^2$ and $m=15M_{\odot}$, as considered in Fig.~\ref{waveformm15}, as well as for an equal-mass case. The blue dashed-dotted and red dashed lines correspond to $q=1/2$ and $q=1/4$, respectively. The black solid lines correspond to $q=1$.
Right: GW phase difference between EdGB and ssGB gravity for the aforementioned binaries. 
}
\label{phasediff}
\end{figure*}

In Fig.~\ref{phasediff}, we show an example of the Fourier-domain GW phase evolution in order to study the effect of dilatonic and shift-symmetric coupling functions on the phasing.  We choose $\alpha=0.01 m^2$ and $m=15 M_\odot$, as it corresponds to the systems considered in Fig.~\ref{waveformm15}. The upper bound on frequency is chosen as $f_{\text{max}}=2(6^{3/2}\pi m)^{-1}\approx586\,\mathrm{Hz}$ and to simplify the comparison, all phases are aligned with the 1PN equal-mass phase in GR at the minimum frequency limit.

As the left panel shows, the difference between EdGB gravity and 1PN GR phase evolution is significant for the $q\neq1$ binaries. The phase difference between EdGB and ssGB gravity is shown on the right panel, where we see that the overall difference is small compared to the difference with GR. For this specific example, the difference is $\mathcal{O}(1)$ cycles and thus is more difficult to distinguish from the deviations from GR.

\subsection{Interpretation of GW constraints for fundamental parameters}
There is an important point to note regarding the interpretation of results in quadratic gravity.
If the action~\eqref{action} 
is taken to be in the Jordan frame, i.e., if matter is minimally coupled to the metric $g_{\mu\nu}$, our results directly provide the waveforms that would be measured by GW detectors and they can be employed to derive observational bounds on quadratic gravity.
However, if Eq.~\eqref{action}
represents the low-energy effective action, 
of a string theory then it is, strictly speaking, in the Einstein frame. That is, the waveforms would have to be transformed to the physical (i.e., Jordan or string) frame to connect between GW detections and observational bounds on string theory.
While there is no distinction between the two frames to leading order in the weak-field approximation, where the conformal factor transforming between them is $\mathcal{A}(\phi)=1+\mathcal{O}(\phi)$, 
we urge the reader (and tester of strong-field gravity) to caution outside this approximation.

\section{Conclusions}\label{Sec:IX}
We have extensively studied the generation of GWs and scalar radiation in inspiralling black-hole binaries in sGB theories, which are characterized by the coupling of a scalar field to higher curvature terms, namely the GB invariant. Using the Direct Integration of Einstein Equations in the post-Newtonian expansion, we have computed the GWs and energy flux to relative 1PN order beyond the quadrupole emission, and the scalar analogue to relative 0.5 PN order.  
We advanced the knowledge of  GW and scalar waveforms in sGB theories beyond the leading results of Ref.~\cite{Yagi:2011xp} where the only effects came from the scalar field, and capture new effects due to curvature nonlinearities.

Furthermore, we compared our results in the small coupling regime against the scalar waveforms obtained with the lower-order post-Newtonian scheme of Ref.~\cite{Yagi:2011xp} and obtained with numerical relativity simulations~\cite{Witek:2018dmd}.

In order to compute the waveforms, we have \textit{skeletonized} the compact bodies to effective point particles with scalar-field dependent masses to consistently include the effect of BH scalar-charges. We derived the two-body equations of motion to 1PN order based on the solutions of near-zone fields and the 1PN Lagrangian given in Ref.~\citep{Julie:2019sab}.
In both the Lagrangian and the waveforms, we see that results differ from those of 1PN ST theory by (i) specific combinations of scalar-charge dependent parameters (see Eq.~\eqref{EOMparams}), and (ii) additional GB coupling-dependent terms which have no analogue in ST theory and encode the impact of higher-curvature contributions.

This has consequences for a future effective-one-body (EOB) description of sGB theories, and suggests that the EOB results for ST theories can not be trivially extended to sGB theories.
In addition, this difference would allow us to distinguish the two classes of theories through the analysis of mixed BH-neutron star binary inspirals which can exhibit only one scalarized body in each theory.

Focusing on compact binary systems in quasi-circular orbits, we have presented the tensor and scalar waveforms in a ready-to-implement form.
That is, the GW polarizations and scalar wave in the time domain, Eqs.~\eqref{PP} to~\eqref{SP},
or the GW phasing in the Fourier domain, Eqs.~\eqref{FphDD} and~\eqref{FphQD}, can be employed directly to construct (phenomenological or EOB) waveform templates. Our results also provide a critical first step for constructing inspiral-merger-ringdown GW templates at high curvature regimes.

We have employed the SPA approximation to derive analytical expressions for the phasing in Fourier domain for systems whose inspiral is driven by the emission of scalar dipolar radiation, as well as those driven by tensor quadrupolar flux.
In a companion paper~\cite{Shiralilou:2020gah}, we have quantified the detectability of ssGB and EdGB phase deviations from GR, varying the binary parameters and the GB coupling. We have shown that the GB phase deviations are potentially detectable by A+LIGO/Virgo/KAGRA sensitivity bands~\cite{TheLIGOScientific:2014jea,TheVirgo:2014hva,Akutsu:2020his}, and thus the analytical results presented here can be used to put tight constraints on sGB theories, either through matching to ppE waveforms or through direct comparison against GW data.
Future, ground- and space-based GW detector networks combining high-frequency observations by, e.g., the Einstein Telescope~\cite{Punturo:2010zza} or Cosmic Explorer~\cite{Evans:2016mbw}, with low-frequency observations by, \textit{e.g.}, the space-based LISA mission~\cite{Audley:2017drz},
open new avenues for multi-band tests of gravity. Our results, in principle, are extendable to cover the phase evolution across the frequency bands and enable novel, multi-band tests of gravity. A detailed analysis is left for future work.

We emphasize that our results and the methods applied here are not restricted to specific choices of the coupling function, nor to the weak coupling limit, and thus can potentially be extended to explore dynamical scalarization or de-scalarization effects of BH binaries~\cite{Silva:2020omi} during the early inspiral phase. In particular, in the current treatment we have neglected the finite-size effects which may account for both the dynamical (de-)scalarization as well as scalar-induced tidal interaction. We leave the
investigation of such effects for future work.



\section{Acknowledgment}
We thank L.~Gualtieri and H.~O.~Silva for useful discussions, and A.~Samajdar for comments on the manuscript. BS, TH and SMN are grateful for financial support from the Nederlandse Organisatie voor Wetenschappelijk Onderzoek (NWO) through the Projectruimte and VIDI grants (Nissanke). TH  and SMN also acknowledges financial support from the NWO sectorplan. NO acknowledges financial support by the CONACyT grants ``Ciencia de frontera" 140630 and 376127, and by the UNAM-PAPIIT grant IA100721. HW acknowledges financial support provided by NSF Grant No. OAC-2004879 and the Royal Society Research Grant No. RGF\textbackslash R1\textbackslash 180073. TH also acknowledges Cost Actions CA18108 - Quantum gravity phenomenology in the multi-messenger approach and CA16104 - Gravitational waves, black holes and fundamental physics. Some calculations used the computer algebra
system Mathematica, in combination with the xAct/xTensor suite~\cite{Brizuela:2008ra}.

\clearpage
\begin{appendices}

\section{Curvature Tensors in terms of gothic metric}\label{gothic}
\subsection{Transformation rules}
In order to re-write curvature tensors in terms of the gothic metric, defined by $\mathfrak{g}^{a b}=\sqrt{-g} g^{a b}$, we first
define its inverse $\mathfrak{g}_{a b}=g_{a b}/\sqrt{-g}$, which is a tensor density of weight -1.
To find Christoffel symbols, we first need to convert the derivative of the metric to the derivative of the gothic one. This is done by differentiating the relation $\mathfrak{g}^{a b} g_{a c}=\sqrt{-g} \delta_{c}^{b}$.
In the case of $b=c$, we have
\begin{equation}\label{derivgoth1}
\begin{split}
0 &=\partial_{c}\left(\mathfrak{g}^{a b} g_{a b}\right)-4 \partial_{c} \sqrt{-g}=g_{a b} \partial_{c} \mathfrak{g}^{a b}+\mathfrak{g}^{a b} g_{a b, c}-2 \sqrt{-g} g^{a b} g_{a b, c} =g_{a b} \partial_{c} \mathfrak{g}^{a b}-\mathfrak{g}^{a b} \partial_{c} g_{a b}\,,
\end{split}
\end{equation}
and in the case of $b\neq c$,
\begin{equation}\label{derivgoth2}
\begin{split}
0 &=g_{b j}\left[\partial_{c}\left(\mathfrak{g}^{a b} g_{a f}\right)-\delta_{f}^{b} \partial_{c} \sqrt{-g}\right]=\sqrt{-g} g_{j f, c}+ (-g) \mathfrak{g}_{a f} \mathfrak{g}_{b j} \partial_{c} \mathfrak{g}^{a b}-\frac{(-g)}{2} \mathfrak{g}_{f j} \mathfrak{g}_{a b} \partial_{c} \mathfrak{g}^{a b} \\
&=g_{j f, c}+\sqrt{-g}\left(\mathfrak{g}_{a f} \mathfrak{g}_{b j}-\frac{1}{2} \mathfrak{g}_{a b} \mathfrak{g}_{f j}\right) \partial_{c} \mathfrak{g}^{a b}\,.
\end{split}
\end{equation}
Equation~\eqref{derivgoth2} can be re-written to relate the partial derivative of the gothic metric to that of its inverse,
\begin{equation}\label{derup}
    \partial_{c}\mathfrak{g}_{jf}=-\mathfrak{g}_{af}\mathfrak{g}_{bj}\partial_{c}\mathfrak{g}^{ab}.
\end{equation}

\subsection{Christoffel symbols and curvature tensors}
By using Eqns.~\eqref{derivgoth1}-\eqref{derup}, we can express quantities in terms of gothic variables. In particular, the Christoffel symbols become
\small
\begin{equation}\label{chrisudd}
\Gamma^{a}_{ij}=\frac{1}{4}\left(2 \mathfrak{g}^{ ah} \mathfrak{g}_{d j} \mathfrak{g}_{gi} \partial_{h} \mathfrak{g}^{dg}-\mathfrak{g}^{ah} \mathfrak{g}_{dg} \mathfrak{g}_{ji} \partial_{h} \mathfrak{g}^{dg}-
2\mathfrak{g}_{e j} \partial_{i} \mathfrak{g}^{ae}+\delta_{j}^{a} \mathfrak{g}_{be} \partial_{i} \mathfrak{g}^{be}-2 \mathfrak{g}_{fi} \partial_{j} \mathfrak{g}^{af}+\delta_{i}^{a} \mathfrak{g}_{cf} \partial_{j} \mathfrak{g}^{cf}\right).
\end{equation}
\normalsize
To derive the curvature tensors, we compute the partial derivative of the Christoffel symbols. In terms of these, the Ricci scalar in gothic formulation is
\small
\begin{equation}\begin{split}
R=-\frac{1}{\sqrt{-\mathfrak{g}}}\Big(& -\frac{1}{4} \mathfrak{g}^{{ab}} \mathfrak{g}_{{cd}}\mathfrak{g}_{{ef}}\partial_{{a}} \mathfrak{g}^{ce}\partial_{{b}} \mathfrak{g}^{{df}}-\frac{1}{8} \mathfrak{g}^{{ab}} \mathfrak{g}_{{cd}} \mathfrak{g}_{{ef}}\partial_{{a}} \mathfrak{g}^{{cd}}\partial_{{b}} \mathfrak{g}^{{ef}}+\partial_{{b}} \partial_{{a}} \mathfrak{g}^{{ab}}\\
&+\frac{1}{2} \mathfrak{g}^{ {ab }} \mathfrak{g}_{ {cd }}\partial_{b} \partial_{{a}} \mathfrak{g}^{ {cd }}-\frac{1}{2}\mathfrak{g}_{ {ab }}\partial_{{c}} \mathfrak{g}^{ {bd }}\partial_{{d}} \mathfrak{g}^{ {ac }}+\frac{1}{2} \mathfrak{g}_{ {ab }}\partial_{c} \mathfrak{g}^{ {ab }}\partial_{{d}} \mathfrak{g}^{ {cd }}\Big)\,.
\end{split}
\end{equation}
\normalsize
The rest of the expressions for curvature tensors can be found in the corresponding Mathematica notebook.

The two important curvature combinations that we need for the field equations are the dual Riemann tensor and the GB scalar. When written in terms of the gothic metric, these quantities become very long. As we work in the weak-field limit of these quantities, we do not report their full expressions here. We consider that it is sufficient to describe their overall structure.
In particular, for the dual Riemann tensor, the metric and its derivatives appear in the following combinations:
\small
\begin{equation}
^*R^{abcd*}\propto \frac{1}{\sqrt{-\mathfrak{{g}}^3}}\Big(
\mathfrak{g}^{..}\mathfrak{g}^{..}\partial_{.}\partial_{.}\mathfrak{g}^{..},\,
\mathfrak{g}^{..}\partial_{.}\mathfrak{g}^{..}\partial_{.}\mathfrak{g}^{..},\,
\mathfrak{g}^{..}\mathfrak{g}^{..}\mathfrak{g}_{..} \partial_{.}\mathfrak{g}^{..}\partial_{.}\mathfrak{g}^{..},\,
\mathfrak{g}^{..}\mathfrak{g}^{..}\mathfrak{g}^{..}\mathfrak{g}_{..} \partial_{.}\partial_{.}\mathfrak{g}^{..},\,
\mathfrak{g}^{..}\mathfrak{g}^{..}\mathfrak{g}^{..}\mathfrak{g}_{..}\mathfrak{g}_{..}\partial_{.}\mathfrak{g}^{..}\partial_{.}\mathfrak{g}^{..}
\Big)\,.
\end{equation}
\normalsize
In the main text, we called this quantity $^*\hat{R}^{abcd*}$. The expression for the $\mathcal{R}_{GB}$ is derived from the above mentioned curvature quantities through Eq.~\eqref{GBscalardef}.

\subsection{Weak-field limit}\label{appAweakfield}
The gauge-fixed GB coupling-dependent quantities $(^*R^{*\alpha a\beta d})\nabla_{ad}f(\phi)$ and $\mathcal{R}^2_{GB}$ enter as part of the source of the wave equations~\eqref{wave} and~\eqref{Sgothic}, respectively. Using Eq.~\eqref{perturbation}, We expand these quantities to second order in the weak-field limit as higher order terms are not relevant for 1PN calculations. It is easy to verify that the expansion of these quantities to first order in the weak-field limit is identically zero. The second order expansions are
\small
\begin{equation}\label{app:ohphi}
\begin{split}
^*&\hat{R}^{*\alpha a \beta d}\nabla_{ad}f(\phi) =f'(\phi_{0}) \nabla_{ad}\delta \phi \Big(-4 \eta^{\alpha e} \eta^{df} \partial_{fe}h^{a\beta} + 4 \eta^{\alpha d} \eta^{ef} \partial_{fe}h^{a\beta} + 4\eta^{\alpha e} \eta^{\beta f} \partial_{fe}h^{ad}\\
&- 4 \eta^{\alpha\beta} \eta^{ef} \partial_{f}\partial_{e}h^{ad}+ 4 \eta^{ae} \eta^{df} \partial_{fe}h^{\alpha\beta} - 4 \eta^{ad} \eta^{ef} \partial_{fe}h^{\alpha\beta} - 4 \eta^{ae} \eta^{\beta f} \partial_{fe}h^{\alpha d} + 4 \eta^{a\beta} \eta^{ef} \partial_{fe}h^{\alpha d}\\
&
+ 2 \eta^{ae} \eta^{\alpha d} \eta^{\beta f} \eta_{gh} \partial_{fe}h^{gh}
- 2 \eta^{ad} \eta^{\alpha e} \eta^{\beta f} \eta_{gh} \partial_{fe}h^{gh} - 2\eta^{ae} \eta^{\alpha\beta}\eta^{df} \eta_{gh} \partial_{fe}h^{gh}
+ 2 \eta^{a\beta} \eta^{\alpha e} \eta^{df} \eta_{gh} \partial_{fe}h^{gh}\\
&+ 2\eta^{ad}\eta^{\alpha\beta}\eta^{ef}\eta_{gh}\partial_{fe}h^{gh}- 2 \eta^{a\beta}\eta^{\alpha d} \eta^{ef} \eta_{gh} \partial_{fe}h^{gh}\Big) 
+\mathcal{O}(h^2\delta\phi^2)\,,
\end{split}
\end{equation}
\normalsize
and
\small
\begin{equation}\label{app:rgbohphi}
\begin{split}
&\hat{\mathcal{R}}^2_{GB}=
2 \partial_{bc} h\,\partial^{bc}h-8 \partial_{bd} h_{c}^{d}\,\partial^{bc} h+2 \partial^{bc} h\,\partial_{cb} h
-4 \partial^{bc}  h\,\partial_{cd} h_{b}^{d}+\partial_{ca}  h^{ a c}\, \partial_{db}  h^{ b d}+8 \partial_{c} \partial^{c} h\,\partial_{db}  h^{ b d}\\
&+8 \partial_{a} \partial^{b} h^{ a c}\,\partial_{db}  h_{c}^{d} 
+8 \partial^{b} \partial_{a} h^{ a c}\,\partial_{db}  h_{c}^{d}+
4 \partial^{bc}  h\,\partial_{db}  h_{c}^{d} +8 \partial_{a} \partial^{b} h^{ a c}\,\partial_{dc}  h_{b}^{d}+8 \partial^{b} \partial_{a} h^{ a c}\,\partial_{dc}  h_{b}^{d}\\
&+8 \partial_{b} \partial^{b} h^{ a c}\,\partial_{d} \partial^{d} h_{a c}+4 \partial_{c} \partial^{c} h_{a}^{ a}\,\partial_{d} \partial^{d} h_{b}^{ b}-24 \partial_{a} \partial^{b} h^{ a c}\,\partial_{d} \partial^{d} h_{c b}
-8 \partial^{b} \partial_{a} h^{ a c}\,\partial_{d} \partial^{d} h_{c b}
\\
&+8 \partial^{bc}  h_{a}^{ a}\,\partial_{d} \partial^{d} h_{c b}+8 \partial_{ca}  h_{b d}\,\partial^{d} \partial^{b} h^{ a c}+8 \partial_{cd}  h_{a b}\,\partial^{db}  h^{ a c}
-8 \partial_{db}  h_{a c}\,\partial^{db} h^{ a c}+8 \partial_{dc}  h_{a b}\,\partial^{db}  h^{ a c}+\mathcal{O}(h^3).
\end{split}
\end{equation}
\normalsize

The expressions, once expanded and simplified using the PN parameter $1/c^2$, give rise to 1PN source terms in the near-zone, which we have reported in Eq.~\eqref{sources} and Eq.~\eqref{sourcesS}, together with the 1PN expansion of matter sources.
\section{Field integrals and calculation techniques}\label{CalcTechs}
In this Appendix we present the calculation of two of the key field integrals for EW moments. The rest of the EW integrals follow the same methodology and reasoning. As many of these integrals involve surface integration over product on unit vectors $\hat{n}^{ij...m}$, it is useful to convert these products to STF products for which the following simplifying relations holds
\small
\begin{equation}\label{STF}
\hat{n}^{\langle L\rangle} \equiv \sum_{p=0}^{\lfloor l / 2\rfloor}(-1)^{p} \frac{(2 l-2 p-1) ! !}{(2 l-1) ! !}\left[\hat{n}^{L-2 P} \delta^{P}+\operatorname{sym}(q)\right]\,.
\end{equation}
\normalsize
Angle braces on indices define a STF tensor. 
We use $\lfloor l / 2\rfloor$ to denote the largest integer less than or equal to $l / 2$. The expression $\operatorname{sym}(q)$ stands for all the distinct terms which result from permuting the indices on $\hat{n}^{L-2 P} \delta^{P}$. As an example, this relation gives
\begin{equation}
\begin{array}{l}
\hat{n}^{\langle i j\rangle}=\hat{n}^{i j}-\frac{1}{3} \delta^{i j},\\
\hat{n}^{\langle i j k\rangle}=\hat{n}^{i j k}-\frac{1}{5}\left(\hat{n}^{i} \delta^{j k}+\hat{n}^{j} \delta^{i k}+\hat{n}^{k} \delta^{i j}\right)
\end{array}
\end{equation}
The STF product of unit vectors are such that their integration over solid angle is zero. Converting back to non-STF form would lead to the following identities
\small
\begin{equation}\label{spatialintegral}
\begin{split}
&\int \hat{n}_{k_{1}...k_{m}} d^2 \Omega=0 \qquad(m \text { odd })\,, \\
&\int \hat{n}_{k_{1}...k_{m}} d^2 \Omega=[4 \pi /(m+1) ! !]\,\times
\left[\delta_{k_{1} k_{2}} \ldots \delta_{k_{m-1} k_{m}}+\text {distinct permutations}\right]\,\,\,( m \text { even })\,.
\end{split}\end{equation}
\normalsize

The simplest integral that appears in the two-index EW moment calculation is $\int (\nabla U)^2 d^3 x$. Applying integration-by-parts on this term gives
\small
\begin{equation}
\int_{\mathcal{M}}(\nabla U)^{2} x^{i j} d^{3} x= \oint_{\partial \mathcal{M}} U U^{, k} x^{i j} d^{2} S^{k}-\int_{\mathcal{M}} U U^{, k k} x^{i j} d^{3} x -2 \int_{\mathcal{M}} U U^{(i} x^{j)} d^{3} x .
\end{equation}
\normalsize
The surface integral is evaluated on the boundary of the near zone $\partial \mathcal{M}$ as a sphere with radius $\mathcal{R}$, such that we can write $x^{i}=\mathcal{R} \hat{n}^{i}$ and $d^{2} S^{k}=\mathcal{R}^{2} \hat{n}^{k}$. To evaluate this term we shall expand $U$ and $U^{,k}$ in inverse powers of $\mathcal{R}$ and ignore the integrands that depend on $\mathcal{R}$. The $\mathcal{R}$-independent piece of this specific surface integral contains the $\mathcal{O}(\mathcal{R}^{-4})$ expansion of $U U^{,k}$ which gives an integrand with odd number of unit vectors and thus vanishes [see Eq.~\eqref{STF}].
For the first volume integral, substituting the Laplacian of $U$ easily shows that
$-\int_{\mathcal{M}} U U^{, k k} x^{i j} d^{3} x=4 \pi \sum_{A, B \neq A} \frac{m_{A} m_{B}}{r} x_{A}^{i j}$. The contribution from the second volume integrals is zero. This can be shown by integrating-by-parts again,
$-2 \int_{\mathcal{M}} U U^{,(i} x^{j)} d^{3} x=-\oint_{\partial \mathcal{M}} U^{2} x^{(i} d^{2} S^{j)}+\int_{\mathcal{M}} U^{2} \delta^{i j} d^{3} x$, which shows that the surface integral vanishes due to an odd number of unit vectors. The volume integral is also ignored as it vanishes through TT projection.

Other important integrals that we encounter in the calculations (such as  $\int U^{,m} U^{,lm} d^3x$) is the $\int U U^{,lm} d^3x$ term, which can be written as~\cite{Will:1996zj}:
\small
\begin{equation}
    \begin{split}
    \int U U^{,lm} d^3x&=\sum_{A, B\neq A}\int \frac{m_A m_B}{\|\mathbf{x}-\mathbf{x}_B\|}\left(\frac{3 (\mathbf{x}-\mathbf{x}_A)^{lm}}{\|\mathbf{x}-\mathbf{x_A}\|^5}-\frac{\delta^{lm}}{\|\mathbf{x}-\mathbf{x}_A\|^3} -\frac{4\pi}{3} \delta^{lm}\delta^3 (\mathbf{x}-\mathbf{x}_B) \right)d^3x\\
    &=\sum_{A, B\neq A}\int \frac{m_A m_B}{\|\mathbf{x}-\mathbf{r}_{AB}\|}\left(\frac{3 y^{\langle lm\rangle}+\delta^{lm}}{y^3}-\frac{\delta^{lm}}{y^3}  \right)d^3y -\sum_{A, B\neq A} \frac{4\pi}{3}\frac{\delta^{lm} m_A m_B}{r}\\
    &-\oint \frac{1}{\|\mathbf{y}-\mathbf{r}_{AB}\|} \frac{y^{lm}}{y^{3}}\left(\mathbf{y} \cdot \mathbf{x}_{A}\right) \mathcal{R}^{2} d^{2} \Omega_{y}+...
    \end{split}
\end{equation}
\normalsize
where in the second line we have changed the integration variables from $\mathbf{x}$ to $\mathbf{y}=\mathbf{x}-\mathbf{x}_{A}$ and have changed $y^{lm}$ to its STF form. There are two cases to consider: $A=B$ and $A \neq B$. For both, the infinite series of surface integrals vanishes as the terms either depend on $\mathcal{R}$ or average to
zero because of an odd number of unit vectors. For the $A = B$ case, $r_{AB}=0$, so the volume integral  can be evaluated easily. When $A \neq B$, we may use of the following expansions:
\small
\begin{equation}
\begin{split}
&\frac{1}{\|\mathbf{y}+\mathbf{r}_{A B}\|}=\sum_{l, m} \frac{4 \pi}{2 l+1} \frac{\left(-r_{<}\right)^{l}}{r_{>}^{l l+1}} Y_{l m}^{*}\left(\hat{\mathbf{n}}_{A B}\right) Y_{l m}(\mathbf{y})\,,\\
&\sum_{m} \int Y_{l m}^{*}\left(\hat{\mathbf{n}}_{A B}\right) Y_{l m}(\hat{\mathbf{y}}) \hat{y}^{\left(L^{\prime}\right\rangle} d^{2} \Omega_{y}=\hat{n}_{A B}^{\langle L\rangle} \delta^{l l^{\prime}}\,,
\end{split}
\end{equation}
\normalsize
where $Y_{l m}$ are the spherical harmonics, and $r_{<(>)}$ denotes the lesser (greater) of $r_{A B}$ and $y$. We substitute the first expansion into the volume integral and use the second identity to evaluate the terms
\small
\begin{equation}
    \sum_{A, B\neq A}  \frac{3m_A m_B}{\|x-r_{AB}\|} \frac{y^{\langle lm \rangle}}{y^3}d^3y =\sum_{A, B\neq A}4\pi m_A m_B \int \sum_{m'}\frac{3}{5}\frac{r_{<}^2}{r_{>}^3}\frac{y^{\langle lm \rangle}}{y^3} Y^{*}_{2m'}(n_{AB})Y_{2m}(y) d^3y\,.
\end{equation}
\normalsize
The radial integral is then evaluated by 
\small
\begin{equation}\label{rint}
\int_{0}^{\mathcal{R}} \frac{r_{<}^{l}}{r_{>}^{l+1}} y^{q} d y=\frac{2 l+1}{(l+q+1)(l-q)} r_{A B}^{q}
\end{equation}
\normalsize
Overall we find
\small
\begin{equation}
    \sum_{A, B\neq A} 3 \frac{m_A m_B}{\|x-r_{AB}\|} \frac{y^{\langle lm \rangle}}{y^3}d^3y = \sum_{A, B\neq A} \frac{2\pi m_A m_B}{r} n_{AB}^{lm}
\end{equation}
\normalsize

\section{Far-zone contribution to waveforms}\label{App:FZ}
In this Appendix, we focus on the solution of Eq.~\eqref{waveall} with far-zone field point and show that the contributions are beyond 1PN order. In the far-zone region, the source terms $\mu_s$ and $\mu^{ij}$ are composed purely of field terms as, by definition, there is no matter source present in this region. It can be shown that (see \textit{e.g.},~\citep{Will:1996zj}) the fields at any intermediate distance R are given by:
\small
\begin{equation}
\label{FZNZmultipole}
h_{\mathcal{NW}}^{\alpha\beta}(x)=\frac{4G}{c^4} \sum_{q=0}^{\infty} \frac{(-1)^{q}}{q!}\left(\frac{ M^{\alpha\beta k_{1} \cdots k_{q}}}{R}\right)_{, k_{1} \cdots
k_{q}},
\qquad\Phi_{\mathcal{NW}}(x)=\frac{2G}{c^4}\sum_{q=0}^{\infty} \frac{(-1)^{q}}{q !}\left(\frac{ M_{s}^{k_{1} \cdots k_{q}}}{R}\right)_{, k_{1} \cdots
k_{q}},
\end{equation}
\normalsize
with a new set of multipole moments defined as:
\small
\begin{equation}
M^{\alpha\beta k_{1} \cdots k_{q}}(\tau) \equiv \int_{\mathcal{M}} \mu^{\alpha\beta}\left(\tau, \mathbf{x}^{\prime}\right) x^{\prime k_{1}} \cdots x^{k_{q}} d^{3} x^{\prime}\,,\qquad
M^{ k_{1} \cdots k_{q}}(\tau) \equiv \int_{\mathcal{M}} \mu_s\left(\tau, \mathbf{x}^{\prime}\right) x^{\prime k_{1}} \cdots x^{k_{q}} d^{3} x^{\prime}\,.
\end{equation}
\normalsize
Note that Eq.~\eqref{FZNZmultipole} reduces to Eqns.~\eqref{EWstart} and \eqref{Swave} in the limit where $R\gg \mathcal{R}$.

As can be seen from the structure of terms in Eq.~\eqref{sources} and Eq.~\eqref{sourcesS}, the sources in the far-zone are composed only of the field components $N$ and $\Phi$. Finding these components by using Eq.~\eqref{FZNZmultipole} requires computing the following multipoles: 
\begin{equation}
\begin{split}
    &M^{00}= m_{A}c^2\left[ 1+\frac{v^2_{A}}{2c^2}-\frac{G m_B(1+\alpha_{A}^{0}\alpha_{B}^{0})}{2rc^2}+{\cal O}(c^{-4})\right]+(A\leftrightarrow B)\,,\\
    &M^{00i}=  m_{A}c^2x^{i}_{A}\left[ 1+\frac{v^2_{A}}{2c^2}-\frac{G m_B(1+\alpha_{A}^{0}\alpha_{B}^{0})}{2rc^2}+{\cal O}(c^{-4})\right]+(A\leftrightarrow B)\,,\\
    &M^{00ij}=m_{A}c^2x^{ij}_{A}\left[1+{\cal O}(c^{-2})\right]+(A\leftrightarrow B)\,,
    \end{split}
\end{equation}
and
\begin{equation}
\begin{split}
    &M_{s}=-m_{A}c^2\alpha^{0}_{A}\left[1-\frac{v^2_{A}}{c^2}-\frac{G m_B\alpha_{AB}}{r\, c^2} +{\cal O}(c^{-4})\right]+(A\leftrightarrow B)\,,\\
   &M_{s}^{i}=-m_{A}c^2x^{i}_{A}\alpha^{0}_{A}\left[1-\frac{v^2_{A}}{c^2}-\frac{G m_B\alpha_{AB}}{r\, c^2}+{\cal O}(c^{-4}) \right]+(A\leftrightarrow B),\\
   &M^{ij}_{s}=-m_{A}c^2\alpha^{0}_{A}x^{ij}_{A}\left[1+{\cal O}(c^{-2})\right]+(A\leftrightarrow B)\,.
\end{split}
\end{equation}
These moments are plugged into Eq.~\eqref{FZNZmultipole} to give far-zone fields, which further generate the wave equations sources. To 1PN order, it is easy to confirm that the expressions for these fields are identical to Eq.~\eqref{leadingNZNZ} by changing $r$ to $R$, and thus give
\small
\begin{equation}\label{FZsource}
\begin{split}
    \mu^{ij}&=
    \frac{G}{4\pi}\left[\frac{m^2}{R^4}+\frac{m\left(m_A\alpha_A^0 +m_B\alpha_B^0\right)}{R^4}\right]\left(\hat{n}^{\langle i j\rangle}-\frac{1}{6} \delta^{i j}\right)\\
    &-\frac{G\alpha f'(\phi_{0})}{\pi}\frac{m \left(m_A\alpha_A^0 +m_B\alpha_B^0\right)}{R^6}\left(\hat{n}^{\langle i j\rangle}+\frac{1}{3} \delta^{i j}\right).
\end{split}
\end{equation}
\normalsize
A similar calculation for $\mu_s$ shows that it equals zero, and thus there are no 0.5PN far-zone contributions to the scalar waveform.\\
In the cases where $\mu^{ij}:= f^{ij}(\tau)\hat{n}^{\langle L\rangle}/(4\pi R^{n})$, the far-zone contribution to the waveform is given by
\small
\begin{equation}
\label{farzone}\begin{split}
h_{\mathcal{W}}^{ij}(x)&=\frac{4G}{Rc^4} n^{\langle L\rangle}\left[\int_{0}^{\mathcal{R}} f^{ij}(\tau-2 s/c) A(s, R) d s\right.\left.+\int_{\mathcal{R}}^{\infty} f^{ij}(\tau-2 s/c) B(s, R) d s\right]\,,
\end{split}
\end{equation}
\normalsize
where we have defined $s=c(\tau-\tau')/2$, and the functions
\small
\begin{equation}
\begin{split}
&A(s, R) \equiv \int_{\mathcal{R}}^{R+s} \frac{P_{l}(\xi)dp}{ p^{n-1}}\,,\qquad\qquad
B(s, R) \equiv \int_{s}^{R+s} \frac{P_{l}(\xi)dp}{p^{n-1}}\,,
\end{split}
\end{equation}
\normalsize
with $P_{l}(\xi)$ being the Legendre polynomials with argument $\xi \equiv (R+2 s)/R-2 s(R+s)/R p$.  Integrating Eq.~\eqref{farzone} using the source of Eq.~\eqref{FZsource},  we see that the leading order contribution to GWs in the far-zone drops as $1/R^2$, which we thus ignore as we are only concerned with terms that are proportional to $1/R$.\\
By computing higher order multipoles and the corresponding far-zone waveform contributions, we have also shown that the leading, non-vanishing, TT terms are 1.5 PN contributions to the waveform. With this, we conclude that the far-zone contribution to GWs accurate to 1PN is zero. 
\section{Calculation of coupling-dependant energy flux terms}\label{GBL}
In order to find the GB coupling-dependent contribution to the tensor energy flux, we first find the time derivative of the GB dependent parts of the waveform, referred to as $P\tilde{Q}^{ij}_{GB}$. Using the expression for $P\tilde{Q}^{ij}_{GB}$ given in Eq.~\eqref{waveformfinale}, we find
\small
\begin{equation}
\begin{split}
    &\frac{d\,P\tilde{Q}^{ij}_{GB}}{dt}=\frac{G\bar{\alpha}m}{r}\frac{\alpha f'(\phi_0)}{\sqrt{\bar{\alpha}}r^2}\Bigg\{8\left(\mathcal{S}_{+}+\frac{\Delta m}{m}\mathcal{S}_{-}\right)\left[ \frac{v^{(i}r^{j)}}{r^2}\left(18\tilde{E}+13\frac{G\bar{\alpha}m}{r}-45\dot{r}^2\right)+\frac{18v^{ij}\dot{r}}{r}\right.\\
    &\quad\left.-\frac{r^{ij}\dot{r}}{r^3}\left(90\tilde{E}+\frac{72G\bar{\alpha}m}{r}-105\dot{r}^2\right)\right]-\frac{16G\bar{\alpha}m}{r^3}\left(3\mathcal{S}_{+}+\frac{\Delta m}{m}\mathcal{S}_-\right)\left(\frac{6\dot{r}r^{ij}}{r}-v^{(i}r^{j)}\right)\\
    &\quad+\frac{6}{r^2}\left[\mathcal{S}_+(1-2\eta)+\mathcal{S}_-(1+2\eta) \right] \Bigg[(\hat{\mathbf{n}}\cdot \mathbf{v})^2\left(\frac{30\dot{r}r^{ij}}{r}-6v^{(i}r^{j)}\right)\\
    &\quad+(\hat{\mathbf{n}}\cdot \mathbf{r})^2\left\{-\frac{\dot{r}r^{ij}}{r^3}\left(210\tilde{E}-315\dot{r}^2+\frac{152\,G\bar{\alpha}m}{r}\right)+\frac{r^{(i}v^{j)}}{r^2}\left(30\tilde{E}-105\dot{r}^2+\frac{21\,G\bar{\alpha}m}{r}\right)+\frac{20\dot{r}v^{ij}}{r}\right\}\\
    &\quad+
    (\hat{\mathbf{n}}\cdot \mathbf{r})(\hat{\mathbf{n}}\cdot \mathbf{v})\left(\frac{60\dot{r}v^{(i}r^{j)}}{r}-12v^{ij}+\frac{r^{ij}}{r^2}\left(60\tilde{E}-210\dot{r}^2+\frac{50G\bar{\alpha}m}{r}\right)\right)
    \Bigg]\Bigg\}\,.
    \end{split}
\end{equation}
\normalsize
We see that each term contains either zero or two unit normal vectors $\hat{\mathbf{n}}$, so we separate the calculation of n-dependent terms from the n-independent ones, and we denote them by $P\tilde{Q}^{ij}_{GB(0)}$ and $P\tilde{Q}^{ij}_{GB(2)}$, respectively. This is particularly essential, as different combinations of unit vectors have different spatial integrals [see Eq.~\eqref{spatialintegral}]. For the n-independent terms, one can show that the spatial integral of Eq.~\eqref{edotsimp} gives
\begin{equation}
    \dot{E}_{GB,(0)}=\frac{\mu^2}{5R^2}<\dot{\tilde{Q}}^{ij}P\dot{\tilde{Q}}^{ij}_{GB(0)}>\,,
\end{equation}
where $\dot{\tilde{Q}}^{ij}=2m\bar{\alpha}(3r^{ij}\dot{r}/r-2v^{(i}r^{j)})/r^3$ is the contribution from the Newtonian order quadrupole.

The n-dependent ones, $P\tilde{Q}^{ij}_{GB(2)}$, now have terms with two, four and six unit vectors to be integrated.
After a long but straightforward calculation we find
\small
\begin{equation}
\begin{split}
   &\dot{E}_{GB,(0)}=\frac{4G\mu^2 f'(\phi_{0})\alpha}{5c^7\sqrt{\bar{\alpha}}r^4}\left(\frac{G\bar{\alpha}m}{r}\right)^2\left[4\left(\mathcal{S}_{+}+\frac{\Delta m}{m}\mathcal{S}_-\right)\left(-4v^2 \left(18\tilde{E}+13\frac{G\bar{\alpha}m}{r}-45\dot{r}^2\right)\right.\right.\\
   &\quad\left.\left.+54\dot{r}^4\right)-\frac{G\bar{\alpha}m}{r}\left(3\mathcal{S}_++\frac{\Delta m}{m}\mathcal{S}_-\right)\left(32v^2+56\dot{r}^2\right)\right]\,,\\
   &\dot{E}_{GB,(2)}=\frac{6G\mu^2f'(\phi_{0})\alpha}{ c^7\sqrt{\bar{\alpha}}r^4}\left(\frac{G\bar{\alpha}m}{r}\right)^2\bigg(\mathcal{S}_+(1-2\eta)+\mathcal{S}_-(1+2\eta)\bigg)\left[ \frac{24}{5}\dot{r}^4-\frac{16}{5}\right.\\
   &\quad\left.-\frac{4}{7}v^2\left(22\tilde{E}+77\frac{G\bar{\alpha}m}{r}-\frac{199}{3}\dot{r}^2+\frac{18}{5}v^2\right)-\frac{   \dot{r}^2}{7}\left(992\tilde{E}+737\frac{G\bar{\alpha}m}{r}-\frac{2404}{3}\dot{r}^2+\frac{8}{5}v^2\right)\right]\,.
   \end{split}
\end{equation}
\normalsize

\section{ Final polarization waveforms and scalar waveform}\label{polarizations}
GW detectors are sensitive to linear combinations of the polarization waveforms $h_{+}$ and $h_{\times}$ through $h(t)=F_{+}h_{+}(t)+F_{\times}h_{\times}(t)$,
with $F_{+}$ and $F_{\times}$ being the so-called pattern functions of the detector, which depend both on the properties of the detector and the position of the source in the sky.
The two GW polarizations can be found straightforwardly using the definition  \eqref{polariz}, and the following combinations resulting from Eq.~\eqref{convwave}:
\small
\begin{equation}
\begin{split}
\left(\hat{n}^{i}_{AB} \hat{n}^{j}_{AB}\right)_{+}&=\frac{1}{4} \sin ^{2}( i)+\frac{1}{4}\left[1+\cos ^{2} (i)\right] \cos (2 \varphi)\,, \qquad\qquad\,\left(\hat{\lambda}^{i} \hat{\lambda}^{j}\right)_{\times}=-\frac{1}{2} \cos (i) \sin (2 \varphi)\,,
 \\
\left(\hat{n}_{AB}^{(i} \hat{\lambda}^{j)}\right)_{+}&=-\frac{1}{4}\left[1+\cos ^{2} (i)\right] \sin( 2 \varphi)\,, \qquad\qquad\qquad\,\,
\left(\hat{n}^{i}_{AB} \hat{n}^{j}_{AB}\right)_{\times}=\frac{1}{2} \cos( i) \sin( 2 \varphi)\,, \\
\left(\hat{\lambda}^{i} \hat{\lambda}^{j}\right)_{+}&=\frac{1}{4} \sin ^{2} (i)-\frac{1}{4}\left[1+\cos ^{2} (i)\right] \cos( 2 \varphi)\,,\,\,\,\quad\quad\quad
\left(\hat{n}_{AB}^{(i} \hat{\lambda}^{j)}\right)_{\times}=\frac{1}{2} \cos(i)\cos (2 \varphi)\,, \\
\hat{\mathbf{n}} \cdot \hat{\mathbf{n}}_{AB}&=\sin (i)\sin (\varphi)\,,\,\,\,\quad\qquad\qquad\qquad\qquad\qquad\qquad\quad\!\!
\hat{\mathbf{n}} \cdot \hat{\boldsymbol{\lambda}}=\sin(i) \cos (\varphi)\,.
\end{split}
\end{equation}
\normalsize
In terms of the PN parameter $x=\left( G\bar{\alpha}m\omega/c^3\right)^{2/3}$, having $\omega$ with 1PN accuracy, the result is
\small
\begin{equation}
    h_{+,\times}=\frac{2G\mu}{R c^2}\left(\frac{G\bar{\alpha}m\omega}{c^3}\right)^{2/3}\left\{H^{0}_{+,\times}+x^{1/2} H^{1/2}_{+,\times}+x H^{1}_{+,\times} +x H^{1}_{+,\times (GB)} \right\}\,,
\end{equation}
\normalsize
with the plus terms being
\small
\begin{equation}\label{PP}
    \begin{split}
    &H^{0}_{+}=- \left[\cos ^2(i)+1\right] \cos (2\phi)\,,\\
    &H^{1/2}_{+}=
    -\frac{\sin(i)}{8} \frac{\Delta m}{m}\left\{\left[5+\cos^2(i)\right] \cos (\varphi)-9\left[1+\cos^2(i)\right] \cos(3\varphi)\right\}\,,\\
    &H^{1}_{+}=
    \frac{1}{6}\left\{\left[19+9 \cos^{2}(i)-2 \cos^{4}(i)\right]-\eta\left[19-11 \cos^{2}(i)-6 \cos^{4}(i)\right]\right\} \cos(2\varphi)\\
    &-\frac{4}{3} \sin^{2}(i)\left[1+\cos^{2}(i)\right](1-3 \eta) \cos(4\varphi)
    +\frac{2}{3}\left(\bar{\gamma}+2\beta_{+}-2\frac{\Delta m}{m}\beta_{-}\right)\left[1+\cos ^{2} (i)\right] \cos (2 \varphi)\,,\\
    &H^{1}_{+(GB)}=
    \frac{c^4}{24G^2} x^2 \frac{\alpha f'(\phi_{0})}{m^2\bar{\alpha}^{5/2}} \Bigg\{32 \left(\frac{\Delta m}{m} \mathcal{S}_{-}+3 \mathcal{S}_{+}\right) \left\{\left[\cos ^2(i)+1\right] \cos (2\varphi)-3 \sin ^2(i)\right\}\\
    &+192\left[\left[\cos (2i)+3\right] \cos (2\varphi) \left(\frac{\Delta m}{m}\mathcal{S}_{-}+2 \mathcal{S}_{+}\right)+\sin ^2(i) \left(\frac{\Delta m}{m}\mathcal{S}_{-}+3 \mathcal{S}_{+}\right)\right]\\
    & +18 \big[(2\eta +1) \mathcal{S}_{-}+(1-2 \eta) \mathcal{S}_{+}\big] \left[2 \sin ^2(2i) \cos (2\varphi)-\sin ^2(i) [\cos (2 i)+3] (3 \cos (4 \varphi)+1)\right]\Bigg\},
    \end{split}
\end{equation}
\normalsize
whereas the cross terms are given by
\small
\begin{equation}\label{CP}
    \begin{split}
    &H^{0}_{\times}= -2 \cos(i) \sin(2\varphi)\,, \\
    &H^{1/2}_{\times}= -\frac{3}{8} \frac{\Delta m}{m}  \sin (2i) \Big[\sin (\varphi)-3 \sin (3 \varphi)\Big]\,,\\
    &H^{1}_{\times}=
    \frac{\cos(i)}{3}\Big[\left(17-4 \cos^{2}(i)\right)-\eta\left(13-12 \cos^{2}(i)\right)\Big] \sin (2 \varphi)-\frac{8}{2}(1-3\eta) \cos(i) \sin^{2}(i) \sin(4\varphi)\\
    &+\frac{4}{3}\cos(i)\sin(2\varphi)\Big[\bar{\gamma}+2\beta_{+}-2\frac{\Delta m}{m}\beta_{-}\Big]\,,\\
   & H^{1}_{\times(GB)}=
   \frac{c^4}{3G^2}x^2\frac{\alpha f'(\phi_{0})}{m^2\bar{\alpha}^{5/2}}  \cos (i) \left\{-27 \sin ^2(i) \sin (4\varphi) \big[(2 \eta+1)\mathcal{S}_{-}+(1-2 \eta) \mathcal{S}_{+}\big]\right.\\
   &\left.+ 2\sin(2\varphi) \left[9 \sin ^2(i) \left[(2 \eta+1)\mathcal{S}_{-}+(1-2 \eta)\mathcal{S}_{+}\right]+52\frac{\Delta m}{m}\mathcal{S}_{-}+108\mathcal{S}_+-3\right]
   \right\}\,.
    \end{split}
\end{equation}
\normalsize

Applying the same procedure on Eq.~\eqref{scalarwavefinale} gives the 0.5PN scalar waveform:
\small
\begin{equation}\label{SP}
\begin{split}
&\Phi=\frac{2G\mu \sqrt{\bar{\alpha}}} {c^2 R}\left(\frac{G m\bar{\alpha}\omega}{c^3}\right)^{2/3}\left\{x^{-1/2} \hat{\Phi}^{-1/2} +\hat{\Phi}^{0}+x^{1/2}\hat{\Phi}^{1/2}+x^{1/2}\hat{\Phi}^{1/2}_{(GB)} \right\}\,,\\
&\hat{\Phi}^{-1/2}=2 \mathcal{S}_{-} \sin (i) \cos (\phi)\,,\\
&\hat{\Phi}^{0}= \left(\mathcal{S}_{+}-\frac{\Delta m}{m} \mathcal{S}_{-}\right) \left(\sin^2(i) \cos(2\varphi)-\frac{1}{2}\right)+\frac{8}{\bar{\gamma} }\left(\beta_{-}\mathcal{S}_{-}+\beta_{+} \mathcal{S}_{+}\right)-2 \mathcal{S}_{+}\,,\\
&\hat{\Phi}^{1/2}=\sin(i) \cos (\phi)\left[ \frac{3}{2} \frac{\Delta m}{m} \mathcal{S}_{+}-\frac{4}{\bar{\gamma}} \frac{\Delta m}{m}\left(\mathcal{S}_{+} \beta_{+}+\mathcal{S}_{-} \beta_{-}\right)+\frac{4}{\bar{\gamma}}\left(\mathcal{S}_{-} \beta_{+}+\mathcal{S}_{+} \beta_{-}\right)\right.
\\
&\left.\quad-\frac{\mathcal{S}_-}{3}\left(\frac{7}{2}+2\bar{\gamma}+4\beta_{+}-5\eta-4\frac{\Delta m}{m}\beta_{-}\right)\right]\\
&\quad+\frac{1}{8} \sin^3(i) \left[\cos(\phi)-9\cos(3\phi)\right]\left(\frac{\Delta m}{m}\mathcal{S}_{+}+(2\eta-1) \mathcal{S}_{-}\right)\,,\\
&\hat{\Phi}^{1/2}_{(GB)}=-\sin(i)\cos(\phi)
\frac{c^4}{G^2 }\frac{\alpha f'(\phi_{0})}{\bar{\alpha}^{5/2}m^2}x^2\left[2 \mathcal{S}_{+}\eta\frac{\Delta m}{m} \left(\frac{\Delta m}{m} \mathcal{S}_{-}+\mathcal{S}_{+}\right)+\frac{16\mathcal{S}_{-}}{3
}\left(3\mathcal{S}_{+}+\frac{\Delta m}{m}\mathcal{S}_{-}\right)
\right]\,.
\end{split}
\end{equation}
\normalsize

\end{appendices}


\begin{thebibliography}{94}%
\makeatletter
\providecommand \@ifxundefined [1]{%
 \@ifx{#1\undefined}
}%
\providecommand \@ifnum [1]{%
 \ifnum #1\expandafter \@firstoftwo
 \else \expandafter \@secondoftwo
 \fi
}%
\providecommand \@ifx [1]{%
 \ifx #1\expandafter \@firstoftwo
 \else \expandafter \@secondoftwo
 \fi
}%
\providecommand \natexlab [1]{#1}%
\providecommand \enquote  [1]{``#1''}%
\providecommand \bibnamefont  [1]{#1}%
\providecommand \bibfnamefont [1]{#1}%
\providecommand \citenamefont [1]{#1}%
\providecommand \href@noop [0]{\@secondoftwo}%
\providecommand \href [0]{\begingroup \@sanitize@url \@href}%
\providecommand \@href[1]{\@@startlink{#1}\@@href}%
\providecommand \@@href[1]{\endgroup#1\@@endlink}%
\providecommand \@sanitize@url [0]{\catcode `\\12\catcode `\$12\catcode
  `\&12\catcode `\#12\catcode `\^12\catcode `\_12\catcode `\%12\relax}%
\providecommand \@@startlink[1]{}%
\providecommand \@@endlink[0]{}%
\providecommand \url  [0]{\begingroup\@sanitize@url \@url }%
\providecommand \@url [1]{\endgroup\@href {#1}{\urlprefix }}%
\providecommand \urlprefix  [0]{URL }%
\providecommand \Eprint [0]{\href }%
\providecommand \doibase [0]{http://dx.doi.org/}%
\providecommand \selectlanguage [0]{\@gobble}%
\providecommand \bibinfo  [0]{\@secondoftwo}%
\providecommand \bibfield  [0]{\@secondoftwo}%
\providecommand \translation [1]{[#1]}%
\providecommand \BibitemOpen [0]{}%
\providecommand \bibitemStop [0]{}%
\providecommand \bibitemNoStop [0]{.\EOS\space}%
\providecommand \EOS [0]{\spacefactor3000\relax}%
\providecommand \BibitemShut  [1]{\csname bibitem#1\endcsname}%
\let\auto@bib@innerbib\@empty
\bibitem [{\citenamefont {Abbott}\ \emph
  {et~al.}(2016{\natexlab{a}})\citenamefont {Abbott} \emph
  {et~al.}}]{Abbott:2016blz}%
  \BibitemOpen
  \bibfield  {author} {\bibinfo {author} {\bibfnamefont {B.}~\bibnamefont
  {Abbott}} \emph {et~al.} (\bibinfo {collaboration} {LIGO Scientific,
  Virgo}),\ }\href {\doibase 10.1103/PhysRevLett.116.061102} {\bibfield
  {journal} {\bibinfo  {journal} {Phys. Rev. Lett.}\ }\textbf {\bibinfo
  {volume} {116}},\ \bibinfo {pages} {061102} (\bibinfo {year}
  {2016}{\natexlab{a}})},\ \Eprint {http://arxiv.org/abs/1602.03837}
  {arXiv:1602.03837 [gr-qc]} \BibitemShut {NoStop}%
\bibitem [{\citenamefont {Abbott}\ \emph
  {et~al.}(2019{\natexlab{a}})\citenamefont {Abbott} \emph
  {et~al.}}]{LIGOScientific:2018mvr}%
  \BibitemOpen
  \bibfield  {author} {\bibinfo {author} {\bibfnamefont {B.}~\bibnamefont
  {Abbott}} \emph {et~al.} (\bibinfo {collaboration} {LIGO Scientific,
  Virgo}),\ }\href {\doibase 10.1103/PhysRevX.9.031040} {\bibfield  {journal}
  {\bibinfo  {journal} {Phys. Rev. X}\ }\textbf {\bibinfo {volume} {9}},\
  \bibinfo {pages} {031040} (\bibinfo {year} {2019}{\natexlab{a}})},\ \Eprint
  {http://arxiv.org/abs/1811.12907} {arXiv:1811.12907 [astro-ph.HE]}
  \BibitemShut {NoStop}%
\bibitem [{\citenamefont {Abbott}\ \emph
  {et~al.}(2020{\natexlab{a}})\citenamefont {Abbott} \emph
  {et~al.}}]{Abbott:2020niy}%
  \BibitemOpen
  \bibfield  {author} {\bibinfo {author} {\bibfnamefont {R.}~\bibnamefont
  {Abbott}} \emph {et~al.} (\bibinfo {collaboration} {LIGO Scientific,
  Virgo}),\ }\href@noop {} {\  (\bibinfo {year} {2020}{\natexlab{a}})},\
  \Eprint {http://arxiv.org/abs/2010.14527} {arXiv:2010.14527 [gr-qc]}
  \BibitemShut {NoStop}%
\bibitem [{\citenamefont {Abbott}\ \emph
  {et~al.}(2020{\natexlab{b}})\citenamefont {Abbott} \emph
  {et~al.}}]{Abbott:2020khf}%
  \BibitemOpen
  \bibfield  {author} {\bibinfo {author} {\bibfnamefont {R.}~\bibnamefont
  {Abbott}} \emph {et~al.} (\bibinfo {collaboration} {LIGO Scientific,
  Virgo}),\ }\href {\doibase 10.3847/2041-8213/ab960f} {\bibfield  {journal}
  {\bibinfo  {journal} {Astrophys. J. Lett.}\ }\textbf {\bibinfo {volume}
  {896}},\ \bibinfo {pages} {L44} (\bibinfo {year} {2020}{\natexlab{b}})},\
  \Eprint {http://arxiv.org/abs/2006.12611} {arXiv:2006.12611 [astro-ph.HE]}
  \BibitemShut {NoStop}%
\bibitem [{\citenamefont {Yunes}\ and\ \citenamefont
  {Siemens}(2013)}]{Yunes:2013dva}%
  \BibitemOpen
  \bibfield  {author} {\bibinfo {author} {\bibfnamefont {N.}~\bibnamefont
  {Yunes}}\ and\ \bibinfo {author} {\bibfnamefont {X.}~\bibnamefont
  {Siemens}},\ }\href {\doibase 10.12942/lrr-2013-9} {\bibfield  {journal}
  {\bibinfo  {journal} {Living Rev. Rel.}\ }\textbf {\bibinfo {volume} {16}},\
  \bibinfo {pages} {9} (\bibinfo {year} {2013})},\ \Eprint
  {http://arxiv.org/abs/1304.3473} {arXiv:1304.3473 [gr-qc]} \BibitemShut
  {NoStop}%
\bibitem [{\citenamefont {Berti}\ \emph {et~al.}(2015)\citenamefont {Berti}
  \emph {et~al.}}]{Berti:2015itd}%
  \BibitemOpen
  \bibfield  {author} {\bibinfo {author} {\bibfnamefont {E.}~\bibnamefont
  {Berti}} \emph {et~al.},\ }\href {\doibase 10.1088/0264-9381/32/24/243001}
  {\bibfield  {journal} {\bibinfo  {journal} {Class. Quant. Grav.}\ }\textbf
  {\bibinfo {volume} {32}},\ \bibinfo {pages} {243001} (\bibinfo {year}
  {2015})},\ \Eprint {http://arxiv.org/abs/1501.07274} {arXiv:1501.07274
  [gr-qc]} \BibitemShut {NoStop}%
\bibitem [{\citenamefont {Yunes}\ \emph {et~al.}(2016)\citenamefont {Yunes},
  \citenamefont {Yagi},\ and\ \citenamefont {Pretorius}}]{Yunes:2016jcc}%
  \BibitemOpen
  \bibfield  {author} {\bibinfo {author} {\bibfnamefont {N.}~\bibnamefont
  {Yunes}}, \bibinfo {author} {\bibfnamefont {K.}~\bibnamefont {Yagi}}, \ and\
  \bibinfo {author} {\bibfnamefont {F.}~\bibnamefont {Pretorius}},\ }\href
  {\doibase 10.1103/PhysRevD.94.084002} {\bibfield  {journal} {\bibinfo
  {journal} {Phys. Rev. D}\ }\textbf {\bibinfo {volume} {94}},\ \bibinfo
  {pages} {084002} (\bibinfo {year} {2016})},\ \Eprint
  {http://arxiv.org/abs/1603.08955} {arXiv:1603.08955 [gr-qc]} \BibitemShut
  {NoStop}%
\bibitem [{\citenamefont {Yagi}\ and\ \citenamefont
  {Stein}(2016)}]{Yagi:2016jml}%
  \BibitemOpen
  \bibfield  {author} {\bibinfo {author} {\bibfnamefont {K.}~\bibnamefont
  {Yagi}}\ and\ \bibinfo {author} {\bibfnamefont {L.~C.}\ \bibnamefont
  {Stein}},\ }\href {\doibase 10.1088/0264-9381/33/5/054001} {\bibfield
  {journal} {\bibinfo  {journal} {Class. Quant. Grav.}\ }\textbf {\bibinfo
  {volume} {33}},\ \bibinfo {pages} {054001} (\bibinfo {year} {2016})},\
  \Eprint {http://arxiv.org/abs/1602.02413} {arXiv:1602.02413 [gr-qc]}
  \BibitemShut {NoStop}%
\bibitem [{\citenamefont {Abbott}\ \emph
  {et~al.}(2019{\natexlab{b}})\citenamefont {Abbott} \emph
  {et~al.}}]{LIGOScientific:2019fpa}%
  \BibitemOpen
  \bibfield  {author} {\bibinfo {author} {\bibfnamefont {B.}~\bibnamefont
  {Abbott}} \emph {et~al.} (\bibinfo {collaboration} {LIGO Scientific,
  Virgo}),\ }\href {\doibase 10.1103/PhysRevD.100.104036} {\bibfield  {journal}
  {\bibinfo  {journal} {Phys. Rev. D}\ }\textbf {\bibinfo {volume} {100}},\
  \bibinfo {pages} {104036} (\bibinfo {year} {2019}{\natexlab{b}})},\ \Eprint
  {http://arxiv.org/abs/1903.04467} {arXiv:1903.04467 [gr-qc]} \BibitemShut
  {NoStop}%
\bibitem [{\citenamefont {Abbott}\ \emph
  {et~al.}(2020{\natexlab{c}})\citenamefont {Abbott} \emph
  {et~al.}}]{Abbott:2020jks}%
  \BibitemOpen
  \bibfield  {author} {\bibinfo {author} {\bibfnamefont {R.}~\bibnamefont
  {Abbott}} \emph {et~al.} (\bibinfo {collaboration} {LIGO Scientific,
  Virgo}),\ }\href@noop {} {\  (\bibinfo {year} {2020}{\natexlab{c}})},\
  \Eprint {http://arxiv.org/abs/2010.14529} {arXiv:2010.14529 [gr-qc]}
  \BibitemShut {NoStop}%
\bibitem [{\citenamefont {Carson}\ and\ \citenamefont
  {Yagi}(2020)}]{Carson:2020rea}%
  \BibitemOpen
  \bibfield  {author} {\bibinfo {author} {\bibfnamefont {Z.}~\bibnamefont
  {Carson}}\ and\ \bibinfo {author} {\bibfnamefont {K.}~\bibnamefont {Yagi}},\
  }\href@noop {} {\  (\bibinfo {year} {2020})},\ \Eprint
  {http://arxiv.org/abs/2011.02938} {arXiv:2011.02938 [gr-qc]} \BibitemShut
  {NoStop}%
\bibitem [{\citenamefont {Abbott}\ \emph
  {et~al.}(2016{\natexlab{b}})\citenamefont {Abbott} \emph
  {et~al.}}]{TheLIGOScientific:2016src}%
  \BibitemOpen
  \bibfield  {author} {\bibinfo {author} {\bibfnamefont {B.~P.}\ \bibnamefont
  {Abbott}} \emph {et~al.} (\bibinfo {collaboration} {LIGO Scientific,
  Virgo}),\ }\href {\doibase 10.1103/PhysRevLett.116.221101} {\bibfield
  {journal} {\bibinfo  {journal} {Phys. Rev. Lett.}\ }\textbf {\bibinfo
  {volume} {116}},\ \bibinfo {pages} {221101} (\bibinfo {year}
  {2016}{\natexlab{b}})},\ \bibinfo {note} {[Erratum: Phys.Rev.Lett. 121,
  129902 (2018)]},\ \Eprint {http://arxiv.org/abs/1602.03841} {arXiv:1602.03841
  [gr-qc]} \BibitemShut {NoStop}%
\bibitem [{\citenamefont {Isi}\ \emph {et~al.}(2019)\citenamefont {Isi},
  \citenamefont {Giesler}, \citenamefont {Farr}, \citenamefont {Scheel},\ and\
  \citenamefont {Teukolsky}}]{Isi:2019aib}%
  \BibitemOpen
  \bibfield  {author} {\bibinfo {author} {\bibfnamefont {M.}~\bibnamefont
  {Isi}}, \bibinfo {author} {\bibfnamefont {M.}~\bibnamefont {Giesler}},
  \bibinfo {author} {\bibfnamefont {W.~M.}\ \bibnamefont {Farr}}, \bibinfo
  {author} {\bibfnamefont {M.~A.}\ \bibnamefont {Scheel}}, \ and\ \bibinfo
  {author} {\bibfnamefont {S.~A.}\ \bibnamefont {Teukolsky}},\ }\href {\doibase
  10.1103/PhysRevLett.123.111102} {\bibfield  {journal} {\bibinfo  {journal}
  {Phys. Rev. Lett.}\ }\textbf {\bibinfo {volume} {123}},\ \bibinfo {pages}
  {111102} (\bibinfo {year} {2019})},\ \Eprint
  {http://arxiv.org/abs/1905.00869} {arXiv:1905.00869 [gr-qc]} \BibitemShut
  {NoStop}%
\bibitem [{\citenamefont {Yunes}\ and\ \citenamefont
  {Pretorius}(2009)}]{Yunes:2009ke}%
  \BibitemOpen
  \bibfield  {author} {\bibinfo {author} {\bibfnamefont {N.}~\bibnamefont
  {Yunes}}\ and\ \bibinfo {author} {\bibfnamefont {F.}~\bibnamefont
  {Pretorius}},\ }\href {\doibase 10.1103/PhysRevD.80.122003} {\bibfield
  {journal} {\bibinfo  {journal} {Phys. Rev. D}\ }\textbf {\bibinfo {volume}
  {80}},\ \bibinfo {pages} {122003} (\bibinfo {year} {2009})},\ \Eprint
  {http://arxiv.org/abs/0909.3328} {arXiv:0909.3328 [gr-qc]} \BibitemShut
  {NoStop}%
\bibitem [{\citenamefont {Will}(2014)}]{Will:2014kxa}%
  \BibitemOpen
  \bibfield  {author} {\bibinfo {author} {\bibfnamefont {C.~M.}\ \bibnamefont
  {Will}},\ }\href {\doibase 10.12942/lrr-2014-4} {\bibfield  {journal}
  {\bibinfo  {journal} {Living Rev. Rel.}\ }\textbf {\bibinfo {volume} {17}},\
  \bibinfo {pages} {4} (\bibinfo {year} {2014})},\ \Eprint
  {http://arxiv.org/abs/1403.7377} {arXiv:1403.7377 [gr-qc]} \BibitemShut
  {NoStop}%
\bibitem [{\citenamefont {Boulware}\ and\ \citenamefont
  {Deser}(1985)}]{Boulware:1985wk}%
  \BibitemOpen
  \bibfield  {author} {\bibinfo {author} {\bibfnamefont {D.~G.}\ \bibnamefont
  {Boulware}}\ and\ \bibinfo {author} {\bibfnamefont {S.}~\bibnamefont
  {Deser}},\ }\href {\doibase 10.1103/PhysRevLett.55.2656} {\bibfield
  {journal} {\bibinfo  {journal} {Phys. Rev. Lett.}\ }\textbf {\bibinfo
  {volume} {55}},\ \bibinfo {pages} {2656} (\bibinfo {year}
  {1985})}\BibitemShut {NoStop}%
\bibitem [{\citenamefont {Gross}\ and\ \citenamefont
  {Sloan}(1987)}]{GROSS198741}%
  \BibitemOpen
  \bibfield  {author} {\bibinfo {author} {\bibfnamefont {D.~J.}\ \bibnamefont
  {Gross}}\ and\ \bibinfo {author} {\bibfnamefont {J.~H.}\ \bibnamefont
  {Sloan}},\ }\href {\doibase https://doi.org/10.1016/0550-3213(87)90465-2}
  {\bibfield  {journal} {\bibinfo  {journal} {Nuclear Physics B}\ }\textbf
  {\bibinfo {volume} {291}},\ \bibinfo {pages} {41 } (\bibinfo {year}
  {1987})}\BibitemShut {NoStop}%
\bibitem [{\citenamefont {Kanti}\ \emph {et~al.}(1996)\citenamefont {Kanti},
  \citenamefont {Mavromatos}, \citenamefont {Rizos}, \citenamefont {Tamvakis},\
  and\ \citenamefont {Winstanley}}]{Kanti:1995vq}%
  \BibitemOpen
  \bibfield  {author} {\bibinfo {author} {\bibfnamefont {P.}~\bibnamefont
  {Kanti}}, \bibinfo {author} {\bibfnamefont {N.}~\bibnamefont {Mavromatos}},
  \bibinfo {author} {\bibfnamefont {J.}~\bibnamefont {Rizos}}, \bibinfo
  {author} {\bibfnamefont {K.}~\bibnamefont {Tamvakis}}, \ and\ \bibinfo
  {author} {\bibfnamefont {E.}~\bibnamefont {Winstanley}},\ }\href {\doibase
  10.1103/PhysRevD.54.5049} {\bibfield  {journal} {\bibinfo  {journal} {Phys.
  Rev. D}\ }\textbf {\bibinfo {volume} {54}},\ \bibinfo {pages} {5049}
  (\bibinfo {year} {1996})},\ \Eprint {http://arxiv.org/abs/hep-th/9511071}
  {arXiv:hep-th/9511071} \BibitemShut {NoStop}%
\bibitem [{\citenamefont {Charmousis}\ \emph
  {et~al.}(2012{\natexlab{a}})\citenamefont {Charmousis}, \citenamefont
  {Copeland}, \citenamefont {Padilla},\ and\ \citenamefont
  {Saffin}}]{Charmousis:2011bf}%
  \BibitemOpen
  \bibfield  {author} {\bibinfo {author} {\bibfnamefont {C.}~\bibnamefont
  {Charmousis}}, \bibinfo {author} {\bibfnamefont {E.~J.}\ \bibnamefont
  {Copeland}}, \bibinfo {author} {\bibfnamefont {A.}~\bibnamefont {Padilla}}, \
  and\ \bibinfo {author} {\bibfnamefont {P.~M.}\ \bibnamefont {Saffin}},\
  }\href {\doibase 10.1103/PhysRevLett.108.051101} {\bibfield  {journal}
  {\bibinfo  {journal} {Phys. Rev. Lett.}\ }\textbf {\bibinfo {volume} {108}},\
  \bibinfo {pages} {051101} (\bibinfo {year} {2012}{\natexlab{a}})},\ \Eprint
  {http://arxiv.org/abs/1106.2000} {arXiv:1106.2000 [hep-th]} \BibitemShut
  {NoStop}%
\bibitem [{\citenamefont {Charmousis}\ \emph
  {et~al.}(2012{\natexlab{b}})\citenamefont {Charmousis}, \citenamefont
  {Copeland}, \citenamefont {Padilla},\ and\ \citenamefont
  {Saffin}}]{Charmousis:2011ea}%
  \BibitemOpen
  \bibfield  {author} {\bibinfo {author} {\bibfnamefont {C.}~\bibnamefont
  {Charmousis}}, \bibinfo {author} {\bibfnamefont {E.~J.}\ \bibnamefont
  {Copeland}}, \bibinfo {author} {\bibfnamefont {A.}~\bibnamefont {Padilla}}, \
  and\ \bibinfo {author} {\bibfnamefont {P.~M.}\ \bibnamefont {Saffin}},\
  }\href {\doibase 10.1103/PhysRevD.85.104040} {\bibfield  {journal} {\bibinfo
  {journal} {Phys. Rev. D}\ }\textbf {\bibinfo {volume} {85}},\ \bibinfo
  {pages} {104040} (\bibinfo {year} {2012}{\natexlab{b}})},\ \Eprint
  {http://arxiv.org/abs/1112.4866} {arXiv:1112.4866 [hep-th]} \BibitemShut
  {NoStop}%
\bibitem [{\citenamefont {Stelle}(1977)}]{PhysRevD.16.953}%
  \BibitemOpen
  \bibfield  {author} {\bibinfo {author} {\bibfnamefont {K.~S.}\ \bibnamefont
  {Stelle}},\ }\href {\doibase 10.1103/PhysRevD.16.953} {\bibfield  {journal}
  {\bibinfo  {journal} {Phys. Rev. D}\ }\textbf {\bibinfo {volume} {16}},\
  \bibinfo {pages} {953} (\bibinfo {year} {1977})}\BibitemShut {NoStop}%
\bibitem [{\citenamefont {Kov\'acs}(2019)}]{Kovacs:2019jqj}%
  \BibitemOpen
  \bibfield  {author} {\bibinfo {author} {\bibfnamefont {A.~D.}\ \bibnamefont
  {Kov\'acs}},\ }\href {\doibase 10.1103/PhysRevD.100.024005} {\bibfield
  {journal} {\bibinfo  {journal} {Phys. Rev. D}\ }\textbf {\bibinfo {volume}
  {100}},\ \bibinfo {pages} {024005} (\bibinfo {year} {2019})},\ \Eprint
  {http://arxiv.org/abs/1904.00963} {arXiv:1904.00963 [gr-qc]} \BibitemShut
  {NoStop}%
\bibitem [{\citenamefont {Kov\'acs}\ and\ \citenamefont
  {Reall}(2020{\natexlab{a}})}]{Kovacs:2020ywu}%
  \BibitemOpen
  \bibfield  {author} {\bibinfo {author} {\bibfnamefont {A.~D.}\ \bibnamefont
  {Kov\'acs}}\ and\ \bibinfo {author} {\bibfnamefont {H.~S.}\ \bibnamefont
  {Reall}},\ }\href {\doibase 10.1103/PhysRevD.101.124003} {\bibfield
  {journal} {\bibinfo  {journal} {Phys. Rev. D}\ }\textbf {\bibinfo {volume}
  {101}},\ \bibinfo {pages} {124003} (\bibinfo {year} {2020}{\natexlab{a}})},\
  \Eprint {http://arxiv.org/abs/2003.08398} {arXiv:2003.08398 [gr-qc]}
  \BibitemShut {NoStop}%
\bibitem [{\citenamefont {Kov\'acs}\ and\ \citenamefont
  {Reall}(2020{\natexlab{b}})}]{Kovacs:2020pns}%
  \BibitemOpen
  \bibfield  {author} {\bibinfo {author} {\bibfnamefont {A.~D.}\ \bibnamefont
  {Kov\'acs}}\ and\ \bibinfo {author} {\bibfnamefont {H.~S.}\ \bibnamefont
  {Reall}},\ }\href {\doibase 10.1103/PhysRevLett.124.221101} {\bibfield
  {journal} {\bibinfo  {journal} {Phys. Rev. Lett.}\ }\textbf {\bibinfo
  {volume} {124}},\ \bibinfo {pages} {221101} (\bibinfo {year}
  {2020}{\natexlab{b}})},\ \Eprint {http://arxiv.org/abs/2003.04327}
  {arXiv:2003.04327 [gr-qc]} \BibitemShut {NoStop}%
\bibitem [{\citenamefont {Antoniou}\ \emph
  {et~al.}(2018{\natexlab{a}})\citenamefont {Antoniou}, \citenamefont
  {Bakopoulos},\ and\ \citenamefont {Kanti}}]{Antoniou:2017acq}%
  \BibitemOpen
  \bibfield  {author} {\bibinfo {author} {\bibfnamefont {G.}~\bibnamefont
  {Antoniou}}, \bibinfo {author} {\bibfnamefont {A.}~\bibnamefont
  {Bakopoulos}}, \ and\ \bibinfo {author} {\bibfnamefont {P.}~\bibnamefont
  {Kanti}},\ }\href {\doibase 10.1103/PhysRevLett.120.131102} {\bibfield
  {journal} {\bibinfo  {journal} {Phys. Rev. Lett.}\ }\textbf {\bibinfo
  {volume} {120}},\ \bibinfo {pages} {131102} (\bibinfo {year}
  {2018}{\natexlab{a}})},\ \Eprint {http://arxiv.org/abs/1711.03390}
  {arXiv:1711.03390 [hep-th]} \BibitemShut {NoStop}%
\bibitem [{\citenamefont {Antoniou}\ \emph
  {et~al.}(2018{\natexlab{b}})\citenamefont {Antoniou}, \citenamefont
  {Bakopoulos},\ and\ \citenamefont {Kanti}}]{Antoniou:2017hxj}%
  \BibitemOpen
  \bibfield  {author} {\bibinfo {author} {\bibfnamefont {G.}~\bibnamefont
  {Antoniou}}, \bibinfo {author} {\bibfnamefont {A.}~\bibnamefont
  {Bakopoulos}}, \ and\ \bibinfo {author} {\bibfnamefont {P.}~\bibnamefont
  {Kanti}},\ }\href {\doibase 10.1103/PhysRevD.97.084037} {\bibfield  {journal}
  {\bibinfo  {journal} {Phys. Rev. D}\ }\textbf {\bibinfo {volume} {97}},\
  \bibinfo {pages} {084037} (\bibinfo {year} {2018}{\natexlab{b}})},\ \Eprint
  {http://arxiv.org/abs/1711.07431} {arXiv:1711.07431 [hep-th]} \BibitemShut
  {NoStop}%
\bibitem [{\citenamefont {Mignemi}\ and\ \citenamefont
  {Stewart}(1993)}]{Mignemi:1992nt}%
  \BibitemOpen
  \bibfield  {author} {\bibinfo {author} {\bibfnamefont {S.}~\bibnamefont
  {Mignemi}}\ and\ \bibinfo {author} {\bibfnamefont {N.~R.}\ \bibnamefont
  {Stewart}},\ }\href {\doibase 10.1103/PhysRevD.47.5259} {\bibfield  {journal}
  {\bibinfo  {journal} {Phys. Rev. D}\ }\textbf {\bibinfo {volume} {47}},\
  \bibinfo {pages} {5259} (\bibinfo {year} {1993})},\ \Eprint
  {http://arxiv.org/abs/hep-th/9212146} {arXiv:hep-th/9212146} \BibitemShut
  {NoStop}%
\bibitem [{\citenamefont {Sotiriou}\ and\ \citenamefont
  {Zhou}(2014)}]{Sotiriou:2014pfa}%
  \BibitemOpen
  \bibfield  {author} {\bibinfo {author} {\bibfnamefont {T.~P.}\ \bibnamefont
  {Sotiriou}}\ and\ \bibinfo {author} {\bibfnamefont {S.-Y.}\ \bibnamefont
  {Zhou}},\ }\href {\doibase 10.1103/PhysRevD.90.124063} {\bibfield  {journal}
  {\bibinfo  {journal} {Phys. Rev. D}\ }\textbf {\bibinfo {volume} {90}},\
  \bibinfo {pages} {124063} (\bibinfo {year} {2014})},\ \Eprint
  {http://arxiv.org/abs/1408.1698} {arXiv:1408.1698 [gr-qc]} \BibitemShut
  {NoStop}%
\bibitem [{\citenamefont {Benkel}\ \emph {et~al.}(2017)\citenamefont {Benkel},
  \citenamefont {Sotiriou},\ and\ \citenamefont {Witek}}]{Benkel:2016rlz}%
  \BibitemOpen
  \bibfield  {author} {\bibinfo {author} {\bibfnamefont {R.}~\bibnamefont
  {Benkel}}, \bibinfo {author} {\bibfnamefont {T.~P.}\ \bibnamefont
  {Sotiriou}}, \ and\ \bibinfo {author} {\bibfnamefont {H.}~\bibnamefont
  {Witek}},\ }\href {\doibase 10.1088/1361-6382/aa5ce7} {\bibfield  {journal}
  {\bibinfo  {journal} {Class. Quant. Grav.}\ }\textbf {\bibinfo {volume}
  {34}},\ \bibinfo {pages} {064001} (\bibinfo {year} {2017})},\ \Eprint
  {http://arxiv.org/abs/1610.09168} {arXiv:1610.09168 [gr-qc]} \BibitemShut
  {NoStop}%
\bibitem [{\citenamefont {Benkel}\ \emph {et~al.}(2016)\citenamefont {Benkel},
  \citenamefont {Sotiriou},\ and\ \citenamefont {Witek}}]{Benkel:2016kcq}%
  \BibitemOpen
  \bibfield  {author} {\bibinfo {author} {\bibfnamefont {R.}~\bibnamefont
  {Benkel}}, \bibinfo {author} {\bibfnamefont {T.~P.}\ \bibnamefont
  {Sotiriou}}, \ and\ \bibinfo {author} {\bibfnamefont {H.}~\bibnamefont
  {Witek}},\ }\href {\doibase 10.1103/PhysRevD.94.121503} {\bibfield  {journal}
  {\bibinfo  {journal} {Phys. Rev. D}\ }\textbf {\bibinfo {volume} {94}},\
  \bibinfo {pages} {121503} (\bibinfo {year} {2016})},\ \Eprint
  {http://arxiv.org/abs/1612.08184} {arXiv:1612.08184 [gr-qc]} \BibitemShut
  {NoStop}%
\bibitem [{\citenamefont {Ripley}\ and\ \citenamefont
  {Pretorius}(2019)}]{Ripley:2019irj}%
  \BibitemOpen
  \bibfield  {author} {\bibinfo {author} {\bibfnamefont {J.~L.}\ \bibnamefont
  {Ripley}}\ and\ \bibinfo {author} {\bibfnamefont {F.}~\bibnamefont
  {Pretorius}},\ }\href {\doibase 10.1088/1361-6382/ab2416} {\bibfield
  {journal} {\bibinfo  {journal} {Class. Quant. Grav.}\ }\textbf {\bibinfo
  {volume} {36}},\ \bibinfo {pages} {134001} (\bibinfo {year} {2019})},\
  \Eprint {http://arxiv.org/abs/1903.07543} {arXiv:1903.07543 [gr-qc]}
  \BibitemShut {NoStop}%
\bibitem [{\citenamefont {Yunes}\ and\ \citenamefont
  {Stein}(2011)}]{Yunes:2011we}%
  \BibitemOpen
  \bibfield  {author} {\bibinfo {author} {\bibfnamefont {N.}~\bibnamefont
  {Yunes}}\ and\ \bibinfo {author} {\bibfnamefont {L.~C.}\ \bibnamefont
  {Stein}},\ }\href {\doibase 10.1103/PhysRevD.83.104002} {\bibfield  {journal}
  {\bibinfo  {journal} {Phys. Rev. D}\ }\textbf {\bibinfo {volume} {83}},\
  \bibinfo {pages} {104002} (\bibinfo {year} {2011})},\ \Eprint
  {http://arxiv.org/abs/1101.2921} {arXiv:1101.2921 [gr-qc]} \BibitemShut
  {NoStop}%
\bibitem [{\citenamefont {Pani}\ and\ \citenamefont
  {Cardoso}(2009)}]{Pani:2009wy}%
  \BibitemOpen
  \bibfield  {author} {\bibinfo {author} {\bibfnamefont {P.}~\bibnamefont
  {Pani}}\ and\ \bibinfo {author} {\bibfnamefont {V.}~\bibnamefont {Cardoso}},\
  }\href {\doibase 10.1103/PhysRevD.79.084031} {\bibfield  {journal} {\bibinfo
  {journal} {Phys. Rev. D}\ }\textbf {\bibinfo {volume} {79}},\ \bibinfo
  {pages} {084031} (\bibinfo {year} {2009})},\ \Eprint
  {http://arxiv.org/abs/0902.1569} {arXiv:0902.1569 [gr-qc]} \BibitemShut
  {NoStop}%
\bibitem [{\citenamefont {Barausse}\ and\ \citenamefont
  {Yagi}(2015)}]{Barausse:2015wia}%
  \BibitemOpen
  \bibfield  {author} {\bibinfo {author} {\bibfnamefont {E.}~\bibnamefont
  {Barausse}}\ and\ \bibinfo {author} {\bibfnamefont {K.}~\bibnamefont
  {Yagi}},\ }\href {\doibase 10.1103/PhysRevLett.115.211105} {\bibfield
  {journal} {\bibinfo  {journal} {Phys. Rev. Lett.}\ }\textbf {\bibinfo
  {volume} {115}},\ \bibinfo {pages} {211105} (\bibinfo {year} {2015})},\
  \Eprint {http://arxiv.org/abs/1509.04539} {arXiv:1509.04539 [gr-qc]}
  \BibitemShut {NoStop}%
\bibitem [{\citenamefont {Kleihaus}\ \emph {et~al.}(2011)\citenamefont
  {Kleihaus}, \citenamefont {Kunz},\ and\ \citenamefont
  {Radu}}]{Kleihaus:2011tg}%
  \BibitemOpen
  \bibfield  {author} {\bibinfo {author} {\bibfnamefont {B.}~\bibnamefont
  {Kleihaus}}, \bibinfo {author} {\bibfnamefont {J.}~\bibnamefont {Kunz}}, \
  and\ \bibinfo {author} {\bibfnamefont {E.}~\bibnamefont {Radu}},\ }\href
  {\doibase 10.1103/PhysRevLett.106.151104} {\bibfield  {journal} {\bibinfo
  {journal} {Phys. Rev. Lett.}\ }\textbf {\bibinfo {volume} {106}},\ \bibinfo
  {pages} {151104} (\bibinfo {year} {2011})},\ \Eprint
  {http://arxiv.org/abs/1101.2868} {arXiv:1101.2868 [gr-qc]} \BibitemShut
  {NoStop}%
\bibitem [{\citenamefont {Kleihaus}\ \emph {et~al.}(2016)\citenamefont
  {Kleihaus}, \citenamefont {Kunz}, \citenamefont {Mojica},\ and\ \citenamefont
  {Radu}}]{Kleihaus:2015aje}%
  \BibitemOpen
  \bibfield  {author} {\bibinfo {author} {\bibfnamefont {B.}~\bibnamefont
  {Kleihaus}}, \bibinfo {author} {\bibfnamefont {J.}~\bibnamefont {Kunz}},
  \bibinfo {author} {\bibfnamefont {S.}~\bibnamefont {Mojica}}, \ and\ \bibinfo
  {author} {\bibfnamefont {E.}~\bibnamefont {Radu}},\ }\href {\doibase
  10.1103/PhysRevD.93.044047} {\bibfield  {journal} {\bibinfo  {journal} {Phys.
  Rev. D}\ }\textbf {\bibinfo {volume} {93}},\ \bibinfo {pages} {044047}
  (\bibinfo {year} {2016})},\ \Eprint {http://arxiv.org/abs/1511.05513}
  {arXiv:1511.05513 [gr-qc]} \BibitemShut {NoStop}%
\bibitem [{\citenamefont {Ripley}\ and\ \citenamefont
  {Pretorius}(2020)}]{Ripley:2019aqj}%
  \BibitemOpen
  \bibfield  {author} {\bibinfo {author} {\bibfnamefont {J.~L.}\ \bibnamefont
  {Ripley}}\ and\ \bibinfo {author} {\bibfnamefont {F.}~\bibnamefont
  {Pretorius}},\ }\href {\doibase 10.1103/PhysRevD.101.044015} {\bibfield
  {journal} {\bibinfo  {journal} {Phys. Rev. D}\ }\textbf {\bibinfo {volume}
  {101}},\ \bibinfo {pages} {044015} (\bibinfo {year} {2020})},\ \Eprint
  {http://arxiv.org/abs/1911.11027} {arXiv:1911.11027 [gr-qc]} \BibitemShut
  {NoStop}%
\bibitem [{\citenamefont {Silva}\ \emph {et~al.}(2018)\citenamefont {Silva},
  \citenamefont {Sakstein}, \citenamefont {Gualtieri}, \citenamefont
  {Sotiriou},\ and\ \citenamefont {Berti}}]{Silva:2017uqg}%
  \BibitemOpen
  \bibfield  {author} {\bibinfo {author} {\bibfnamefont {H.~O.}\ \bibnamefont
  {Silva}}, \bibinfo {author} {\bibfnamefont {J.}~\bibnamefont {Sakstein}},
  \bibinfo {author} {\bibfnamefont {L.}~\bibnamefont {Gualtieri}}, \bibinfo
  {author} {\bibfnamefont {T.~P.}\ \bibnamefont {Sotiriou}}, \ and\ \bibinfo
  {author} {\bibfnamefont {E.}~\bibnamefont {Berti}},\ }\href {\doibase
  10.1103/PhysRevLett.120.131104} {\bibfield  {journal} {\bibinfo  {journal}
  {Phys. Rev. Lett.}\ }\textbf {\bibinfo {volume} {120}},\ \bibinfo {pages}
  {131104} (\bibinfo {year} {2018})},\ \Eprint
  {http://arxiv.org/abs/1711.02080} {arXiv:1711.02080 [gr-qc]} \BibitemShut
  {NoStop}%
\bibitem [{\citenamefont {Doneva}\ and\ \citenamefont
  {Yazadjiev}(2018)}]{Doneva:2017bvd}%
  \BibitemOpen
  \bibfield  {author} {\bibinfo {author} {\bibfnamefont {D.~D.}\ \bibnamefont
  {Doneva}}\ and\ \bibinfo {author} {\bibfnamefont {S.~S.}\ \bibnamefont
  {Yazadjiev}},\ }\href {\doibase 10.1103/PhysRevLett.120.131103} {\bibfield
  {journal} {\bibinfo  {journal} {Phys. Rev. Lett.}\ }\textbf {\bibinfo
  {volume} {120}},\ \bibinfo {pages} {131103} (\bibinfo {year} {2018})},\
  \Eprint {http://arxiv.org/abs/1711.01187} {arXiv:1711.01187 [gr-qc]}
  \BibitemShut {NoStop}%
\bibitem [{\citenamefont {Berti}\ \emph {et~al.}(2020)\citenamefont {Berti},
  \citenamefont {Collodel}, \citenamefont {Kleihaus},\ and\ \citenamefont
  {Kunz}}]{Berti:2020kgk}%
  \BibitemOpen
  \bibfield  {author} {\bibinfo {author} {\bibfnamefont {E.}~\bibnamefont
  {Berti}}, \bibinfo {author} {\bibfnamefont {L.~G.}\ \bibnamefont {Collodel}},
  \bibinfo {author} {\bibfnamefont {B.}~\bibnamefont {Kleihaus}}, \ and\
  \bibinfo {author} {\bibfnamefont {J.}~\bibnamefont {Kunz}},\ }\href@noop {}
  {\  (\bibinfo {year} {2020})},\ \Eprint {http://arxiv.org/abs/2009.03905}
  {arXiv:2009.03905 [gr-qc]} \BibitemShut {NoStop}%
\bibitem [{\citenamefont {Silva}\ \emph {et~al.}(2020)\citenamefont {Silva},
  \citenamefont {Witek}, \citenamefont {Elley},\ and\ \citenamefont
  {Yunes}}]{Silva:2020omi}%
  \BibitemOpen
  \bibfield  {author} {\bibinfo {author} {\bibfnamefont {H.~O.}\ \bibnamefont
  {Silva}}, \bibinfo {author} {\bibfnamefont {H.}~\bibnamefont {Witek}},
  \bibinfo {author} {\bibfnamefont {M.}~\bibnamefont {Elley}}, \ and\ \bibinfo
  {author} {\bibfnamefont {N.}~\bibnamefont {Yunes}},\ }\href@noop {} {\
  (\bibinfo {year} {2020})},\ \Eprint {http://arxiv.org/abs/2012.10436}
  {arXiv:2012.10436 [gr-qc]} \BibitemShut {NoStop}%
\bibitem [{\citenamefont {Dima}\ \emph {et~al.}(2020)\citenamefont {Dima},
  \citenamefont {Barausse}, \citenamefont {Franchini},\ and\ \citenamefont
  {Sotiriou}}]{Dima:2020yac}%
  \BibitemOpen
  \bibfield  {author} {\bibinfo {author} {\bibfnamefont {A.}~\bibnamefont
  {Dima}}, \bibinfo {author} {\bibfnamefont {E.}~\bibnamefont {Barausse}},
  \bibinfo {author} {\bibfnamefont {N.}~\bibnamefont {Franchini}}, \ and\
  \bibinfo {author} {\bibfnamefont {T.~P.}\ \bibnamefont {Sotiriou}},\ }\href
  {\doibase 10.1103/PhysRevLett.125.231101} {\bibfield  {journal} {\bibinfo
  {journal} {Phys. Rev. Lett.}\ }\textbf {\bibinfo {volume} {125}},\ \bibinfo
  {pages} {231101} (\bibinfo {year} {2020})},\ \Eprint
  {http://arxiv.org/abs/2006.03095} {arXiv:2006.03095 [gr-qc]} \BibitemShut
  {NoStop}%
\bibitem [{\citenamefont {Herdeiro}\ \emph {et~al.}(2021)\citenamefont
  {Herdeiro}, \citenamefont {Radu}, \citenamefont {Silva}, \citenamefont
  {Sotiriou},\ and\ \citenamefont {Yunes}}]{Herdeiro:2020wei}%
  \BibitemOpen
  \bibfield  {author} {\bibinfo {author} {\bibfnamefont {C.~A.~R.}\
  \bibnamefont {Herdeiro}}, \bibinfo {author} {\bibfnamefont {E.}~\bibnamefont
  {Radu}}, \bibinfo {author} {\bibfnamefont {H.~O.}\ \bibnamefont {Silva}},
  \bibinfo {author} {\bibfnamefont {T.~P.}\ \bibnamefont {Sotiriou}}, \ and\
  \bibinfo {author} {\bibfnamefont {N.}~\bibnamefont {Yunes}},\ }\href
  {\doibase 10.1103/PhysRevLett.126.011103} {\bibfield  {journal} {\bibinfo
  {journal} {Phys. Rev. Lett.}\ }\textbf {\bibinfo {volume} {126}},\ \bibinfo
  {pages} {011103} (\bibinfo {year} {2021})},\ \Eprint
  {http://arxiv.org/abs/2009.03904} {arXiv:2009.03904 [gr-qc]} \BibitemShut
  {NoStop}%
\bibitem [{\citenamefont {Collodel}\ \emph {et~al.}(2020)\citenamefont
  {Collodel}, \citenamefont {Kleihaus}, \citenamefont {Kunz},\ and\
  \citenamefont {Berti}}]{Collodel:2019kkx}%
  \BibitemOpen
  \bibfield  {author} {\bibinfo {author} {\bibfnamefont {L.~G.}\ \bibnamefont
  {Collodel}}, \bibinfo {author} {\bibfnamefont {B.}~\bibnamefont {Kleihaus}},
  \bibinfo {author} {\bibfnamefont {J.}~\bibnamefont {Kunz}}, \ and\ \bibinfo
  {author} {\bibfnamefont {E.}~\bibnamefont {Berti}},\ }\href {\doibase
  10.1088/1361-6382/ab74f9} {\bibfield  {journal} {\bibinfo  {journal} {Class.
  Quant. Grav.}\ }\textbf {\bibinfo {volume} {37}},\ \bibinfo {pages} {075018}
  (\bibinfo {year} {2020})},\ \Eprint {http://arxiv.org/abs/1912.05382}
  {arXiv:1912.05382 [gr-qc]} \BibitemShut {NoStop}%
\bibitem [{\citenamefont {Doneva}\ and\ \citenamefont
  {Yazadjiev}(2021)}]{Doneva:2021dqn}%
  \BibitemOpen
  \bibfield  {author} {\bibinfo {author} {\bibfnamefont {D.~D.}\ \bibnamefont
  {Doneva}}\ and\ \bibinfo {author} {\bibfnamefont {S.~S.}\ \bibnamefont
  {Yazadjiev}},\ }\href {\doibase 10.1103/PhysRevD.103.064024} {\bibfield
  {journal} {\bibinfo  {journal} {Phys. Rev. D}\ }\textbf {\bibinfo {volume}
  {103}},\ \bibinfo {pages} {064024} (\bibinfo {year} {2021})},\ \Eprint
  {http://arxiv.org/abs/2101.03514} {arXiv:2101.03514 [gr-qc]} \BibitemShut
  {NoStop}%
\bibitem [{\citenamefont {Doneva}\ \emph {et~al.}(2020)\citenamefont {Doneva},
  \citenamefont {Collodel}, \citenamefont {Kr\"uger},\ and\ \citenamefont
  {Yazadjiev}}]{Doneva:2020nbb}%
  \BibitemOpen
  \bibfield  {author} {\bibinfo {author} {\bibfnamefont {D.~D.}\ \bibnamefont
  {Doneva}}, \bibinfo {author} {\bibfnamefont {L.~G.}\ \bibnamefont
  {Collodel}}, \bibinfo {author} {\bibfnamefont {C.~J.}\ \bibnamefont
  {Kr\"uger}}, \ and\ \bibinfo {author} {\bibfnamefont {S.~S.}\ \bibnamefont
  {Yazadjiev}},\ }\href {\doibase 10.1103/PhysRevD.102.104027} {\bibfield
  {journal} {\bibinfo  {journal} {Phys. Rev. D}\ }\textbf {\bibinfo {volume}
  {102}},\ \bibinfo {pages} {104027} (\bibinfo {year} {2020})},\ \Eprint
  {http://arxiv.org/abs/2008.07391} {arXiv:2008.07391 [gr-qc]} \BibitemShut
  {NoStop}%
\bibitem [{\citenamefont {East}\ and\ \citenamefont
  {Ripley}(2021{\natexlab{a}})}]{East:2021bqk}%
  \BibitemOpen
  \bibfield  {author} {\bibinfo {author} {\bibfnamefont {W.~E.}\ \bibnamefont
  {East}}\ and\ \bibinfo {author} {\bibfnamefont {J.~L.}\ \bibnamefont
  {Ripley}},\ }\href@noop {} {\  (\bibinfo {year} {2021}{\natexlab{a}})},\
  \Eprint {http://arxiv.org/abs/2105.08571} {arXiv:2105.08571 [gr-qc]}
  \BibitemShut {NoStop}%
\bibitem [{\citenamefont {Annulli}(2021)}]{Annulli:2021lmn}%
  \BibitemOpen
  \bibfield  {author} {\bibinfo {author} {\bibfnamefont {L.}~\bibnamefont
  {Annulli}},\ }\href@noop {} {\  (\bibinfo {year} {2021})},\ \Eprint
  {http://arxiv.org/abs/2105.08728} {arXiv:2105.08728 [gr-qc]} \BibitemShut
  {NoStop}%
\bibitem [{\citenamefont {Kuan}\ \emph {et~al.}(2021)\citenamefont {Kuan},
  \citenamefont {Doneva},\ and\ \citenamefont {Yazadjiev}}]{Kuan:2021lol}%
  \BibitemOpen
  \bibfield  {author} {\bibinfo {author} {\bibfnamefont {H.-J.}\ \bibnamefont
  {Kuan}}, \bibinfo {author} {\bibfnamefont {D.~D.}\ \bibnamefont {Doneva}}, \
  and\ \bibinfo {author} {\bibfnamefont {S.~S.}\ \bibnamefont {Yazadjiev}},\
  }\href@noop {} {\  (\bibinfo {year} {2021})},\ \Eprint
  {http://arxiv.org/abs/2103.11999} {arXiv:2103.11999 [gr-qc]} \BibitemShut
  {NoStop}%
\bibitem [{\citenamefont {Yagi}(2012)}]{Yagi:2012gp}%
  \BibitemOpen
  \bibfield  {author} {\bibinfo {author} {\bibfnamefont {K.}~\bibnamefont
  {Yagi}},\ }\href {\doibase 10.1103/PhysRevD.86.081504} {\bibfield  {journal}
  {\bibinfo  {journal} {Phys. Rev. D}\ }\textbf {\bibinfo {volume} {86}},\
  \bibinfo {pages} {081504} (\bibinfo {year} {2012})},\ \Eprint
  {http://arxiv.org/abs/1204.4524} {arXiv:1204.4524 [gr-qc]} \BibitemShut
  {NoStop}%
\bibitem [{\citenamefont {Nair}\ \emph {et~al.}(2019)\citenamefont {Nair},
  \citenamefont {Perkins}, \citenamefont {Silva},\ and\ \citenamefont
  {Yunes}}]{Nair:2019iur}%
  \BibitemOpen
  \bibfield  {author} {\bibinfo {author} {\bibfnamefont {R.}~\bibnamefont
  {Nair}}, \bibinfo {author} {\bibfnamefont {S.}~\bibnamefont {Perkins}},
  \bibinfo {author} {\bibfnamefont {H.~O.}\ \bibnamefont {Silva}}, \ and\
  \bibinfo {author} {\bibfnamefont {N.}~\bibnamefont {Yunes}},\ }\href
  {\doibase 10.1103/PhysRevLett.123.191101} {\bibfield  {journal} {\bibinfo
  {journal} {Phys. Rev. Lett.}\ }\textbf {\bibinfo {volume} {123}},\ \bibinfo
  {pages} {191101} (\bibinfo {year} {2019})},\ \Eprint
  {http://arxiv.org/abs/1905.00870} {arXiv:1905.00870 [gr-qc]} \BibitemShut
  {NoStop}%
\bibitem [{\citenamefont {Perkins}\ \emph {et~al.}(2021)\citenamefont
  {Perkins}, \citenamefont {Nair}, \citenamefont {Silva},\ and\ \citenamefont
  {Yunes}}]{Perkins:2021mhb}%
  \BibitemOpen
  \bibfield  {author} {\bibinfo {author} {\bibfnamefont {S.~E.}\ \bibnamefont
  {Perkins}}, \bibinfo {author} {\bibfnamefont {R.}~\bibnamefont {Nair}},
  \bibinfo {author} {\bibfnamefont {H.~O.}\ \bibnamefont {Silva}}, \ and\
  \bibinfo {author} {\bibfnamefont {N.}~\bibnamefont {Yunes}},\ }\href@noop {}
  {\  (\bibinfo {year} {2021})},\ \Eprint {http://arxiv.org/abs/2104.11189}
  {arXiv:2104.11189 [gr-qc]} \BibitemShut {NoStop}%
\bibitem [{\citenamefont {Wang}\ \emph {et~al.}(2021)\citenamefont {Wang},
  \citenamefont {Tang}, \citenamefont {Li}, \citenamefont {Han},\ and\
  \citenamefont {Fan}}]{Wang:2021yll}%
  \BibitemOpen
  \bibfield  {author} {\bibinfo {author} {\bibfnamefont {H.-T.}\ \bibnamefont
  {Wang}}, \bibinfo {author} {\bibfnamefont {S.-P.}\ \bibnamefont {Tang}},
  \bibinfo {author} {\bibfnamefont {P.-C.}\ \bibnamefont {Li}}, \bibinfo
  {author} {\bibfnamefont {M.-Z.}\ \bibnamefont {Han}}, \ and\ \bibinfo
  {author} {\bibfnamefont {Y.-Z.}\ \bibnamefont {Fan}},\ }\href@noop {} {\
  (\bibinfo {year} {2021})},\ \Eprint {http://arxiv.org/abs/2104.07590}
  {arXiv:2104.07590 [gr-qc]} \BibitemShut {NoStop}%
\bibitem [{\citenamefont {Witek}\ \emph {et~al.}(2019)\citenamefont {Witek},
  \citenamefont {Gualtieri}, \citenamefont {Pani},\ and\ \citenamefont
  {Sotiriou}}]{Witek:2018dmd}%
  \BibitemOpen
  \bibfield  {author} {\bibinfo {author} {\bibfnamefont {H.}~\bibnamefont
  {Witek}}, \bibinfo {author} {\bibfnamefont {L.}~\bibnamefont {Gualtieri}},
  \bibinfo {author} {\bibfnamefont {P.}~\bibnamefont {Pani}}, \ and\ \bibinfo
  {author} {\bibfnamefont {T.~P.}\ \bibnamefont {Sotiriou}},\ }\href {\doibase
  10.1103/PhysRevD.99.064035} {\bibfield  {journal} {\bibinfo  {journal} {Phys.
  Rev. D}\ }\textbf {\bibinfo {volume} {99}},\ \bibinfo {pages} {064035}
  (\bibinfo {year} {2019})},\ \Eprint {http://arxiv.org/abs/1810.05177}
  {arXiv:1810.05177 [gr-qc]} \BibitemShut {NoStop}%
\bibitem [{\citenamefont {Okounkova}(2020)}]{Okounkova:2020rqw}%
  \BibitemOpen
  \bibfield  {author} {\bibinfo {author} {\bibfnamefont {M.}~\bibnamefont
  {Okounkova}},\ }\href {\doibase 10.1103/PhysRevD.102.084046} {\bibfield
  {journal} {\bibinfo  {journal} {Phys. Rev. D}\ }\textbf {\bibinfo {volume}
  {102}},\ \bibinfo {pages} {084046} (\bibinfo {year} {2020})},\ \Eprint
  {http://arxiv.org/abs/2001.03571} {arXiv:2001.03571 [gr-qc]} \BibitemShut
  {NoStop}%
\bibitem [{\citenamefont {Kovacs}(2021)}]{Kovacs:2021lgk}%
  \BibitemOpen
  \bibfield  {author} {\bibinfo {author} {\bibfnamefont {A.~D.}\ \bibnamefont
  {Kovacs}},\ }\href@noop {} {\  (\bibinfo {year} {2021})},\ \Eprint
  {http://arxiv.org/abs/2103.06895} {arXiv:2103.06895 [gr-qc]} \BibitemShut
  {NoStop}%
\bibitem [{\citenamefont {East}\ and\ \citenamefont
  {Ripley}(2021{\natexlab{b}})}]{East:2020hgw}%
  \BibitemOpen
  \bibfield  {author} {\bibinfo {author} {\bibfnamefont {W.~E.}\ \bibnamefont
  {East}}\ and\ \bibinfo {author} {\bibfnamefont {J.~L.}\ \bibnamefont
  {Ripley}},\ }\href {\doibase 10.1103/PhysRevD.103.044040} {\bibfield
  {journal} {\bibinfo  {journal} {Phys. Rev. D}\ }\textbf {\bibinfo {volume}
  {103}},\ \bibinfo {pages} {044040} (\bibinfo {year} {2021}{\natexlab{b}})},\
  \Eprint {http://arxiv.org/abs/2011.03547} {arXiv:2011.03547 [gr-qc]}
  \BibitemShut {NoStop}%
\bibitem [{\citenamefont {Bl\'azquez-Salcedo}\ \emph
  {et~al.}(2016)\citenamefont {Bl\'azquez-Salcedo}, \citenamefont {Macedo},
  \citenamefont {Cardoso}, \citenamefont {Ferrari}, \citenamefont {Gualtieri},
  \citenamefont {Khoo}, \citenamefont {Kunz},\ and\ \citenamefont
  {Pani}}]{Blazquez-Salcedo:2016enn}%
  \BibitemOpen
  \bibfield  {author} {\bibinfo {author} {\bibfnamefont {J.~L.}\ \bibnamefont
  {Bl\'azquez-Salcedo}}, \bibinfo {author} {\bibfnamefont {C.~F.~B.}\
  \bibnamefont {Macedo}}, \bibinfo {author} {\bibfnamefont {V.}~\bibnamefont
  {Cardoso}}, \bibinfo {author} {\bibfnamefont {V.}~\bibnamefont {Ferrari}},
  \bibinfo {author} {\bibfnamefont {L.}~\bibnamefont {Gualtieri}}, \bibinfo
  {author} {\bibfnamefont {F.~S.}\ \bibnamefont {Khoo}}, \bibinfo {author}
  {\bibfnamefont {J.}~\bibnamefont {Kunz}}, \ and\ \bibinfo {author}
  {\bibfnamefont {P.}~\bibnamefont {Pani}},\ }\href {\doibase
  10.1103/PhysRevD.94.104024} {\bibfield  {journal} {\bibinfo  {journal} {Phys.
  Rev. D}\ }\textbf {\bibinfo {volume} {94}},\ \bibinfo {pages} {104024}
  (\bibinfo {year} {2016})},\ \Eprint {http://arxiv.org/abs/1609.01286}
  {arXiv:1609.01286 [gr-qc]} \BibitemShut {NoStop}%
\bibitem [{\citenamefont {Pierini}\ and\ \citenamefont
  {Gualtieri}(2021)}]{Pierini:2021jxd}%
  \BibitemOpen
  \bibfield  {author} {\bibinfo {author} {\bibfnamefont {L.}~\bibnamefont
  {Pierini}}\ and\ \bibinfo {author} {\bibfnamefont {L.}~\bibnamefont
  {Gualtieri}},\ }\href@noop {} {\  (\bibinfo {year} {2021})},\ \Eprint
  {http://arxiv.org/abs/2103.09870} {arXiv:2103.09870 [gr-qc]} \BibitemShut
  {NoStop}%
\bibitem [{\citenamefont {Yagi}\ \emph {et~al.}(2012)\citenamefont {Yagi},
  \citenamefont {Stein}, \citenamefont {Yunes},\ and\ \citenamefont
  {Tanaka}}]{Yagi:2011xp}%
  \BibitemOpen
  \bibfield  {author} {\bibinfo {author} {\bibfnamefont {K.}~\bibnamefont
  {Yagi}}, \bibinfo {author} {\bibfnamefont {L.~C.}\ \bibnamefont {Stein}},
  \bibinfo {author} {\bibfnamefont {N.}~\bibnamefont {Yunes}}, \ and\ \bibinfo
  {author} {\bibfnamefont {T.}~\bibnamefont {Tanaka}},\ }\href {\doibase
  10.1103/PhysRevD.85.064022} {\bibfield  {journal} {\bibinfo  {journal} {Phys.
  Rev. D}\ }\textbf {\bibinfo {volume} {85}},\ \bibinfo {pages} {064022}
  (\bibinfo {year} {2012})},\ \bibinfo {note} {[Erratum: Phys.Rev.D 93, 029902
  (2016)]},\ \Eprint {http://arxiv.org/abs/1110.5950} {arXiv:1110.5950 [gr-qc]}
  \BibitemShut {NoStop}%
\bibitem [{\citenamefont {Juli\'e}\ and\ \citenamefont
  {Berti}(2019)}]{Julie:2019sab}%
  \BibitemOpen
  \bibfield  {author} {\bibinfo {author} {\bibfnamefont {F.-L.}\ \bibnamefont
  {Juli\'e}}\ and\ \bibinfo {author} {\bibfnamefont {E.}~\bibnamefont
  {Berti}},\ }\href {\doibase 10.1103/PhysRevD.100.104061} {\bibfield
  {journal} {\bibinfo  {journal} {Phys. Rev. D}\ }\textbf {\bibinfo {volume}
  {100}},\ \bibinfo {pages} {104061} (\bibinfo {year} {2019})},\ \Eprint
  {http://arxiv.org/abs/1909.05258} {arXiv:1909.05258 [gr-qc]} \BibitemShut
  {NoStop}%
\bibitem [{\citenamefont {{Epstein}}\ and\ \citenamefont
  {{Wagoner}}(1975)}]{EWpaper}%
  \BibitemOpen
  \bibfield  {author} {\bibinfo {author} {\bibfnamefont {R.}~\bibnamefont
  {{Epstein}}}\ and\ \bibinfo {author} {\bibfnamefont {R.~V.}\ \bibnamefont
  {{Wagoner}}},\ }\href {\doibase 10.1086/153561} {\bibfield  {journal}
  {\bibinfo  {journal} {\apj}\ }\textbf {\bibinfo {volume} {197}},\ \bibinfo
  {pages} {717} (\bibinfo {year} {1975})}\BibitemShut {NoStop}%
\bibitem [{\citenamefont {Will}\ and\ \citenamefont
  {Wiseman}(1996)}]{Will:1996zj}%
  \BibitemOpen
  \bibfield  {author} {\bibinfo {author} {\bibfnamefont {C.~M.}\ \bibnamefont
  {Will}}\ and\ \bibinfo {author} {\bibfnamefont {A.~G.}\ \bibnamefont
  {Wiseman}},\ }\href {\doibase 10.1103/PhysRevD.54.4813} {\bibfield  {journal}
  {\bibinfo  {journal} {Phys. Rev. D}\ }\textbf {\bibinfo {volume} {54}},\
  \bibinfo {pages} {4813} (\bibinfo {year} {1996})},\ \Eprint
  {http://arxiv.org/abs/gr-qc/9608012} {arXiv:gr-qc/9608012} \BibitemShut
  {NoStop}%
\bibitem [{\citenamefont {Pati}\ and\ \citenamefont
  {Will}(2000)}]{Pati:2000vt}%
  \BibitemOpen
  \bibfield  {author} {\bibinfo {author} {\bibfnamefont {M.~E.}\ \bibnamefont
  {Pati}}\ and\ \bibinfo {author} {\bibfnamefont {C.~M.}\ \bibnamefont
  {Will}},\ }\href {\doibase 10.1103/PhysRevD.62.124015} {\bibfield  {journal}
  {\bibinfo  {journal} {Phys. Rev. D}\ }\textbf {\bibinfo {volume} {62}},\
  \bibinfo {pages} {124015} (\bibinfo {year} {2000})},\ \Eprint
  {http://arxiv.org/abs/gr-qc/0007087} {arXiv:gr-qc/0007087} \BibitemShut
  {NoStop}%
\bibitem [{\citenamefont {Pati}\ and\ \citenamefont
  {Will}(2002)}]{Pati:2002ux}%
  \BibitemOpen
  \bibfield  {author} {\bibinfo {author} {\bibfnamefont {M.~E.}\ \bibnamefont
  {Pati}}\ and\ \bibinfo {author} {\bibfnamefont {C.~M.}\ \bibnamefont
  {Will}},\ }\href {\doibase 10.1103/PhysRevD.65.104008} {\bibfield  {journal}
  {\bibinfo  {journal} {Phys. Rev. D}\ }\textbf {\bibinfo {volume} {65}},\
  \bibinfo {pages} {104008} (\bibinfo {year} {2002})},\ \Eprint
  {http://arxiv.org/abs/gr-qc/0201001} {arXiv:gr-qc/0201001} \BibitemShut
  {NoStop}%
\bibitem [{\citenamefont {{Landau}}\ and\ \citenamefont
  {{Lifshitz}}(1975)}]{1975ctf..book.....L}%
  \BibitemOpen
  \bibfield  {author} {\bibinfo {author} {\bibfnamefont {L.~D.}\ \bibnamefont
  {{Landau}}}\ and\ \bibinfo {author} {\bibfnamefont {E.~M.}\ \bibnamefont
  {{Lifshitz}}},\ }\href@noop {} {\emph {\bibinfo {title} {{The classical
  theory of fields}}}}\ (\bibinfo {year} {1975})\BibitemShut {NoStop}%
\bibitem [{\citenamefont {Blanchet}(2014)}]{Blanchet:2013haa}%
  \BibitemOpen
  \bibfield  {author} {\bibinfo {author} {\bibfnamefont {L.}~\bibnamefont
  {Blanchet}},\ }\href {\doibase 10.12942/lrr-2014-2} {\bibfield  {journal}
  {\bibinfo  {journal} {Living Rev. Rel.}\ }\textbf {\bibinfo {volume} {17}},\
  \bibinfo {pages} {2} (\bibinfo {year} {2014})},\ \Eprint
  {http://arxiv.org/abs/1310.1528} {arXiv:1310.1528 [gr-qc]} \BibitemShut
  {NoStop}%
\bibitem [{\citenamefont {Damour}\ and\ \citenamefont
  {Esposito-Farese}(1993)}]{Damour:1993hw}%
  \BibitemOpen
  \bibfield  {author} {\bibinfo {author} {\bibfnamefont {T.}~\bibnamefont
  {Damour}}\ and\ \bibinfo {author} {\bibfnamefont {G.}~\bibnamefont
  {Esposito-Farese}},\ }\href {\doibase 10.1103/PhysRevLett.70.2220} {\bibfield
   {journal} {\bibinfo  {journal} {Phys. Rev. Lett.}\ }\textbf {\bibinfo
  {volume} {70}},\ \bibinfo {pages} {2220} (\bibinfo {year}
  {1993})}\BibitemShut {NoStop}%
\bibitem [{\citenamefont {Bernard}(2018)}]{Bernard:2018hta}%
  \BibitemOpen
  \bibfield  {author} {\bibinfo {author} {\bibfnamefont {L.}~\bibnamefont
  {Bernard}},\ }\href {\doibase 10.1103/PhysRevD.98.044004} {\bibfield
  {journal} {\bibinfo  {journal} {Phys. Rev. D}\ }\textbf {\bibinfo {volume}
  {98}},\ \bibinfo {pages} {044004} (\bibinfo {year} {2018})},\ \Eprint
  {http://arxiv.org/abs/1802.10201} {arXiv:1802.10201 [gr-qc]} \BibitemShut
  {NoStop}%
\bibitem [{\citenamefont {Mirshekari}\ and\ \citenamefont
  {Will}(2013)}]{Mirshekari:2013vb}%
  \BibitemOpen
  \bibfield  {author} {\bibinfo {author} {\bibfnamefont {S.}~\bibnamefont
  {Mirshekari}}\ and\ \bibinfo {author} {\bibfnamefont {C.~M.}\ \bibnamefont
  {Will}},\ }\href {\doibase 10.1103/PhysRevD.87.084070} {\bibfield  {journal}
  {\bibinfo  {journal} {Phys. Rev. D}\ }\textbf {\bibinfo {volume} {87}},\
  \bibinfo {pages} {084070} (\bibinfo {year} {2013})},\ \Eprint
  {http://arxiv.org/abs/1301.4680} {arXiv:1301.4680 [gr-qc]} \BibitemShut
  {NoStop}%
\bibitem [{\citenamefont {Lang}(2014)}]{Lang:2013fna}%
  \BibitemOpen
  \bibfield  {author} {\bibinfo {author} {\bibfnamefont {R.~N.}\ \bibnamefont
  {Lang}},\ }\href {\doibase 10.1103/PhysRevD.89.084014} {\bibfield  {journal}
  {\bibinfo  {journal} {Phys. Rev. D}\ }\textbf {\bibinfo {volume} {89}},\
  \bibinfo {pages} {084014} (\bibinfo {year} {2014})},\ \Eprint
  {http://arxiv.org/abs/1310.3320} {arXiv:1310.3320 [gr-qc]} \BibitemShut
  {NoStop}%
\bibitem [{\citenamefont {Sennett}\ \emph {et~al.}(2016)\citenamefont
  {Sennett}, \citenamefont {Marsat},\ and\ \citenamefont
  {Buonanno}}]{Sennett:2016klh}%
  \BibitemOpen
  \bibfield  {author} {\bibinfo {author} {\bibfnamefont {N.}~\bibnamefont
  {Sennett}}, \bibinfo {author} {\bibfnamefont {S.}~\bibnamefont {Marsat}}, \
  and\ \bibinfo {author} {\bibfnamefont {A.}~\bibnamefont {Buonanno}},\ }\href
  {\doibase 10.1103/PhysRevD.94.084003} {\bibfield  {journal} {\bibinfo
  {journal} {Phys. Rev. D}\ }\textbf {\bibinfo {volume} {94}},\ \bibinfo
  {pages} {084003} (\bibinfo {year} {2016})},\ \Eprint
  {http://arxiv.org/abs/1607.01420} {arXiv:1607.01420 [gr-qc]} \BibitemShut
  {NoStop}%
\bibitem [{\citenamefont {Lang}(2015)}]{Lang:2014osa}%
  \BibitemOpen
  \bibfield  {author} {\bibinfo {author} {\bibfnamefont {R.~N.}\ \bibnamefont
  {Lang}},\ }\href {\doibase 10.1103/PhysRevD.91.084027} {\bibfield  {journal}
  {\bibinfo  {journal} {Phys. Rev. D}\ }\textbf {\bibinfo {volume} {91}},\
  \bibinfo {pages} {084027} (\bibinfo {year} {2015})},\ \Eprint
  {http://arxiv.org/abs/1411.3073} {arXiv:1411.3073 [gr-qc]} \BibitemShut
  {NoStop}%
\bibitem [{\citenamefont {Shiralilou}\ \emph {et~al.}(2020)\citenamefont
  {Shiralilou}, \citenamefont {Hinderer}, \citenamefont {Nissanke},
  \citenamefont {Ortiz},\ and\ \citenamefont {Witek}}]{Shiralilou:2020gah}%
  \BibitemOpen
  \bibfield  {author} {\bibinfo {author} {\bibfnamefont {B.}~\bibnamefont
  {Shiralilou}}, \bibinfo {author} {\bibfnamefont {T.}~\bibnamefont
  {Hinderer}}, \bibinfo {author} {\bibfnamefont {S.}~\bibnamefont {Nissanke}},
  \bibinfo {author} {\bibfnamefont {N.}~\bibnamefont {Ortiz}}, \ and\ \bibinfo
  {author} {\bibfnamefont {H.}~\bibnamefont {Witek}},\ }\href@noop {} {\
  (\bibinfo {year} {2020})},\ \Eprint {http://arxiv.org/abs/2012.09162}
  {arXiv:2012.09162 [gr-qc]} \BibitemShut {NoStop}%
\bibitem [{\citenamefont {Tahura}\ and\ \citenamefont
  {Yagi}(2018)}]{Tahura:2018zuq}%
  \BibitemOpen
  \bibfield  {author} {\bibinfo {author} {\bibfnamefont {S.}~\bibnamefont
  {Tahura}}\ and\ \bibinfo {author} {\bibfnamefont {K.}~\bibnamefont {Yagi}},\
  }\href {\doibase 10.1103/PhysRevD.98.084042} {\bibfield  {journal} {\bibinfo
  {journal} {Phys. Rev. D}\ }\textbf {\bibinfo {volume} {98}},\ \bibinfo
  {pages} {084042} (\bibinfo {year} {2018})},\ \bibinfo {note} {[Erratum:
  Phys.Rev.D 101, 109902 (2020)]},\ \Eprint {http://arxiv.org/abs/1809.00259}
  {arXiv:1809.00259 [gr-qc]} \BibitemShut {NoStop}%
\bibitem [{\citenamefont {Carson}\ \emph {et~al.}(2020)\citenamefont {Carson},
  \citenamefont {Seymour},\ and\ \citenamefont {Yagi}}]{Carson:2019fxr}%
  \BibitemOpen
  \bibfield  {author} {\bibinfo {author} {\bibfnamefont {Z.}~\bibnamefont
  {Carson}}, \bibinfo {author} {\bibfnamefont {B.~C.}\ \bibnamefont {Seymour}},
  \ and\ \bibinfo {author} {\bibfnamefont {K.}~\bibnamefont {Yagi}},\ }\href
  {\doibase 10.1088/1361-6382/ab6a1f} {\bibfield  {journal} {\bibinfo
  {journal} {Class. Quant. Grav.}\ }\textbf {\bibinfo {volume} {37}},\ \bibinfo
  {pages} {065008} (\bibinfo {year} {2020})},\ \Eprint
  {http://arxiv.org/abs/1907.03897} {arXiv:1907.03897 [gr-qc]} \BibitemShut
  {NoStop}%
\bibitem [{\citenamefont {Juli\'e}\ and\ \citenamefont
  {Berti}(2020)}]{Julie:2020vov}%
  \BibitemOpen
  \bibfield  {author} {\bibinfo {author} {\bibfnamefont {F.-L.}\ \bibnamefont
  {Juli\'e}}\ and\ \bibinfo {author} {\bibfnamefont {E.}~\bibnamefont
  {Berti}},\ }\href {\doibase 10.1103/PhysRevD.101.124045} {\bibfield
  {journal} {\bibinfo  {journal} {Phys. Rev. D}\ }\textbf {\bibinfo {volume}
  {101}},\ \bibinfo {pages} {124045} (\bibinfo {year} {2020})},\ \Eprint
  {http://arxiv.org/abs/2004.00003} {arXiv:2004.00003 [gr-qc]} \BibitemShut
  {NoStop}%
\bibitem [{\citenamefont {Metsaev}\ and\ \citenamefont
  {Tseytlin}(1987)}]{Metsaev:1987zx}%
  \BibitemOpen
  \bibfield  {author} {\bibinfo {author} {\bibfnamefont {R.~R.}\ \bibnamefont
  {Metsaev}}\ and\ \bibinfo {author} {\bibfnamefont {A.~A.}\ \bibnamefont
  {Tseytlin}},\ }\href {\doibase 10.1016/0550-3213(87)90077-0} {\bibfield
  {journal} {\bibinfo  {journal} {Nucl. Phys. B}\ }\textbf {\bibinfo {volume}
  {293}},\ \bibinfo {pages} {385} (\bibinfo {year} {1987})}\BibitemShut
  {NoStop}%
\bibitem [{\citenamefont {{Eardley}}(1975{\natexlab{a}})}]{Eardley}%
  \BibitemOpen
  \bibfield  {author} {\bibinfo {author} {\bibfnamefont {D.~M.}\ \bibnamefont
  {{Eardley}}},\ }\href {\doibase 10.1086/181744} {\bibfield  {journal}
  {\bibinfo  {journal} {The Astrophysical Journal}\ }\textbf {\bibinfo {volume} {196}},\ \bibinfo
  {pages} {L59} (\bibinfo {year} {1975}{\natexlab{a}})}\BibitemShut {NoStop}%
\bibitem [{\citenamefont
  {{Eardley}}(1975{\natexlab{b}})}]{1975ApJ...196L..59E}%
  \BibitemOpen
  \bibfield  {author} {\bibinfo {author} {\bibfnamefont {D.~M.}\ \bibnamefont
  {{Eardley}}},\ }\href {\doibase 10.1086/181744} {\bibfield  {journal}
  {\bibinfo  {journal} {The Astrophysical Journal}\ }\textbf {\bibinfo {volume} {196}},\ \bibinfo
  {pages} {L59} (\bibinfo {year} {1975}{\natexlab{b}})}\BibitemShut {NoStop}%
\bibitem [{\citenamefont {Damour}\ and\ \citenamefont
  {Esposito-Farese}(1992{\natexlab{a}})}]{Damour:1992we}%
  \BibitemOpen
  \bibfield  {author} {\bibinfo {author} {\bibfnamefont {T.}~\bibnamefont
  {Damour}}\ and\ \bibinfo {author} {\bibfnamefont {G.}~\bibnamefont
  {Esposito-Farese}},\ }\href {\doibase 10.1088/0264-9381/9/9/015} {\bibfield
  {journal} {\bibinfo  {journal} {Class. Quant. Grav.}\ }\textbf {\bibinfo
  {volume} {9}},\ \bibinfo {pages} {2093} (\bibinfo {year}
  {1992}{\natexlab{a}})}\BibitemShut {NoStop}%
\bibitem [{\citenamefont {Juli\'e}(2018)}]{Julie:2017rpw}%
  \BibitemOpen
  \bibfield  {author} {\bibinfo {author} {\bibfnamefont {F.-L.}\ \bibnamefont
  {Juli\'e}},\ }\href {\doibase 10.1088/1475-7516/2018/01/026} {\bibfield
  {journal} {\bibinfo  {journal} {JCAP}\ }\textbf {\bibinfo {volume} {01}},\
  \bibinfo {pages} {026} (\bibinfo {year} {2018})},\ \Eprint
  {http://arxiv.org/abs/1711.10769} {arXiv:1711.10769 [gr-qc]} \BibitemShut
  {NoStop}%
\bibitem [{\citenamefont {Flanagan}\ and\ \citenamefont
  {Hinderer}(2008)}]{Flanagan:2007ix}%
  \BibitemOpen
  \bibfield  {author} {\bibinfo {author} {\bibfnamefont {E.~E.}\ \bibnamefont
  {Flanagan}}\ and\ \bibinfo {author} {\bibfnamefont {T.}~\bibnamefont
  {Hinderer}},\ }\href {\doibase 10.1103/PhysRevD.77.021502} {\bibfield
  {journal} {\bibinfo  {journal} {Phys. Rev. D}\ }\textbf {\bibinfo {volume}
  {77}},\ \bibinfo {pages} {021502} (\bibinfo {year} {2008})},\ \Eprint
  {http://arxiv.org/abs/0709.1915} {arXiv:0709.1915 [astro-ph]} \BibitemShut
  {NoStop}%
\bibitem [{\citenamefont {Flanagan}(1998)}]{Flanagan:1997fn}%
  \BibitemOpen
  \bibfield  {author} {\bibinfo {author} {\bibfnamefont {E.~E.}\ \bibnamefont
  {Flanagan}},\ }\href {\doibase 10.1103/PhysRevD.58.124030} {\bibfield
  {journal} {\bibinfo  {journal} {Phys. Rev. D}\ }\textbf {\bibinfo {volume}
  {58}},\ \bibinfo {pages} {124030} (\bibinfo {year} {1998})},\ \Eprint
  {http://arxiv.org/abs/gr-qc/9706045} {arXiv:gr-qc/9706045} \BibitemShut
  {NoStop}%
\bibitem [{\citenamefont {Damour}\ and\ \citenamefont
  {Esposito-Farese}(1992{\natexlab{b}})}]{Damour:1992w}%
  \BibitemOpen
  \bibfield  {author} {\bibinfo {author} {\bibfnamefont {T.}~\bibnamefont
  {Damour}}\ and\ \bibinfo {author} {\bibfnamefont {G.}~\bibnamefont
  {Esposito-Farese}},\ }\href {\doibase 10.1088/0264-9381/9/9/015} {\bibfield
  {journal} {\bibinfo  {journal} {Class. Quant. Grav.}\ }\textbf {\bibinfo
  {volume} {9}},\ \bibinfo {pages} {2093} (\bibinfo {year}
  {1992}{\natexlab{b}})}\BibitemShut {NoStop}%
\bibitem [{\citenamefont {Buonanno}\ \emph {et~al.}(2009)\citenamefont
  {Buonanno}, \citenamefont {Iyer}, \citenamefont {Ochsner}, \citenamefont
  {Pan},\ and\ \citenamefont {Sathyaprakash}}]{Buonanno:2009zt}%
  \BibitemOpen
  \bibfield  {author} {\bibinfo {author} {\bibfnamefont {A.}~\bibnamefont
  {Buonanno}}, \bibinfo {author} {\bibfnamefont {B.}~\bibnamefont {Iyer}},
  \bibinfo {author} {\bibfnamefont {E.}~\bibnamefont {Ochsner}}, \bibinfo
  {author} {\bibfnamefont {Y.}~\bibnamefont {Pan}}, \ and\ \bibinfo {author}
  {\bibfnamefont {B.}~\bibnamefont {Sathyaprakash}},\ }\href {\doibase
  10.1103/PhysRevD.80.084043} {\bibfield  {journal} {\bibinfo  {journal} {Phys.
  Rev. D}\ }\textbf {\bibinfo {volume} {80}},\ \bibinfo {pages} {084043}
  (\bibinfo {year} {2009})},\ \Eprint {http://arxiv.org/abs/0907.0700}
  {arXiv:0907.0700 [gr-qc]} \BibitemShut {NoStop}%
\bibitem [{\citenamefont {Damour}\ \emph {et~al.}(2012)\citenamefont {Damour},
  \citenamefont {Nagar},\ and\ \citenamefont {Villain}}]{Damour:2012yf}%
  \BibitemOpen
  \bibfield  {author} {\bibinfo {author} {\bibfnamefont {T.}~\bibnamefont
  {Damour}}, \bibinfo {author} {\bibfnamefont {A.}~\bibnamefont {Nagar}}, \
  and\ \bibinfo {author} {\bibfnamefont {L.}~\bibnamefont {Villain}},\ }\href
  {\doibase 10.1103/PhysRevD.85.123007} {\bibfield  {journal} {\bibinfo
  {journal} {Phys. Rev. D}\ }\textbf {\bibinfo {volume} {85}},\ \bibinfo
  {pages} {123007} (\bibinfo {year} {2012})},\ \Eprint
  {http://arxiv.org/abs/1203.4352} {arXiv:1203.4352 [gr-qc]} \BibitemShut
  {NoStop}%
\bibitem [{\citenamefont {Aasi}\ \emph {et~al.}(2015)\citenamefont {Aasi} \emph
  {et~al.}}]{TheLIGOScientific:2014jea}%
  \BibitemOpen
  \bibfield  {author} {\bibinfo {author} {\bibfnamefont {J.}~\bibnamefont
  {Aasi}} \emph {et~al.} (\bibinfo {collaboration} {LIGO Scientific}),\ }\href
  {\doibase 10.1088/0264-9381/32/7/074001} {\bibfield  {journal} {\bibinfo
  {journal} {Class. Quant. Grav.}\ }\textbf {\bibinfo {volume} {32}},\ \bibinfo
  {pages} {074001} (\bibinfo {year} {2015})},\ \Eprint
  {http://arxiv.org/abs/1411.4547} {arXiv:1411.4547 [gr-qc]} \BibitemShut
  {NoStop}%
\bibitem [{\citenamefont {Acernese}\ \emph {et~al.}(2015)\citenamefont
  {Acernese} \emph {et~al.}}]{TheVirgo:2014hva}%
  \BibitemOpen
  \bibfield  {author} {\bibinfo {author} {\bibfnamefont {F.}~\bibnamefont
  {Acernese}} \emph {et~al.} (\bibinfo {collaboration} {VIRGO}),\ }\href
  {\doibase 10.1088/0264-9381/32/2/024001} {\bibfield  {journal} {\bibinfo
  {journal} {Class. Quant. Grav.}\ }\textbf {\bibinfo {volume} {32}},\ \bibinfo
  {pages} {024001} (\bibinfo {year} {2015})},\ \Eprint
  {http://arxiv.org/abs/1408.3978} {arXiv:1408.3978 [gr-qc]} \BibitemShut
  {NoStop}%
\bibitem [{\citenamefont {Akutsu}\ \emph {et~al.}(2020)\citenamefont {Akutsu}
  \emph {et~al.}}]{Akutsu:2020his}%
  \BibitemOpen
  \bibfield  {author} {\bibinfo {author} {\bibfnamefont {T.}~\bibnamefont
  {Akutsu}} \emph {et~al.} (\bibinfo {collaboration} {KAGRA}),\ }\href@noop {}
  {\  (\bibinfo {year} {2020})},\ \Eprint {http://arxiv.org/abs/2005.05574}
  {arXiv:2005.05574 [physics.ins-det]} \BibitemShut {NoStop}%
\bibitem [{\citenamefont {Punturo}\ \emph {et~al.}(2010)\citenamefont {Punturo}
  \emph {et~al.}}]{Punturo:2010zza}%
  \BibitemOpen
  \bibfield  {author} {\bibinfo {author} {\bibfnamefont {M.}~\bibnamefont
  {Punturo}} \emph {et~al.},\ }\href {\doibase 10.1088/0264-9381/27/8/084007}
  {\bibfield  {journal} {\bibinfo  {journal} {Class. Quant. Grav.}\ }\textbf
  {\bibinfo {volume} {27}},\ \bibinfo {pages} {084007} (\bibinfo {year}
  {2010})}\BibitemShut {NoStop}%
\bibitem [{\citenamefont {Abbott}\ \emph {et~al.}(2017)\citenamefont {Abbott}
  \emph {et~al.}}]{Evans:2016mbw}%
  \BibitemOpen
  \bibfield  {author} {\bibinfo {author} {\bibfnamefont {B.~P.}\ \bibnamefont
  {Abbott}} \emph {et~al.} (\bibinfo {collaboration} {LIGO Scientific}),\
  }\href {\doibase 10.1088/1361-6382/aa51f4} {\bibfield  {journal} {\bibinfo
  {journal} {Class. Quant. Grav.}\ }\textbf {\bibinfo {volume} {34}},\ \bibinfo
  {pages} {044001} (\bibinfo {year} {2017})},\ \Eprint
  {http://arxiv.org/abs/1607.08697} {arXiv:1607.08697 [astro-ph.IM]}
  \BibitemShut {NoStop}%
\bibitem [{\citenamefont {Amaro-Seoane}\ \emph {et~al.}(2017)\citenamefont
  {Amaro-Seoane} \emph {et~al.}}]{Audley:2017drz}%
  \BibitemOpen
  \bibfield  {author} {\bibinfo {author} {\bibfnamefont {P.}~\bibnamefont
  {Amaro-Seoane}} \emph {et~al.} (\bibinfo {collaboration} {LISA}),\
  }\href@noop {} {\  (\bibinfo {year} {2017})},\ \Eprint
  {http://arxiv.org/abs/1702.00786} {arXiv:1702.00786 [astro-ph.IM]}
  \BibitemShut {NoStop}%
\bibitem [{\citenamefont {Brizuela}\ \emph {et~al.}(2009)\citenamefont
  {Brizuela}, \citenamefont {Martin-Garcia},\ and\ \citenamefont
  {Mena~Marugan}}]{Brizuela:2008ra}%
  \BibitemOpen
  \bibfield  {author} {\bibinfo {author} {\bibfnamefont {D.}~\bibnamefont
  {Brizuela}}, \bibinfo {author} {\bibfnamefont {J.~M.}\ \bibnamefont
  {Martin-Garcia}}, \ and\ \bibinfo {author} {\bibfnamefont {G.~A.}\
  \bibnamefont {Mena~Marugan}},\ }\href {\doibase 10.1007/s10714-009-0773-2}
  {\bibfield  {journal} {\bibinfo  {journal} {Gen. Rel. Grav.}\ }\textbf
  {\bibinfo {volume} {41}},\ \bibinfo {pages} {2415} (\bibinfo {year}
  {2009})},\ \Eprint {http://arxiv.org/abs/0807.0824} {arXiv:0807.0824 [gr-qc]}
  \BibitemShut {NoStop}%
\end{thebibliography}

%

\end{document}